\documentclass[12pt]{article}
\usepackage{graphicx}
\usepackage{amsmath}

\begin{document}

\hfill VPI-IPNAS-08-01

\hfill ILL-TH-08-1

\vspace{0.5in}

\begin{center}
{\large\bf A-twisted Landau-Ginzburg models}

\vspace{0.1in}

Josh Guffin$^1$, Eric Sharpe$^2$ \\

$\,$

\begin{tabular}{cc}
  \begin{tabular}{rl}
	 $^1$ \hspace*{-5.5mm} & Department of Physics\\
	                       & University of Illinois, Urbana-Champaign \\
	                       & 1110 West Green Street \\
	                       & Urbana, IL  61801-3080\\
  \end{tabular} &
  \begin{tabular}{rl}
	 $^2$ \hspace*{-5mm} & Physics Department\\
	                     & Robeson Hall (0435) \\
	                     & Virginia Tech\\
	                     & Blacksburg, VA  24061\\
  \end{tabular}
\end{tabular}

{\tt guffin@uiuc.edu}, 
{\tt ersharpe@vt.edu} \\

$\,$

\end{center}

%In this paper we discuss correlation functions in certain A-twisted
%Landau-Ginzburg models.  B-twisted Landau-Ginzburg models have been
%discussed extensively in the literature, but virtually no work has been
%done on A-twisted theories.
%We study here examples of Landau-Ginzburg models over topologically
%nontrivial spaces, not just vector spaces, and away from large-radius
%limits, so that one expects nontrivial curve corrections.  By studying
%examples of Landau-Ginzburg models which are in the same universality class
%as nonlinear sigma models on nontrivial Calabi-Yau's, we are able to get
%some nontrivial tests of our methods, as well as a physical realization of
%some very simple examples of virtual fundamental class computations. 
%
In this paper we discuss correlation functions in certain A-twisted
Landau-Ginzburg models.  Although B-twisted Landau-Ginzburg models have been
discussed extensively in the literature, virtually no work has been
done on A-twisted theories.
In particular, we study examples of Landau-Ginzburg models over topologically
nontrivial spaces -- not just vector spaces -- away from large-radius
limits, so that one expects nontrivial curve corrections.  By studying
examples of Landau-Ginzburg models in the same universality class
as nonlinear sigma models on nontrivial Calabi-Yaus, we obtain 
nontrivial tests of our methods as well as a physical realization of
some simple examples of virtual fundamental class computations.

\begin{flushleft}
January 2008
\end{flushleft}

\newpage

\tableofcontents

\newpage

\section{Introduction}

Landau-Ginzburg models have become a standard set of models in string
compactifications.  Their B model topological twistings have been studied
extensively.  However, there has been very little work on A model
topological twistings of Landau-Ginzburg models.  In this paper we study
such A-twistings.

We are interested in Landau-Ginzburg theories defined over topologically
nontrivial noncompact spaces, rather than merely Landau-Ginzburg models
over vector spaces, and we will not work at infinite-radius limits of the
underlying space.  For such models, away from large-$|r|$ limits,
one does expect nontrivial curve corrections,
hence A-model twistings should give interesting
information.

Let us speak to a potential language confusion.  In the physics literature,
the term ``Landau-Ginzburg'' has often only been applied to (ungauged)
linear sigma models on vector spaces with a superpotential.  The phrase
``hybrid Landau-Ginzburg'' has been used to describe phases of gauged
linear sigma models (GLSMs) involving nonlinear sigma models (NLSMs) with
superpotential on stacks (as we shall describe in more detail later in this
paper).  NLSMs with superpotentials on general spaces have
only rarely been considered in the physics community.  In the math
community, on the other hand, the language usage is often different:  the term
``Landau-Ginzburg'' is often used to describe a NLSM with superpotential on
any space, not just a vector space, and the term ``hybrid Landau-Ginzburg''
does not seem to be used at all.  To be clear, in this paper, we shall
use the term ``Landau-Ginzburg'' to describe NLSMs with superpotential
on general spaces and stacks.  We will also sometimes use the term ``hybrid
Landau-Ginzburg,'' though exclusively to describe NLSMs with superpotentials
on stacks arising as phases of GLSMs.

We will check our methods by comparing correlation functions in A-twisted
Landau-Ginzburg models to those in NLSMs in the same universality class.
In doing so, we will find some interesting results, including a physical
realization of some of Kontsevich's tricks for computing Gromov-Witten
invariants and simple virtual fundamental class
computations.  To physically realize the virtual fundamental class
computations appearing in Gromov-Witten theory, one would need to consider
theories coupled to worldsheet gravity, and we do not perform such a
coupling.  Nevertheless, our results suggest that if one were to closely
analyze the physics of Landau-Ginzburg theories coupled to topological
gravity, one should be able to give a physical derivation of the virtual
fundamental class constructions used in Gromov-Witten theory. 

We begin in section~\ref{stdbrev} with a review of untwisted and B-twisted
Landau-Ginzburg models over general spaces.  As quantum field
theories, these are supersymmetric NLSMs with superpotential.  We describe
the chiral rings of the B-twisted theories in detail as certain
hypercohomology groups, and then check that the general description
correctly specializes in various well-known examples.

In section~\ref{atwist22} we discuss A-twisted Landau-Ginzburg models.
The A-twisting of the underlying NLSM does not
naively lift to the theory with the superpotential -- twisting the
Yukawa couplings breaks worldsheet Lorentz invariance.  There are at
least two workarounds.  We spend the bulk of the section
discussing shifts the twist using an isometry of the underlying space
that lifts to the superpotential.  We apply this method to A twist
examples of Landau-Ginzburg models in the same universality class
as NLSMs on nontrivial Calabi-Yaus.
Since the topological field theory is invariant across
universality classes of renormalization group flow, the correlation
functions of the (A-twisted)
Landau-Ginzburg theory must match those of the
corresponding (A-twisted) NLSM, and this provides
some thorough checks of our methods.  
The Landau-Ginzburg correlation functions give the same results but
the details of the computations are different.  In particular,
the Landau-Ginzburg computations give, to our knowledge, the
first physical realization of some
very simple virtual fundamental class computations, and also
nearly\footnote{
The trick in question, turning an integral over a moduli space of
curves in a hypersurface in a toric variety
into an integral over a moduli space of curves
in the ambient toric variety, was also physically realized in
\cite{schwarz}, which described A-twisted sigma models on supermanifolds.
However, although it was argued there that some subset of the
A model correlation functions on the supermanifold matched certain of
those of A-twisted NLSMs, it is not clear whether
all A model data, much less the full conformal field theory, should match.
In the present case, the CFT associated to the Landau-Ginzburg model does
match that of the NLSM.
In addition, the paper \cite{daveronen} made an ansatz for computing
quintic correlation functions in terms of those on ${\bf P}^4$, that
is mathematically equivalent to the trick in question; however, they did not
claim to have any sort of physical derivation of that ansatz.
} the first physical realizations of
some tricks of Kontsevich \cite{kont2}. 
We also describe related
computations in corresponding GLSMs -- although
not directly pertinent, the same methods we apply to Landau-Ginzburg
models can also be applied to GLSMs, and so provide
some additional indirect tests.

In section~\ref{lgstx} we very briefly outline how the hybrid
Landau-Ginzburg phases appearing in limits of generic GLSMs
can be understood as Landau-Ginzburg models over stacks rather than
spaces.  A detailed analysis of those theories will appear elsewhere.

Finally, in appendix~\ref{alta}, we describe one alternative A twist to
that discussed in the bulk of this paper, and then in appendix~\ref{hypbmodel}
we prove a result on hypercohomology that is used in chiral ring computations.

Although A-twisted Landau-Ginzburg models have been discussed only rarely
in the literature, they have been discussed previously a nonzero number of
times.  For example, \cite{ito1} writes down the action for an A-twisted
Landau-Ginzburg model over a vector space, in the special case that the
worldsheet is a complex plane (avoiding various subtleties in defining the
twist on more general worldsheets).  More recently, as this paper was being
finished, the papers \cite{fjr1,fjr2} appeared, which describe a
mathematical ansatz for correlation functions in A-twisted Landau-Ginzburg
models.  On the one hand, \cite{fjr1,fjr2} also couples to topological
gravity, which we do not; on the other hand, \cite{fjr1,fjr2} only consider
Landau-Ginzburg models over (orbifolds of) vector spaces, whereas we are
interested in more general cases.  Briefly, our concerns and results seem
to be orthogonal to those of \cite{fjr1,fjr2}.

In the upcoming work \cite{jstoappear}, we shall discuss A-twisted
heterotic Landau-Ginzburg models.

\section{B-twisted Landau-Ginzburg models}  \label{stdbrev}

\subsection{Review of untwisted Landau-Ginzburg models}

A Landau-Ginzburg model is a NLSM together with a 
superpotential.  To define such a model,
one must specify both a complex Riemannian manifold and a
holomorphic function -- the superpotential -- over that Riemannian manifold.

The most general Landau-Ginzburg model (over a space) that one can write
down has the following action:
\begin{eqnarray*}
\lefteqn{ 
\frac{1}{\alpha'} \int_{\Sigma} d^2z \left(
g_{\mu \nu} \partial \phi^{\mu} \overline{\partial} \phi^{\nu}
\: + \: i B_{\mu \nu} \partial \phi^{\mu} \overline{\partial}
\phi^{\nu} \: + \:
\frac{i}{2} g_{\mu \nu} \psi_-^{\mu} D_z \psi_-^{\nu} \: + \:
\frac{i}{2} g_{\mu \nu} \psi_+^{\mu} D_{\overline{z}} \psi_+^{\nu}
\right. } \\
& & \hspace*{1.0in}  \left. 
+ \:
R_{i \overline{\jmath} k \overline{l}} \psi_+^i \psi_+^{\overline{\jmath}} 
\psi_-^k
\psi_-^{\overline{l}}
+ \: 2 g^{i \overline{\jmath}} \partial_i W \partial_{\overline{\jmath}}
\overline{W} \: + \:  \psi_+^i \psi_-^j D_i \partial_j W \: + \:
 \psi_+^{\overline{\imath}} \psi_-^{\overline{\jmath}} D_{\overline{\imath}}
\partial_{\overline{\jmath}} \overline{W} \right),
\end{eqnarray*}
where $W$ is  the superpotential and
\begin{displaymath}
D_i \partial_j W \: = \: \partial_i \partial_j W \: - \:
\Gamma_{ij}^k \partial_k W.
\end{displaymath}
The fermions couple to the bundles
\begin{align*}
\psi_+^i \: &\in \: \Gamma_{ C^{\infty} }\left( K_{\Sigma}^{1/2} \otimes
\phi^* T^{1,0}X \right)                                         & \psi_-^i
\: &\in \: \Gamma_{ C^{\infty} }\left( \overline{K}_{\Sigma}^{1/2} \otimes \left( \phi^* T^{0,1} X\right)^{\vee} \right) \\
\psi_+^{\overline{\imath}} \: &\in \: \Gamma_{ C^{\infty} }\left(
K_{\Sigma}^{1/2} \otimes \left( \phi^* T^{1,0}X \right)^{\vee} \right) &
\psi_-^{\overline{\imath}} \: &\in \: \Gamma_{ C^{\infty} } \left(
\overline{K}_{\Sigma}^{1/2} \otimes \phi^* T^{0,1} X \right),
\end{align*}
where $K_{\Sigma}$ denotes the canonical bundle on the worldsheet
$\Sigma$.
The bosonic potential is of the form $\sum_i | \partial_i W |^2$.
The action possesses the supersymmetry transformations:
\begin{eqnarray*}
\delta \phi^i & = & i \alpha_- \psi_+^i \: + \: i \alpha_+ \psi_-^i \\
\delta \phi^{\overline{\imath}} & = & i \tilde{\alpha}_- \psi_+^{
\overline{\imath}} \: + \: i \tilde{\alpha}_+ \psi_-^{\overline{\imath}} \\
\delta \psi_+^i & = & - \tilde{\alpha}_- \partial \phi^i \: - \:
i \alpha_+ \psi_-^j \Gamma^i_{j m} \psi_+^m \: - \:
i \alpha_+ g^{i \overline{\jmath}} \partial_{\overline{\jmath}} \overline{W}\\
\delta \psi_+^{\overline{\imath}} & = & - \alpha_- \partial 
\phi^{\overline{\imath}}
\: - \: i \tilde{\alpha}_+ \psi_-^{\overline{\jmath}} 
\Gamma^{\overline{\imath}}_{\overline{\jmath} \overline{m} }
\psi_+^{\overline{m}} 
\: - \: i \tilde{\alpha}_+ g^{\overline{\imath} j} \partial_j W \\
\delta \psi_-^i & = & - \tilde{\alpha}_+ \overline{\partial} \phi^i \: - \:
i \alpha_- \psi_+^j \Gamma^i_{j m} \psi_-^m 
\: + \: i \alpha_- g^{i \overline{\jmath}} \partial_{\overline{\jmath}}
\overline{W} \\
\delta \psi_-^{\overline{\imath}} & = & - \alpha_+ \overline{\partial}
\phi^{\overline{\imath}} \: - \: 
i \tilde{\alpha}_- \psi_+^{\overline{\jmath}} 
\Gamma^{\overline{\imath}}_{\overline{\jmath} \overline{m}}
\psi_-^{\overline{m}} 
\: + \: i \tilde{\alpha}_- g^{\overline{\imath} j} \partial_j W.
\end{eqnarray*}
The ordinary NLSM is classically scale-invariant,
but notice that the action above is not scale-invariant even
classically when
$W \neq 0$.  This means that 
Landau-Ginzburg models are not themselves conformal field theories.
However, we can use them to define conformal field theories by
applying renormalization group  
flow -- the endpoint of which
is a (possibly trivial) conformal field theory.

\subsection{Review of B-twisted Landau-Ginzburg models}
\label{genlbtwiststudy}

In this section, we will review B-twisted Landau-Ginzburg models,
as originally described in \cite{vafatoplg}.
The B twist for the theory with superpotential is defined by taking the
fermions to be the sections of the bundles
\begin{align*}
\psi_+^i \: &\in \: \Gamma_{ C^{\infty} }\left( K_{\Sigma} \otimes \phi^*
T^{1,0}X \right)                      & \psi_-^i \: &\in \: \Gamma_{ C^{\infty} }\left( \overline{K}_{\Sigma} \otimes \left( \phi^* T^{0,1} X\right)^{\vee} \right) \\
\psi_+^{\overline{\imath}} \: &\in \: \Gamma_{ C^{\infty} }\left( \left(
\phi^* T^{1,0}X \right)^{\vee} \right) & \psi_-^{\overline{\imath}} \: &\in
\: \Gamma_{ C^{\infty} } \left( \phi^* T^{0,1} X \right).
\end{align*}
For the ordinary B-twisted (2,2) NLSM, $X$ is constrained to have its
canonical bundle square to the trivial bundle, a condition often stated as
the sufficient -- and in particular, not necessary -- condition that $X$
must be Calabi-Yau \cite{wittentft,bchir}.  The same condition is present
here: in order to make sense of the path integral measure in the B model
with superpotential, we must also demand that $K_X^{\otimes 2}$ is
trivial\footnote{ One way to see this is to think about the path integral
measure in the Landau-Ginzburg theory:  the result follows from demanding
that the path integral measure be a scalar, not a section of a bundle, just
as in the usual story for NLSMs without a superpotential.  In (2,2) GLSMs,
we can reason as follows.  Since the superpotential is gauge-invariant, it
cannot contribute to the one-loop anomaly diagrams, unless it gave a mass
to one of the fields.  But then, it would have to give a mass to a pair of
fields, which would necessarily be of equal and opposite charge, hence the
linear anomalies could not be affected by the superpotential.  Similarly,
back in a NLSM, if $X$ were the total space of a nontrivial vector bundle,
then to give a mass to some of the fiber directions would require that the
various bundle factors be dual to one another, hence the anomaly condition
would reduce to that of the base.  }.

It is convenient (and conventional) to define
\begin{align*}
\eta^{\overline{\imath}} & = \psi_+^{\overline{\imath}} \: + \: \psi_-^{\overline{\imath}} \\
\theta_i                 & = g_{i \overline{\jmath}} \left( \psi_+^{\overline{\jmath}} \: - \: \psi_-^{\overline{\jmath}} \right) \\
\rho_z^i                 & = \psi_+^i \\
\rho_{\overline{z}}^i    & = \psi_-^i.
\end{align*}
The scalar supersymmetry transformation parameters are $\tilde{\alpha}_+$
and $\tilde{\alpha}_-$, as before.  Taking
$\tilde{\alpha}_+ = \tilde{\alpha}_- = \alpha$ as before, the BRST
transformations are now
\begin{align*}
\delta \phi^i                   & = 0 \\
\delta \phi^{\overline{\imath}} & = i \alpha \eta^{\overline{\imath}} \\
\delta \eta^{\overline{\imath}} & = 0 \\
\delta \theta_i                 & = - 2 i \alpha \partial_i W \\
\delta \rho^i                   & = - \alpha d \phi^i.
\end{align*}
These are almost the same as for the ordinary B model,
except that $\theta_i$ is no longer BRST-invariant.

The chiral ring for this theory can be built in almost the same
fashion as for the ordinary B model, {\it i.e.} states are BRST-closed
(mod BRST-exact) products of the
form
\begin{displaymath}
b(\phi)^{j_1 \cdots j_m}_{\overline{\imath}_1 \cdots \overline{\imath}_n}
\eta^{\overline{\imath}_1} \cdots \eta^{\overline{\imath}_n}
\theta_{j_1} \cdots \theta_{j_m}.
\end{displaymath}
Mathematically, because $\theta$ is no longer BRST-invariant,
these can no longer be interpreted as sheaf cohomology of
exterior powers of the tangent bundle, but rather
\cite{tonypriv622}
are
hypercohomology of the complex
\begin{equation}   \label{bmodcpx}
\ldots \: \stackrel{i_{dW}}{\longrightarrow} \: \Lambda^{3}TX \:
\stackrel{i_{dW}}{\longrightarrow} \:
\Lambda^{2}TX \: \stackrel{i_{dW}}{\longrightarrow}  \:
TX \: \stackrel{i_{dW}}{\longrightarrow} \:
{\cal O}_{X}.
\end{equation}
This hypercohomology can be described 
as the endpoint of a spectral sequence, with $E_2^{p,q}$'s given
by degree $p$ sheaf cohomology on $X$ valued in the
$q$th cohomology sheaf of the complex above, {\it i.e.},
\begin{displaymath}
E_2^{p,q} \: = \: H^p(X, {\cal H}^q).
\end{displaymath} 
However, there is an easier way to understand these groups.
First, in the special case that $Y$ is a fat point
(for example, if $X$ is a vector space and $W$ a quasi-homogeneous
polynomial in the chiral superfields), then the complex~(\ref{bmodcpx})
is exact, and the hypercohomology immediately reduces to $H^*(Y,
\Lambda^* TY)$, or more simply (since $Y$ is a point),
$H^0(Y, {\cal O}_Y)$.  We shall see in the next section that this
matches the standard result for the chiral ring of Landau-Ginzburg
models over vector spaces with quasi-homogeneous superpotentials.
In appendix~\ref{hypbmodel} it is
shown for more general cases that
that
the hypercohomology of the complex~(\ref{bmodcpx}) is 
given by
\begin{displaymath}
H^*\left(Y, \Lambda^* TY  \otimes \Lambda^{top} N_{Y/X}^{\vee} \right),
\end{displaymath}
where $N_{Y/X}$ is the normal bundle to $Y$ in $X$.
We shall see in the next section that in cases in which the
Landau-Ginzburg model RG flows to a nontrivial NLSM, the bundle $\Lambda^{top}
N_{Y/X}$ is trivial, and so the sheaf cohomology above reduces
to the chiral ring of the B-twisted NLSM on $Y$.

Unlike the ordinary B model, the B-twisted NLSM with a superpotential is
not BRST exact.  Rather, the stress tensor is exact, implying that the
theory is indeed independent of the worldsheet metric.  In particular,
rescaling the worldsheet metric is equivalent to adding BRST-exact terms to
the action (see \cite{vafatoplg} for more information.)
Under such a rescaling $z \mapsto \lambda z$, the 
superpotential-dependent terms in the 
action become
\begin{eqnarray*}
\lefteqn{
\frac{1}{\alpha'} \int_{\Sigma} d^2z \left(
2 \lambda^2 g^{i \overline{\jmath}} \partial_i W 
\partial_{\overline{\jmath}} \overline{W} \: + \:
 \psi_+^i \psi_-^j D_i \partial_j W \: + \:
\lambda^2 \psi_+^{\overline{\imath}} \psi_-^{\overline{\jmath}}
D_{\overline{\imath}} \partial_{\overline{\jmath}} \overline{W}
\right)
} \\
& = & \frac{1}{\alpha'} \int_{\Sigma} d^2z \left(
2 \lambda^2 g^{i \overline{\jmath}} \partial_i W 
\partial_{\overline{\jmath}} \overline{W} \: + \:
 \rho_z^i \rho_{\overline{z}}^j D_i \partial_j W \: + \:
\frac{1}{2} \lambda^2 \eta^{\overline{\imath}} g^{\overline{\jmath} k}
\theta_k
D_{\overline{\imath}} \partial_{\overline{\jmath}} \overline{W}
\right)
\end{eqnarray*}
in the B-twisted theory (where we have used the fact that
$\psi_{\pm}^i$ are worldsheet vectors and $\psi_{\pm}^{\overline{\imath}}$
are worldsheet scalars).
Since $\lambda$ is arbitrary, we can take a large $\lambda$ limit from which
we see that contributions to correlation functions will arise from
fields $\phi$ such that $\partial W = 0$, which (since $W$ is
holomorphic), is equivalent to configurations such that $d W = 0$.

We may thus restrict to degree zero maps,
and the path integral reduces to ordinary and Grassmann integrals over
bosonic and fermionic zero modes. For large $\lambda$, the path integral reduces to
the product of an integral over bosonic zero modes 
\begin{equation}
\int_{ X } d \phi \exp\left( - 2 \frac{A}{\alpha'} 
\sum_i | \lambda \partial_i W |^2 
\right)
\end{equation}
(for the NLSM on the target $X = {\bf C}^n$) and an integral over the fermionic zero modes
\begin{equation}
  \begin{split}
	 \int \prod \left (d \eta^{\overline{\imath}} \right) \left( d \theta_i \right)
	 \exp\left( - \frac{1}{2} \frac{A}{\alpha'} 
	 \lambda^2 \eta^{\overline{\imath}} g^{\overline{\jmath}k}\theta_k
	 D_{\overline{\imath}} \partial_{\overline{\jmath}} \overline{W}
	 \right)
	 &\int \prod \left( \sqrt{\alpha'} d \rho_z^i \right)
	 \left( \sqrt{\alpha'} d \rho_{\overline{z}}^i \right)
	  \\
	 & \cdot
	 \exp\left( - \frac{1}{\alpha'} \int_{\Sigma} \rho_z^i
	 \rho_{\overline{z}}^j D_i \partial_j W \right),
	 \label{eq:fermiZeroModeIntegral}
  \end{split}
\end{equation}
where $A$ is the area of the worldsheet $\Sigma$.
The zero mode $i$ indices are implicitly contracted with a holomorphic
top-form\footnote{For simplicity, we assume that $X$ is
Calabi-Yau, rather than merely obeying $K_X^{\otimes 2} \cong {\cal O}$.
As discussed in \cite{bchir}, the difference is minor in any event.} 
$\Omega_X$ on $X$, so that for example
\begin{displaymath}
\int \prod ( d \eta^{\overline{\imath}}) \: = \:
\int \overline{\Omega}_{X \overline{\imath}_1 \cdots \overline{\imath}_N }
d \eta^{\overline{\imath}_1} \cdots d \eta^{\overline{\imath}_N}.
\end{displaymath}
The $\eta^{\overline{\imath}}$ and $\theta_i$ zero modes couple to a trivial
bundle, spanning a $\mbox{dim }X$-dimensional vector space,
while the $\rho_z^i$ and $\rho_{\overline{z}}^i$ zero modes are respectively holomorphic
and antiholomorphic sections of the worldsheet canonical bundle tensored
with the (trivial) pullback of the tangent bundle of $X$.
Since only the $\eta^{\overline{\imath}}$, $\theta_i$ zero 
modes are constants,
only in the first exponential factor of equation
\eqref{eq:fermiZeroModeIntegral}
may be trivially integrated over the worldsheet $\Sigma$ 
to give a factor of the worldsheet area $A$.

Now, let us evaluate the bosonic factor
\begin{displaymath}
\int_X d \phi \exp\left( - 2 \frac{A}{\alpha'}\sum_i | \lambda \partial_i W |^2 
\right).
\end{displaymath}
We will argue in a moment that the method of steepest descent will give an
exact answer for this integral, not just the leading order term in an
asymptotic series, \index{asymptotic series} because we can make $\lambda$
arbitrarily large.  To see this, expand
\begin{displaymath}
\phi \: = \: \phi_0 \: + \: \delta \phi,
\end{displaymath}
where $\phi_0$ is a constant map that solves $dW = 0$,
and $\delta \phi$ is a perturbation to another (constant) map.
Then we see that the potential
term in the action can be expanded as
\begin{displaymath}
\frac{1}{\alpha'} \int_{\Sigma} d^2z \left[
- 2 \lambda^2 g^{i \overline{\jmath}} (D_k \partial_i W )(\phi_0)
(D_{\overline{k}} \partial_{\overline{\jmath}} \overline{W} )(\phi_0)
(\delta \phi^k) (\delta \phi^{\overline{k}} ) 
\: + \: {\cal O}((\delta \phi)^3) \right].
\end{displaymath}
Now, define  
\begin{displaymath}
\tilde{\phi} \: = \: \frac{\lambda}{\sqrt{\alpha'}} \delta \phi.
\end{displaymath}
We see that the expansion of the potential term in the action now has
the form
\begin{displaymath}
\int_{\Sigma} d^2z \left[
- 2 g^{i \overline{\jmath}} (D_k \partial_i W)(\phi_0)
(D_{\overline{k}} \partial_{\overline{\imath}} \overline{W} )(\phi_0)
\tilde{\phi}^k \tilde{\phi}^{\overline{k}} \: + \:
\frac{\sqrt{\alpha'}}{\lambda} {\cal O}((\tilde{\phi})^3) \right].
\end{displaymath}
Furthermore, the terms of greater than quadratic order are suppressed by
factors proportional to $\lambda^{-1}$.  Since we can make $\lambda$
arbitrarily large without affecting correlation functions in the topological
subsector, the method of steepest descent gives an exact answer to the
integral in the $\lambda \mapsto \infty$ limit.

In any event, given the general analysis above, we can evaluate the
bosonic zero mode integral
\begin{displaymath}
\int_X d \phi \exp\left( - 2 \frac{A}{\alpha'}\sum_i | \lambda \partial_i W |^2 
\right).
\end{displaymath}
This is simply a multi-variable Gaussian, along directions transverse
to $Y \equiv \{ dW = 0 \}$, plus an integral along $Y$.
If we define
\begin{displaymath}
H \: = \: \det \left( D_i \partial_j W \right), \: \: \:
\overline{H} \: = \: \det \left( D_{\overline{\imath}}
\partial_{\overline{\jmath}} \overline{W} \right),
\end{displaymath}
we can read off, more or less immediately, that\footnote{
We will assume, throughout this review of B-twisted Landau-Ginzburg
models, that the Hessian $H$ is nonvanishing everywhere
along the $\{ dW = 0 \}$ locus, since our goal is primarily
to reproduce the results of \cite{vafatoplg}, albeit in a slightly
more general context.  This is usually not true for examples of Landau-Ginzburg
models over vector spaces with quasi-homogeneous superpotentials,
an important class of examples that one must deal with in a slightly
different fashion.
} 
\begin{align*}
\int_X d \phi \exp\left( - 2 \frac{A}{\alpha'}\sum_i | 
\lambda \partial_i W |^2 \right)
&=
\int_{ {\cal N}_{Y/X}} d \phi \exp\left( - 2 \frac{A}{\alpha'}\sum_i | 
\lambda \partial_i W |^2 \right) \\
&= \int_{Y} \pi^n \left( \frac{ \alpha' }{ 2 A } \right)^n
\lambda^{-2n} H^{-1} \overline{H}^{-1}.
\end{align*}
Here, ${\cal N}_{Y/X}$ is the total space of the normal bundle to $Y = \{
dW = 0 \}$ in $X$, the Hessians $H$ and $\overline{H}$ are to be evaluated
at each point of $Y$, and $n$ is the codimension of $Y$ in $X$.

Next, let us evaluate the factor corresponding to fermionic zero modes:
\begin{eqnarray*}
\lefteqn{
\int \prod \left (d \eta^{\overline{\imath}} \right)
\left( d \theta_i \right)
\exp\left( - \frac{1}{2} \frac{A}{\alpha'} 
\lambda^2 \eta^{\overline{\imath}} g^{\overline{\jmath}k}\theta_k
D_{\overline{\imath}} \partial_{\overline{\jmath}} \overline{W}
\right)
\int \prod \left( \sqrt{\alpha'} d \rho_z^i \right)
\left( \sqrt{\alpha'} d \rho_{\overline{z}}^i \right)
} \\
& & \hspace*{2.5in} \cdot
\exp\left( - \frac{1}{\alpha'} \int_{\Sigma} \rho_z^i
\rho_{\overline{z}}^j D_i \partial_j W \right).
\end{eqnarray*}
Given the zero mode counting outlined earlier, if $\mbox{dim }Y = 0$,
then we see immediately
that these factors by themselves would give
\begin{align*}
\int \prod \left (d \eta^{\overline{\imath}} \right)
\left( d \theta_i \right) \exp\left( - \frac{1}{2} \frac{A}{\alpha'} 
\lambda^2 \eta^{\overline{\imath}} g^{\overline{\jmath}k}\theta_k
D_{\overline{\imath}} \partial_{\overline{\jmath}} \overline{W}
\right) & = \frac{1}{n!}\left(- \lambda^2 \frac{A}{\alpha'} \right)^n \overline{H} \\
\int \prod \left( \sqrt{\alpha'} d \rho_z^i \right)
\left( \sqrt{\alpha'} d \rho_{\overline{z}}^i \right)
\exp\left( - \frac{1}{\alpha'} \int_{\Sigma} \rho_z^i
\rho_{\overline{z}}^j D_i \partial_j W \right)
 & = \frac{1}{(ng)!}\left( \frac{-}{\alpha'} \right)^{ng} (\alpha')^{ng}
 H^g,
\end{align*}
where $g$ is the genus of the worldsheet $\Sigma$.
Proceeding a little more carefully, in a more general case where
$\mbox{dim }Y$ need not vanish, one finds
at worldsheet genus zero that
\begin{equation}  \label{btwistgen}
\langle {\cal O}_1 \cdots {\cal O}_n \rangle \: = \:
\int_Y d\phi \frac{ 
\omega_{1 \overline{\imath} }^j \cdots \omega_{n \overline{\imath}}^j
(\Omega_X)^{\overline{\imath} \cdots \overline{\imath} \overline{k} \cdots
\overline{k} } (\Omega_X)_{j \cdots j m \cdots m}
\left( D_{\overline{k}} \partial^m \overline{W} \right) \cdots
\left( D_{\overline{k}} \partial^m \overline{W} \right)
}{
H \overline{H} }, 
\end{equation}
where each ${\cal O}_i$ is represented by the operator
\begin{displaymath}
\omega_{i \overline{\imath}_1 \cdots \overline{\imath}_k}^{
j_1 \cdots j_m} \eta^{\overline{\imath}_1} \cdots
\eta^{\overline{\imath}_k} \theta_{j_1} \cdots \theta_{j_m}.
\end{displaymath}
In writing the expression above, it is implicitly assumed 
that $K_X$ is trivial; if $K_X^{ \otimes 2}$ rather than $K_X$ is trivial,
one merely replaces the product of two $\Omega_X$'s with a trivialization
of $K_X^{\otimes 2}$.
We have been sloppy about factors of $i$ and $\pi$; more
important are the facts that, first, all factors of
$\alpha'$, $A$, and $\lambda$ cancel out as expected for the
topological field theory, and second, the contribution from 
$\overline{H}$ will also cancel out, giving a result that depends
only upon the holomorphic quantity $H$.

\subsection{Consistency checks}

As our first consistency check, we consider the special case that the
Landau-Ginzburg model is defined by a quasi-homogeneous potential over
a vector space ${\bf C}^n$.
First, we compare the chiral ring derived from our general
considerations to the standard results.
In this case,
the spectral sequence above degenerates:
each ${\cal H}^q$ has support only at a point, so the hypercohomology
at degree $p$ is given by 
\begin{displaymath}
H^0\left(pt, {\cal H}^q \right).
\end{displaymath}
As discussed earlier, the hypercohomology is the same as
$H^*(Y, \Lambda^* TY)$, where $Y = \{ dW = 0 \}$.
In this case, $Y$ is a zero-dimensional scheme\footnote{
On an unrelated note, for a description of how schemes can be used
to describe certain D-branes, see \cite{dks}.
}, so the only nonzero group
is
\begin{displaymath}
H^0\left(Y, {\cal O}_Y \right) \: = \:
{\bf C}[x_1, \cdots, x_n] / ( \partial W),
\end{displaymath}
which matches the standard physics result \cite{vafatoplg} for this case.

As a further check, in this special case equation~(\ref{btwistgen})
reduces to the standard result for B-twisted Landau-Ginzburg model
correlation functions in \cite{vafatoplg}.
Here, the correlators are polynomials with no $\eta$ or $\theta$ factors,
and the $\overline{H}$ in the denominator of equation~(\ref{btwistgen})
cancels the
factors of $\overline{\partial}^2 \overline{W}$ in the numerator. 
Equation~(\ref{btwistgen}) thus reduces to the standard result
\begin{displaymath}
\langle {\cal O}_1 \cdots {\cal O}_n \rangle \: = \:
\sum_Y \frac{ \omega_1  \cdots  \omega_n }{H^{1-g}}
\end{displaymath}
for $g=0$ (since $Y$ is dimension zero, the integral has been
replaced by a sum.) 
Wedge products are omitted: each of the correlators above correspond to
degree zero forms, for which wedge products are equivalent to ordinate
products.

As another check of this result, we can compare correlation functions of
certain Landau-Ginzburg models and the NLSMs to which they flow.
The topological subsector should be preserved
by renormalization group flow, hence the correlation functions should match.
We will use such examples at several points in this paper to check our
computations, so let us elaborate on their details.

In particular, we consider the quintic.
Now, to be clear, the Landau-Ginzburg model we shall study is
{\it not} defined by a quintic polynomial on the orbifold
$[ {\bf C}^5/{\bf Z}_5]$.  We emphasize this, because it was sometimes
mistakenly claimed in older literature that that Landau-Ginzburg model
flows to a NLSM on the quintic.  The correct statement
is that that Landau-Ginzburg model defines a distinct CFT at its
RG endpoint, one which is on the same K\"ahler moduli space as a
NLSM on the corresponding quintic.  Although this has
been known for many years, some authors persist to this day in incorrectly
claiming
that the Landau-Ginzburg model above flows to a NLSM on
the quintic under the renormalization group.

Instead, we consider a different Landau-Ginzburg model, one which
(unlike the one above) should flow under the renormalization group to
a NLSM on the quintic.
Consider a Landau-Ginzburg model defined on $X$ given by the
total space of the line bundle ${\cal O}(-5) \rightarrow {\bf P}^4$,
with superpotential $W = p G(\phi)$, where $p$ is a fiber coordinate
on ${\cal O}(-5)$, and $G(\phi)$ is the defining polynomial of the quintic,
but expressed in local coordinates on ${\bf P}^4$.
Across coordinate patches, both $p$ and $G(\phi)$ transform non-trivially,
but in opposite ways, so that the product $p G(\phi)$ is a well-defined
function globally. 
The F-terms in this Landau-Ginzburg model are of the form\footnote{
After a mild coordinate change at any point to remove cross terms.
}
\begin{displaymath}
|G(\phi)|^2 \: + \: \sum_i \left| p \frac{\partial G}{\partial x_i}
\right|^2.
\end{displaymath}
By assuming that the quintic is smooth and following the same F-term analysis
as for GLSMs, we find that this Landau-Ginzburg model
should flow in the IR to a theory defined over $p = G = 0$,
which is exactly the quintic inside $X$.

This Landau-Ginzburg model is very reminiscent of the structure of the
GLSM for the quintic, and indeed, this is no coincidence.
Naively,
the Landau-Ginzburg model above represents an intermediate step
along renormalization group flow between the large-radius region of the
quintic GLSM and the NLSM.

Let us perform consistency checks on a more general family that includes
the quintic above as a special case.  Consider Landau-Ginzburg models
over spaces $X$ of the form
\begin{displaymath}
X \: = \: \mbox{Tot} \left( {\cal E} \: \stackrel{\pi}{
\longrightarrow} \: B \right),
\end{displaymath}
for ${\cal E}$ a holomorphic vector bundle on $B$,
with a superpotential $W = p_{\alpha} s^{\alpha}$. Here
the $p_{\alpha}$ are fiber coordinates on ${\cal E}$ and
$s^{\alpha}$ denotes the local form of a holomorphic section of  ${\cal E}^{\vee}$.
Let $Y$ denote the vanishing locus of $s$, {\it i.e.},
$Y = \{ s^{\alpha} = 0 \}$.
From the form of the F terms implied by the superpotential,
this Landau-Ginzburg model appears to flow\footnote{Note that in
general a B-twisted Landau-Ginzburg model on a noncompact Calabi-Yau
need {\it not} flow to a NLSM on the subspace
$\{ dW = 0 \}$.  For example, consider a Landau-Ginzburg model on
${\bf C}^n$ with $W$ a generic homogeneous polynomial of degree $d$.
Only in the special case that $d=N+1$ could one hope that the
locus $\{ dW = 0 \}$ would be a local model of a Calabi-Yau;
in all other cases, a B-twisted NLSM on
$\{ dW = 0 \}$ could not even exist, much less be the endpoint
of RG flow from a Landau-Ginzburg model.}
to a NLSM on $Y$.  Let us check our general picture of 
B-twisted Landau-Ginzburg models by comparing Landau-Ginzburg
computations in this theory to computations in the corresponding
NLSM.

First, let us compare anomalies.
In order for the B-twisted Landau-Ginzburg model to be consistent, the space
$X$ must have $K_X^{\otimes 2} \cong {\cal O}_X$ (this is the 
condition that is more commonly claimed to be that
$X$ must be Calabi-Yau).
Given this condition on $X$, it is straightforward to check that $Y$ must also satisfy anomaly
cancellation in the B-twisted NLSM.
First, from the short exact sequence
\begin{displaymath}
0 \: \longrightarrow \: \pi^* {\cal E}^{\vee}
 \: \longrightarrow \: TX \: \longrightarrow
\: \pi^* TB \: \longrightarrow \: 0,
\end{displaymath}
we have that $K_X \cong \pi^* K_B \otimes \pi^* \Lambda^{top} {\cal E}$.
Similarly, there is a short exact sequence
\begin{displaymath}
0 \: \longrightarrow \: TY \: \longrightarrow \: i^* TB \: \longrightarrow
\: i^* {\cal E} \: \longrightarrow \: 0,
\end{displaymath}
where $i: Y \hookrightarrow B$ is the inclusion map,
and we have used the fact that the normal bundle
${\cal N}_{Y/B} \cong i^* {\cal E}$.
From this short exact sequence, we find
$K_Y \cong i^* K_B \otimes i^* \Lambda^{top} {\cal E}$.
Let $i_0: B \hookrightarrow X$ be inclusion along the zero section of
${\cal E}^{\vee}$, so that $\pi \circ i_0 \circ i = i$.
Putting these together, we find that
\begin{eqnarray*}
(i_0 \circ i)^* K_X & \cong & (i_0 \circ i)^* \pi^* K_B \otimes
(i_0 \circ i)^* \pi^* \Lambda^{top} {\cal E} \\
& \cong & i^* K_B \otimes i^* \Lambda^{top} {\cal E} \\
& \cong & K_Y \otimes i^* \Lambda^{top} {\cal E}^{\vee}
\otimes i^* \Lambda^{top} {\cal E} \\
& \cong & K_Y.
\end{eqnarray*}

As a result, we see that if $K_X^{\otimes 2} \cong {\cal O}_X$, then
$K_Y^{\otimes 2} \cong {\cal O}_Y$.  Thus, if the B-twisted Landau-Ginzburg
model is anomaly-free, then so is the B-twisted NLSM to
which it flows.  This also serves as a form of 't Hooft anomaly matching, a
consistency check on our claim that these two theories are related by
renormalization group flow.

From the general discussion in the last section and appendix~\ref{hypbmodel}, 
the additive part of the chiral ring in this
theory, as defined by the hypercohomology of a sequence discussed earlier,
is given by 
\begin{displaymath}
H^*\left(Y, \Lambda^* TY  \otimes \Lambda^{top} N_{Y/X}^{\vee} \right),
\end{displaymath}
where $N_{Y/X}$ is the normal bundle to $Y$ in $X$.
However, in the present case, we can show that $N_{Y/X}$ has trivial
determinant.  To see this, first note from the short exact sequence
\begin{displaymath}
0 \: \longrightarrow \: TY \: \longrightarrow \: TX|_Y 
\: \longrightarrow \: N_{Y/X} \: \longrightarrow \: 0
\end{displaymath}
one has $K_Y \cong K_X|_Y \otimes \Lambda^{top} N_{Y/X}$.
We showed above that $K_Y \cong K_X |_Y$, hence,
$\Lambda^{top} N_{Y/X}$ is trivial, and so we see that the additive
part of the chiral ring in this B-twisted Landau-Ginzburg model is given by
\begin{displaymath}
H^*\left(Y, \Lambda^* TY \right)
\end{displaymath}
matching the BRST cohomology of the B-twisted NLSM on $Y$.

Next, let us check that correlation
functions in the Landau-Ginzburg model match those in the
NLSM to which it flows.
First, recall the expression given in equation~(\ref{btwistgen}) for genus zero correlation functions, 
\begin{displaymath}  
\langle {\cal O}_1 \cdots {\cal O}_n \rangle \: = \:
\int_Y d\phi \frac{ 
\omega_{1 \overline{\imath} }^j \cdots \omega_{n \overline{\imath}}^j
(\Omega_X)^{\overline{\imath} \cdots \overline{\imath} \overline{k} \cdots
\overline{k} } (\Omega_X)_{j \cdots j m \cdots m}
\left( D_{\overline{k}} \partial^m \overline{W} \right) \cdots
\left( D_{\overline{k}} \partial^m \overline{W} \right)
}{
H \overline{H} }.
\end{displaymath}
To rewrite this in terms of data on $Y$,
we will need an expression for the holomorphic top-form on $Y$.
From \cite{gh}[p. 147], for a hypersurface $V = \{f=0\}$ in 
a manifold $M$, for any $n$ there is an isomorphism
\begin{displaymath}
\Omega_M^n(V) \: \stackrel{\sim}{\longrightarrow} \:
\Omega_V^{n-1},
\end{displaymath}
which for any $k$ is given in local coordinates by
\begin{displaymath}
\frac{ \omega_{i_1 \cdots i_n} dz^{i_1} \wedge \cdots \wedge
dz^{i_n} }{ f } \: \longrightarrow \:
(-)^k \left( \omega_{ i_1 \cdots \hat{i_k} \cdots i_n} \right)
\frac{ dz^{i_1} \wedge \cdots \wedge \hat{ dz^{i_k} } \wedge \cdots
\wedge dz^{i_n} }{\partial f / \partial x^{i_k} }.
\end{displaymath}
The fact that this holds for any $k$ follows from the identity
\begin{displaymath}
\sum_i \partial_i f dx^i \: = \: 0.
\end{displaymath}
Specifically, if one solves for one of the $dx$'s using the expression
above and plugs in, then the result is the same expression back again
but with a different $k$.
In the present case, we have a local complete intersection
$Y = \{s=0\}$, for $s \in \Gamma({\cal E}^{\vee})$.
If we let $X$ denote the total space of $\pi: {\cal E} \rightarrow X$,
with $p_1, \cdots, p_k$ local coordinates along the fibers, then
\begin{displaymath}
Y \: = \: \{ p_1 = \cdots = p_k = 0 = s^1 = \cdots = s^k \}.
\end{displaymath}
Iterating the expression above, we find
\begin{displaymath}
\left( \Omega_Y \right)_{i_1 \cdots i_{n-2k}} \: = \:
( \pm) \frac{  \left( \Omega_X \right)_{i_{2k+1}\cdots i_n} }{
(\partial_{i_1}s^1 ) \cdots (\partial_{i_k} s^k) },
\end{displaymath}
where the $\Omega$'s are the holomorphic top-forms on $X$ and $Y$,
and $k$ is the rank of ${\cal E}$.
As a result, we can write
\begin{eqnarray*}
\lefteqn{
\int_Y d\phi \frac{ 
\omega_{1 \overline{\imath} }^j \cdots \omega_{n \overline{\imath}}^j
(\Omega_X)^{\overline{\imath} \cdots \overline{\imath} \overline{k} \cdots
\overline{k} } (\Omega_X)_{j \cdots j m \cdots m}
\left( D_{\overline{k}} \partial^m \overline{W} \right) \cdots
\left( D_{\overline{k}} \partial^m \overline{W} \right)
}{
H \overline{H} } 
} \\
& \hspace*{1.5in}  = &
\int_Y d \phi \;
\omega_{1 \overline{\imath} }^j \cdots \omega_{n \overline{\imath}}^j
(\Omega_Y)^{\overline{\imath} \cdots \overline{\imath} }
(\Omega_Y)_{j \cdots j }
\end{eqnarray*}
(using the previous conventions for indices of the $\omega$'s).
Thus, we see that correlation functions in the B-twisted
Landau-Ginzburg model
are given by
\begin{displaymath}  
\langle {\cal O}_1 \cdots {\cal O}_n \rangle \: = \:
\int_Y d \phi \;
\omega_{1 \overline{\imath} }^j \cdots \omega_{n \overline{\imath}}^j
(\Omega_Y)^{\overline{\imath} \cdots \overline{\imath} }
(\Omega_Y)_{j \cdots j },
\end{displaymath}
which precisely matches\footnote{
As in expression~(\ref{btwistgen}), the expression written implicitly
assumes that $K_Y$ is trivial, not just $K_Y^{\otimes 2}$.
In the more general case where $K_Y^{\otimes 2}$ is trivial,
one merely replaces the product of $\Omega_Y$'s with a trivialization
of $K_Y^{\otimes 2}$.
} correlation functions in the B-twisted 
NLSM, verifying that they lie in the same universality
class.

Another set of consistency checks one could explore involves
matrix factorizations.  
For example, for the Landau-Ginzburg model on the total space
of the line bundle ${\cal O}(-5) \rightarrow {\bf P}^4$, that
RG flows to a NLSM on a quintic in ${\bf P}^4$,
the matrix factorizations in the Landau-Ginzburg model should
be equivalent to sheaf theory on the quintic, by virtue of the fact
that the two topological field theories are in the same universality
class.  More generally, in a Landau-Ginzburg model on the total space
of a holomorphic vector bundle ${\cal E} \rightarrow X$
with suitable superpotential,
the matrix factorizations should be equivalent to the sheaf
theory on the zero locus of ${\cal E}$ defined by the superpotential,
again by virtue of the fact that the two topological field theories
are related by the renormalization group.
To the best of our knowledge, these physically-obvious statements
have not been proven mathematically.

Another consistency check involving matrix factorizations was
outlined in \cite{cdhps}.  There, it was pointed out that in a topological
field theory, the Born-Oppenheimer approximation should be exact,
so that in a B-twisted Landau-Ginzburg model on a space which is the
total space of a bundle, the matrix factorizations should behave
nicely fiberwise -- matrix factorizations in the theory should
be understandable in terms of matrix factorizations of the Landau-Ginzburg
models along the fibers.  Phrased more invariantly,
matrix factorizations should behave well in families.
This should tie into the previous observation
concerning matrix factorizations, and also has not to our knowledge
been proven mathematically.

\section{A-twisted Landau-Ginzburg models}   \label{atwist22}

\subsection{Making sense of the twist}   \label{atwiststart}

The A twist of the NLSM without superpotential
is defined by
taking $\psi_+^i$, $\psi_-^{\overline{\imath}}$ to be worldsheet scalars,
and the other two fermions to be worldsheet vectors.
In other words, in the A-twisted NLSM without
superpotential is defined by taking the fermions to be sections of the
following bundles:
\begin{align*}
\psi_+^i \: &\in \: \Gamma_{ C^{\infty} }\left( \phi^* T^{1,0}X \right)                                                            & \psi_-^i \: &\in \: \Gamma_{ C^{\infty} }\left( \overline{K}_{\Sigma} \otimes \left( \phi^* T^{0,1} X\right)^{\vee} \right) \\
\psi_+^{\overline{\imath}} \: &\in \: \Gamma_{ C^{\infty} }\left( K_{\Sigma} \otimes \left( \phi^* T^{1,0}X \right)^{\vee} \right) & \psi_-^{\overline{\imath}} \: &\in \: \Gamma_{ C^{\infty} } \left( \phi^* T^{0,1} X \right).
\end{align*}
As a result, the two scalar supersymmetry transformations are
$\alpha_-$ and $\tilde{\alpha}_+$.

Now, when we try to perform the same twist in a NLSM
with superpotential, we run into a problem.
Specifically, the Yukawa coupling terms involving the superpotential,
\begin{displaymath}
 \psi_+^i \psi_-^j D_i \partial_j W \: + \:
 \psi_+^{\overline{\imath}} \psi_-^{\overline{\jmath}} D_{\overline{\imath}}
\partial_{\overline{\jmath}} \overline{W}, 
\end{displaymath}
are no longer Lorentz-invariant on the worldsheet after this twist
 -- instead of becoming functions or (1,1) forms after the twist,
as happened in the B-twist, they are now (1,0) and (0,1)
forms.

There are several possible ways to solve this problem,
and so, several different notions of an ``A-twisted Landau-Ginzburg model.''
For example, Appendix~\ref{alta} contains a discussion of a notion of A-twist
that is different from what we shall use in most of this paper.
We are going to primarily work with an A twist such that
an A-twisted Landau-Ginzburg model on the total
space of a holomorphic vector bundle ${\cal E}^{\vee} \rightarrow B$,
with superpotential of the form $p_{\alpha} s^{\alpha}$,
$S^{\alpha}$ a section of ${\cal E}$ and $p_{\alpha}$ fiber coordinates,
will flow\footnote{We will always assume that not only is the vanishing
locus smooth, but so are the components of the section.}
to a NLSM on the zero locus $\{ s^{\alpha} = 0 \}$
of the section
$s^{\alpha}$ of ${\cal E}$.
In particular, one example we will discuss in detail is an
A-twisted Landau-Ginzburg model on the total space of
${\cal O}(-5) \rightarrow {\bf P}^4$ that flows to an A-twisted
NLSM on a quintic Calabi-Yau.  The superpotential
in this Landau-Ginzburg model is $W = pG$, for $p$ a fiber coordinate
and $G$ a generic section of ${\cal O}(5)$, and the quintic is described
as the zero locus of $G$ in ${\bf P}^4$.

In particular, since renormalization group flow preserves topological
subsectors, this means that a good notion of Gromov-Witten theory
for Landau-Ginzburg models (``LG-GW invariants'')
should have the property that the
LG-GW invariants of a Landau-Ginzburg model on
the total space of ${\cal E}^{\vee} \rightarrow B$ with
superpotential as above,
should match the ordinary Gromov-Witten invariants of the space
$\{ s^{\alpha} = 0 \} \subset B$.
We have heard several mathematicians speak in general terms of
LG-GW invariants, though very little seems to be
written up (see \cite{fjr1,fjr2} for a very recent example,
and the introduction of this paper for a brief discussion).
However, this prediction seems 
to be in agreement with general mathematical ideas about the structure of
LG-GW invariants:  as explained to us \cite{clarkepriv},
any non-constant compact holomorphic curve must land in a fiber of
the superpotential $W$, and to be invariant under gradient flow of
the real part of $W$, $\Re W$, 
the curve must lie in the critical locus of $W$ which is
precisely the zero section.  Since the curve-counting must then match that
of the zero section, the idea that LG-GW invariants of the Landau-Ginzburg
model above should match the ordinary Gromov-Witten invariants of the
zero section seems very natural.

In order for our A twist to have the universality property described above,
we need to twist by an R-symmetry -- however, the R-symmetry of NLSMs
does not lift to a symmetry of Landau-Ginzburg theories,
as we have seen, so we must modify the left and right R-charges of the
fields.
It will turn out that
for the case of the total space of a holomorphic vector bundle ${\cal E}^{\vee} \rightarrow B$,
with superpotential of the form $p_{\alpha} s^{\alpha}$,
we will twist the chiral superfields
describing local coordinates on $B$ differently from those
describing the fibers of ${\cal E}^{\vee}$.

To construct the R-symmetry for a Landau-Ginzburg model
on $X$ with superpotential $W: X \rightarrow {\bf C}$, we need for
$X$ to admit a $U(1)$ isometry for which $W$ is ``quasi-homogeneous,''
meaning that if $\alpha \in U(1)$, then
$\alpha^* W = \exp(i \theta) W$ where $\theta$ is defined by 
$\alpha = \exp(i \theta)$. In other words, the superpotential
must be {\it nearly} invariant under the isometry:  pulling back $W$ 
along the $U(1)$ isometry only has the effect of multiplying $W$ by a phase
factor.  Phrased yet another way still, the superpotential $W$ must
have charge 1 under this isometry.  Let $J$ denote the current generating
this isometry.

We should note immediately that not every example will possess such
an isometry.  For example, the `Toda duals' to the A model on ${\bf P}^n$
do not \cite{horivafa}[section 3.1], as one
should expect -- this obstruction to the existence
of the A twist of the Landau-Ginzburg theory is mirror to the fact that the
B model cannot be defined on ${\bf P}^n$, since the square of its canonical bundle
is not trivial.
Shortly, we shall see an additional constraint:
not only must an isometry exist, but it must satisfy a certain ``integral
charge'' condition in order to be able to perform the A-twist.

Given such an isometry, we can now see how to fix the twist so that the
two-dimensional action will be Lorentz-invariant.
First, let us slightly rephrase the ordinary A-twist, in a manner that
will make the resolution of the problem more clear.
The ordinary A-model twist\footnote{
The ordinary B-model twist involves tensoring with $K_{\Sigma}^{+(1/2)Q_R}
\otimes \overline{K}_{\Sigma}^{+(1/2)Q_L}$.
} modifies the bundles to which the fields couple
by tensoring with 
\begin{displaymath}
K_{\Sigma}^{-(1/2)Q_R} \overline{K}_{\Sigma}^{+(1/2)Q_L}
\end{displaymath}
where the fields have $Q_R$, $Q_L$ eigenvalues as below:
\begin{center}
\begin{tabular}{c|cc}
Field & $Q_R$ & $Q_L$ \\ \hline
$\phi^i$ & $0$ & $0$ \\
$\phi^{\overline{\imath}}$ & $0$ & $0$ \\
$\psi_+^i$ & $1$ & $0$ \\
$\psi_+^{\overline{\imath}}$ & $-1$ & $0$ \\
$\psi_-^i$ & $0$ & $1$ \\
$\psi_-^{\overline{\imath}}$ & $0$ & $-1$ 
\end{tabular}
\end{center}
For example, $\psi_+^i$, which in the untwisted theory is a $C^{\infty}$
section of $K_{\Sigma}^{1/2} \otimes \phi^* T^{1,0}X$,
in the twisted theory becomes a $C^{\infty}$ section of
\begin{displaymath}
K_{\Sigma}^{1/2} \otimes K_{\Sigma}^{-1/2} \otimes \overline{K}_{\Sigma}^0
\otimes \phi^* T^{1,0}X \: \cong \:
\phi^* T^{1,0}X.
\end{displaymath}
The twist above generates non-Lorentz-invariant terms in the theory with
a superpotential ultimately because of Yukawa couplings of the form
\begin{displaymath}
\psi_+^i \psi_-^j D_i \partial_j W.
\end{displaymath}
These terms are not invariant under the R-symmetries of the $W=0$ theory
defined by $Q_R$, $Q_L$.  However, if we define new charges by
$Q'_R = Q_R - Q$, $Q'_L = Q_L-Q$, where $Q$ is the eigenvalue of the
isometry generator $J$, then the Yukawa terms above will be invariant.

So, the new twisting will be defined by tensoring fields with
\begin{displaymath}
K_{\Sigma}^{-(1/2)Q'_R} \overline{K}_{\Sigma}^{+(1/2)Q'_L}
\end{displaymath}
for $Q'_R$, $Q'_R$ defined above, and the result will be a 
Lorentz-invariant action.

In cases in which the Landau-Ginzburg model flows in the IR to a 
NLSM on some Calabi-Yau, we believe that this
R-symmetry, defined by $Q'_R$, $Q'_L$, is the one that flows in the
IR to the R-symmetry of the NLSM.
We believe this because we get the correct chiral ring and we do not
see signs of any non-unitarity that one might expect if one were to
use the wrong R-symmetry, as would happen if there were an 
accidental $U(1)$ that appeared in the IR that mixed with the
present symmetry.  Furthermore, this is the standard trick to
obtain such R-symmetries --  for example, see \cite{shamited}.

One example of a Landau-Ginzburg model in which such an isometry
exists was discussed in \cite{shamited}.  There, a Landau-Ginzburg
model over $X = {\bf C}^5$ was considered, with a superpotential
$W$ defined by a degree-five homogeneous polynomial in the fields.
The $U(1)$ isometry was the simple $\phi^i \mapsto
\exp(2 \pi i \alpha/5) \phi^i$, for $\alpha \in [0,1]$,
under which $W \mapsto \exp(i \alpha)W$.
In the language above, $Q(\phi^i, \psi_+^i, \psi_-^i) = 1/5$ and
$Q(\phi^{\overline{\imath}}, \psi_+^{\overline{\imath}},
\psi_-^{\overline{\imath}}) = - 1/5$, and so one computes
\begin{center}
\begin{tabular}{c|cc}
Field & $Q'_R$ & $Q'_L$ \\ \hline
$\phi^i$ & $-1/5$ & $-1/5$ \\
$\phi^{\overline{\imath}}$ & $1/5$ & $1/5$ \\
$\psi_+^i$ & $4/5$ & $-1/5$ \\
$\psi_+^{\overline{\imath}}$ & $-4/5$ & $1/5$ \\
$\psi_-^i$ & $-1/5$ & $4/5$ \\
$\psi_-^{\overline{\imath}}$ & $1/5$ & $-4/5$
\end{tabular}
\end{center}
which (up to a meaningless overall sign) matches table 2 of
\cite{shamited}.

If we were to try to push the A twist further in this example from
\cite{shamited}, we would quickly run into a problem with integrality
of charges:  we would need to make sense of {\it e.g.} 
$K_{\Sigma}^{(1/2)(1/5)}$,
which is only well-defined on worldsheets of special genera.
Here, one would need to require that $5$ divide $g-1$, so
the A twist of this example could only be defined on worldsheets
of genus $1$, $6$, $11$, and so forth.  Performing an orbifold
does not improve matters.\footnote{
All bundles on a stack can be understood as bundles on the atlas, that
are `well-behaved' with respect to certain identifications.
For a $[X/{\bf Z}_5]$ orbifold, for example, all bundles on the orbifold
are ${\bf Z}_5$-equivariant bundles on $X$.  As a result, if there is no
way to make sense of $L^{1/5}$ on $X$, then it also can not make sense
over the stack $[X/{\bf Z}_5]$.  This does not preclude the possibility
of interesting bundles on stacks:  gerbes, for example, often have
bundles which cannot be understood as bundles over the base.
In that case, however, all the bundles can be understood as honest
bundles on the atlas -- but only some of those descend to honest bundles
on the base of the gerbe.
} 

On the other hand, if one were to couple to topological gravity,
which we do not do in this paper, then one would be able to get 
nonvanishing correlation functions in more genera.  
There, because one works over moduli spaces of punctured
Riemann surfaces, special operator insertions can modify the canonical
class of the worldsheet.  In the present case, for operator insertions
generating factors of ${\cal O}(m_i)$, the condition for the twist to
be defined would become that $10$ divide $2g - 2 - \sum_i m_i$,
so for any genus $g$, one can find suitable operator insertions ($m_i$'s)
so that some correlation functions would be nonvanishing.  Such roots of
canonical bundles are discussed in detail in \cite{ed-milnor,ed-nmat}.
Furthermore, we believe this is the strategy implicitly being applied
in \cite{fjr1,fjr2}, as they are concerned with A-twisted Landau-Ginzburg
models of precisely the form above, coupled to topological gravity.
As we do not couple to topological gravity, the condition for the twist
to make sense imposes a very strong constraint on worldsheet genus,
and so we will usually exclude these cases.

The reader might object that there are other symmetries of the Yukawa
coupling $\psi_+^i \psi_-^j D_i \partial_j W$ that one could twist by
instead; however, those other symmetries will usually not be R-symmetries
and so will not generate a scalar supercharge.
For example, in the case of the Landau-Ginzburg model above on ${\bf C}^5$
with quintic superpotential, one could imagine twisting by
\begin{center}
\begin{tabular}{c|cc}
Field & $Q'_R$ & $Q'_L$ \\ \hline
$\phi^i$ & $1$ & $1$ \\
$\psi_+^i$ & $-2$ & $-1$ \\
$\psi_-^i$ & $-1$ & $-2$
\end{tabular}
\end{center}
with charges of the complex conjugate fields determined by negation.
However, although this is a symmetry of the Yukawa coupling, and the
action more generally, it does not obey the conditions $Q'_R(\phi^i) = Q'_R(\psi_-^i)$,
$Q'_L(\phi^i) = Q'_L(\psi_+^i)$, $Q'_R(\psi_+^i) = Q'_R(\phi^i) + 1$,
or $Q'_L(\psi_-^i) = Q'_L(\phi^i) + 1$, which are required for this to
be an R-symmetry.  Imposing those conditions is equivalent to twisting by
a $U(1)$ isometry.

Regardless of whether it is an R-symmetry, if we were to twist using the $Q'_R$, $Q'_L$ above by tensoring 
$K^{-(1/2)Q'_R} \overline{K}^{+(1/2)Q'_L}$ with those bundles to which our fields couple,
then it is straightforward to check that there is no scalar supercharge:
\begin{displaymath}
\alpha_- \: \in  \: \Gamma(K^{-3}), \: \: \:
\alpha_+ \: \in \: \Gamma(\overline{K}^2), \: \: \:
\tilde{\alpha}_- \: \in \: \Gamma(K^2), \: \: \:
\tilde{\alpha}_+ \: \in \: \Gamma(\overline{K}^{-3}).
\end{displaymath}
As a result, one cannot obtain a topological field theory in this fashion.

Let us return to the example of $X$ given by the total space of the
vector bundle $\pi: {\cal E}^{\vee} \rightarrow B$,
with superpotential of the form $W = p_{\alpha} s^{\alpha}$.
Here also, there exists an isometry under which $W$ is
quasi-homogeneous, namely the isometry that rotates the $p_{\alpha}$
by phases.  In other words, $Q(p_{\alpha}, \psi_+^p, \psi_-^p) = 1$,
$Q(\phi^i, \psi_+^i, \psi_-^i) = 0$ 
where $\phi^i$ is a local coordinate along the base.
(So long as we restrict to local coordinate patches on $X$ of the form
$U \times F$ for $U$ open in $B$ and $F$ a fiber of ${\cal E}^{\vee}$,
and coordinate transformations which are linear on fibers,
{\it i.e.} $p'_{\alpha} = \Lambda(\phi) p_{\alpha}$,
the isometry defined above makes sense across patches.)
In this case, we compute
\begin{center}
  \begin{tabular}{c|cc|c|cc}
	 Field & $Q'_R$ & $Q'_L$  & Field & $Q'_R$ & $Q'_L$ \\ \hline
	 $\phi^i$ & $0$ & $0$ & $p$ & $-1$ & $-1$ \\
	 $\psi_+^i$ & $1$ & $0$ & $\psi_+^p$ & $0$ & $-1$ \\
	 $\psi_-^i$ & $0$ & $1$ &
	 $\psi_-^p$ & $-1$ & $0$ 
  \end{tabular}
\end{center}
 The charges of the complex conjugates are minus those above, and so for 
brevity are omitted.

From this we can read off the twistings, which we have collected
in Table~\ref{table:quinticFieldsAndBundles}.
Therein,
we have implicitly used the fact that as $C^{\infty}$ bundles,
a holomorphic vector bundle ${\cal E}$ and the dual $\overline{ {\cal E} }^{\vee}$
of its antiholomorphic complex conjugate bundle are isomorphic:  ${\cal E} \cong \overline{ {\cal E} }^{\vee}$.

\newcommand\tb{\rule{0pt}{2.6ex} \rule[-1.5ex]{0pt}{0pt}}

\begin{table}
  \begin{center}
	\begin{tabular}[h]{|llll|}
	  \hline
		\tb & $\!\!$Untwisted Bundle & \multicolumn{2}{l|}{$\!\!$Twisted Bundle}                                                                                                                                                                                                                                                                    \\ 
	  \hline
		\tb $\!\!\psi_+^i$                   & $\!\!K_{\Sigma}^{1/2} \otimes \phi^* T^{1,0}B$                                       & $\!\!K_{\Sigma}^{1/2} \otimes K_{\Sigma}^{-(1/2)(1)} \otimes \overline{K}_{\Sigma}^{+(1/2)(0)} \otimes \phi^* T^{1,0}B$                                        & $\!\!\!\!\!\!\cong\phi^* T^{1,0}B$                                                     \\ 
		\tb $\!\!\psi_+^{\overline{\imath}}$ & $\!\!K_{\Sigma}^{1/2} \otimes \left( \phi^* T^{1,0}B \right)^{\vee}$                 & $\!\!K_{\Sigma}^{1/2} \otimes K_{\Sigma}^{-(1/2)(-1)} \otimes \overline{K}_{\Sigma}^{+(1/2)(0)} \otimes \left( \phi^* T^{1,0}B \right)^{\vee}$                 & $\!\!\!\!\!\!\cong K_{\Sigma} \otimes \left( \phi^* T^{1,0}B \right)^{\vee}\!\!$            \\ 
		\tb $\!\!\psi_-^i$                   & $\!\!\overline{K}_{\Sigma}^{1/2} \otimes \left( \phi^* T^{0,1}B \right)^{\vee}$      & $\!\!\overline{K}_{\Sigma}^{1/2} \otimes K_{\Sigma}^{-(1/2)(0)} \otimes \overline{K}_{\Sigma}^{+(1/2)(1) } \otimes \left( \phi^* T^{0,1}B \right)^{\vee}$      & $\!\!\!\!\!\!\cong\overline{K}_{\Sigma} \otimes \left( \phi^* T^{0,1}B \right)^{\vee}\!\!$ \\ 
		\tb $\!\!\psi_-^{\overline{\imath}}$ & $\!\!\overline{K}_{\Sigma}^{1/2} \otimes \phi^* T^{0,1}B$                            & $\!\!\overline{K}_{\Sigma}^{1/2} \otimes K_{\Sigma}^{-(1/2)(0)} \otimes \overline{K}_{\Sigma}^{+(1/2)(-1)} \otimes \phi^* T^{0,1}B$                            & $\!\!\!\!\!\!\cong\phi^* T^{0,1}B$                                                     \\ 
	  \hline
		\tb $\!\!\psi_+^p$                   & $\!\!K_{\Sigma}^{1/2} \otimes \phi^* T^{1,0}_{\pi}$                                  & $\!\!K_{\Sigma}^{1/2} \otimes K_{\Sigma}^{-(1/2)(0)}\otimes \overline{K}_{\Sigma}^{+(1/2)(-1)} \otimes \phi^* T^{1,0}_{\pi}$                                   & $\!\!\!\!\!\!\cong K_{\Sigma} \otimes \phi^* T^{1,0}_{\pi}$                             \\ 
		\tb $\!\!\psi_+^{\overline{p}}$      & $\!\!K_{\Sigma}^{1/2} \otimes \left( \phi^* T^{1,0}_{\pi} \right)^{\vee}$            & $\!\!K_{\Sigma}^{1/2} \otimes K_{\Sigma}^{-(1/2)(0)} \otimes \overline{K}_{\Sigma}^{+(1/2)(1)} \otimes \left( \phi^* T^{1,0}_{\pi} \right)^{\vee}$             & $\!\!\!\!\!\!\cong\left( \phi^* T^{1,0}_{\pi} \right)^{\vee}$                          \\ 
		\tb $\!\!\psi_-^p$                   & $\!\!\overline{K}_{\Sigma}^{1/2} \otimes \left( \phi^* T^{0,1}_{\pi} \right)^{\vee}$ & $\!\!\overline{K}_{\Sigma}^{1/2} \otimes K_{\Sigma}^{-(1/2)(-1)} \otimes \overline{K}_{\Sigma}^{+(1/2)(0)} \otimes \left( \phi^* T^{0,1}_{\pi} \right)^{\vee}$ & $\!\!\!\!\!\!\cong\left( \phi^* T^{0,1}_{\pi} \right)^{\vee}$                          \\ 
		\tb $\!\!\psi_-^{\overline{p}}$      & $\!\!\overline{K}_{\Sigma}^{1/2} \otimes \phi^* T^{0,1}_{\pi}$                       & $\!\!\overline{K}_{\Sigma}^{1/2} \otimes K_{\Sigma}^{-(1/2)(+1)} \otimes \overline{K}_{\Sigma}^{+(1/2)(0)} \otimes \phi^* T^{0,1}_{\pi}$                       & $\!\!\!\!\!\!\cong\overline{K}_{\Sigma} \otimes \phi^* T^{0,1}_{\pi}$                  \\
		\tb $\!\!p$                          & $\!\!\phi^* T_{\pi}$                                                                 & $\!\!K_{\Sigma}^{-1/2(-1)} \otimes \overline{K}_{\Sigma}^{+1/2(-1)} \otimes \phi^* T_{\pi}$                                                                    & $\!\!\!\!\!\!\cong K_{\Sigma} \otimes \phi^* T_{\pi}$                                   \\ 
	  \hline
	\end{tabular}
	\caption{Various fields in the Landau-Ginzburg description of \protect \( {\cal O}(-5)\rightarrow {\mathbf P}^4 \protect \).}
	\label{table:quinticFieldsAndBundles}
  \end{center}
\end{table}

Note that in this twisting, we must twist a bosonic field, the $p$
field.  Let us take a moment to discuss some details of this.
Because the theory has (2,2) supersymmetry, all target-space-metric-dependent
terms are determined by a K\"ahler potential, and because of the existence
of the isometry, the $p$ field enters the K\"ahler potential only in
the combination $|p|^2$.  Mechanically, we perform the twist by
Taylor expanding\footnote{Of course, not all smooth functions are
real analytic.  However, in order to make sense of NLSMs
as quantum field theories, there is always an implicit
assumption that metrics and so forth all admit Taylor series expansions
with nonzero radius of convergence, so we are not assuming anything that
is not routinely assumed by others.} the K\"ahler potential in powers of
$|p|^2$ and then replacing each $|p|^2$ with $g_{\Sigma}^{z \overline{z}}
p_z \overline{p}_{\overline{z}}$, where $g_{\Sigma}^{z \overline{z}}$
is the inverse of the worldsheet metric.  Since we are just manipulating
the K\"ahler potential, the result is guaranteed to still possess
(2,2) supersymmetry.  Moreover, because the kinetic terms for the $p$
field are quadratic in derivatives rather than linear, we do not have to
worry about picture-changing or other subtleties of
linear bosonic kinetic terms described in \cite{fms}.

In passing, note that if the superpotential were $W = p^k G$ for $k>1$,
where $G$ is a section of a line bundle, then we would not be
able to perform the A twist above because the charge-integrality condition
would fail.

These constraints on the space $X$ and the superpotential $W: X \rightarrow
{\bf C}$ -- namely existence of a suitable isometry plus the integral
charge condition -- might conceivably be strong enough to insure that the
only theories one can consider (for this notion of A-twisting) are ones in
which $X$ is the total space of a vector bundle and $W = p G(\phi)$, for
$G$ a section of the dual bundle, theories which RG flow to NLSMs on the
locus $\{ G = 0 \}$.  Certainly we have not been able to find any twistable
examples not of this form.  Even if this is true, however, these methods
are still useful in that they give alternative physical computations of
A-twisted NLSM correlation functions, as well as insight into direct
computations in GLSMs.  

Given the nature of the twist, namely that we want to make one of the
bosonic fields a section of a nontrivial line bundle, it would be
extremely convenient if the total space of a vector bundle were to
admit a metric
with the following two properties:
\begin{enumerate}
\item It should be block-diagonal:  using a splitting defined by a connection
(defined by a holomorphic structure plus fiber metric), the metric can be
written in block-diagonal form with one block for the metric along the
fibers and another for the metric along the base, with no
mixed fiber/base metric components.
\item The metric should be independent of position along fiber directions.
This would help simplify the meaning of ``twisted $p$ fields,'' as metric
components would have no $p$-dependence.
\end{enumerate}

Unfortunately, atlases with patches covered by
such metrics only seem to exist for flat\footnote{
In the case of a trivial bundle, metrics of this form have appeared previously
as, for example, ten-dimensional metrics about branes.
In such cases, the metric possesses a translation invariance parallel
to the brane (just as the desired metric form 
above possesses translation invariance
along fibers), and the warp factor corresponds to the metric components
along the fiber.
} vector bundles.
For example, if we were to start in one coordinate patch with a metric
of this form, then even the most nearly trivial coordinate transformation,
$x' = x(x)$, $p' = \Lambda p$ where $\Lambda$ are the transition functions
for the bundle, would result in off-diagonal metric components
proportional to $p \partial \Lambda / \partial x$, which could only
vanish for flat bundles.
We can also see the problem more formally as follows \cite{tonypriv}.
We would like that for each point $x \in X$, there exists an open
neighborhood $x \in U \subset X$ such that when we restrict the short
exact sequence
\begin{displaymath}
0 \: \longrightarrow \: \pi^* {\cal E}^{\vee} \: \longrightarrow
\: TX \: \longrightarrow \: \pi^* TB \: \longrightarrow \: 0
\end{displaymath}
to $U$, we can find a holomorphic splitting
\begin{displaymath}
s_U: \: \pi^* TB|_U \: \longrightarrow \: TX|_U,
\end{displaymath}
so that $s_U(\pi^* TB|_U)$ is orthogonal to $\pi^* {\cal E}^{\vee}$.
However, this places a global condition on ${\cal E}$:
the $C^{\infty}$ orthogonal complement of $\pi^* {\cal E} \subset TX$
is in fact a holomorphic subbundle of $TX$,
so that the sequence above
splits globally holomorphically on $X$.  The obstruction to splitting
this sequence is the Atiyah class of ${\cal E}^{\vee}$, and its
vanishing implies that all the Chern classes of ${\cal E}$ must vanish.

Now that we have a consistent twist in hand,
let us turn to other properties
of the action.  B-twisted Landau-Ginzburg models famously do not
have BRST-exact actions -- only their stress tensors are BRST exact,
which suffices for computations.  A-twisted Landau-Ginzburg models,
on the other hand, do have BRST-exact actions, unlike their B-twisted
counterparts.  If we let $a$ be an index running over all holomorphic
indices, regardless of twist, then the action for an A-twisted theory
can be written in the form
\begin{displaymath}
Q_{BRST} \cdot \left[ g_{a \overline{b}} \left( \psi_+^{\overline{b}} 
\overline{\partial} \phi^a \: + \: \psi_-^a \partial \phi^{\overline{b}}
\right) \: - \: i \left( \psi_+^{\overline{a}} \partial_{\overline{a}}
\overline{W} \: - \: \psi_-^a \partial_a W \right)
\right],
\end{displaymath}
where $Q_{BRST}$ is the BRST operator.
Up to boundary pieces, the terms of the base NLSM action are
the BRST variation of
\begin{displaymath}
g_{a \overline{b}} \left( \psi_+^{\overline{b}} 
\overline{\partial} \phi^a \: + \: \psi_-^a \partial \phi^{\overline{b}}
\right),
\end{displaymath}
and the superpotential terms are the BRST variation of
\begin{displaymath}
\psi_+^{\overline{a}} \partial_{\overline{a}}
\overline{W} \: - \: \psi_-^a \partial_a W.
\end{displaymath}

\subsection{Nonperturbative sectors}

In an ordinary NLSM, the nonperturbative sectors
are given by holomorphic maps from the worldsheet into the target space.
Let us take a moment to carefully work through the nonperturbative
sectors of a NLSM with superpotential.

The nonperturbative sectors should become manifest by rewriting
the bosonic part of the action as the absolute value square of
something.  Here, it is straightforward algebra to write
\begin{eqnarray*}
\lefteqn{
g_{\mu \nu} \partial \phi^{\mu} \overline{\partial} \phi^{\nu}
\: + \: i 
B_{\mu \nu} \partial \phi^{\mu} \overline{\partial} \phi^{\nu}
\: + \: 2 g^{i \overline{\jmath}} \partial_i W \partial_{\overline{\jmath}}
\overline{W} }
\\
& = & 2 g_{i \overline{\jmath}} \overline{\partial} \phi^i
\partial \phi^{\overline{\jmath}} \: + \: 2 g^{i \overline{\jmath}}
\partial_i W \partial_{\overline{\jmath}} \overline{W} \\
& & \: + \: 
\left( g_{i \overline{\jmath}} \: + \: i B_{i \overline{\jmath}} \right)
\left( \partial \phi^i \overline{\partial} \phi^{\overline{\jmath}}
\: - \: \overline{\partial} \phi^i \partial \phi^{\overline{\jmath}}
\right) \\
& = & 2 g_{i \overline{\jmath}} \left(
\overline{\partial} \phi^i \: - \: i g^{i \overline{k}} \partial_{\overline{k}}
\overline{W} \right) \left(
\partial \phi^{\overline{\jmath}} \: + \: i g^{\overline{\jmath} m}
\partial_{m} W \right) \\
& & \: + \:
\left( g_{i \overline{\jmath}} \: + \: i B_{i \overline{\jmath}} \right)
\left( \partial \phi^i \overline{\partial} \phi^{\overline{\jmath}}
\: - \: \overline{\partial} \phi^i \partial \phi^{\overline{\jmath}}
\right)
\: + \: 2 i \left( \partial \overline{W} \: - \: \overline{\partial} W
\right).
\end{eqnarray*}

In the last equation above, the terms
\begin{displaymath}
\left( g_{i \overline{\jmath}} \: + \: i B_{i \overline{\jmath}} \right)
\left( \partial \phi^i \overline{\partial} \phi^{\overline{\jmath}}
\: - \: \overline{\partial} \phi^i \partial \phi^{\overline{\jmath}}
\right)
\: + \: 2 i \left( \partial \overline{W} \: - \: \overline{\partial} W
\right)
\end{displaymath}
are purely topological -- the first set of terms are
the pullback of the complexified K\"ahler parameter, and the
second set, proportional to $\partial \overline{W} - \overline{\partial} W$,
give another topological class described in {\it e.g.} \cite{hiv}.
The non-topological part of the kinetic terms can be written as the
absolute-value-square of the quantity $\overline{\partial} \phi^i - i
g^{i \overline{\jmath}} \partial_{\overline{\jmath}} \overline{W}$,
which suggests that instead of working on moduli spaces of holomorphic
curves, we should work on moduli spaces of solutions to the partial
differential equation
\begin{equation}  \label{wittenequation}
\overline{\partial} \phi^i \: - \: i
g^{i \overline{\jmath}} \partial_{\overline{\jmath}} \overline{W}
\: = \: 0.
\end{equation}
This was called the `Witten equation' in \cite{fjr1,fjr2}.  
Further support from this hypothesis follows from the fact that
if we apply the usual topological-field-theory idea of localization,
then since there are BRST variations of the form
\begin{eqnarray*}
\delta \psi_+^{\overline{a}} & = & - \alpha \left(
\partial \phi^{\overline{a}} \: + \: i g^{\overline{a} b} \partial_b W 
\right)  \: + \: \mbox{3-fermi terms}\\
\delta \psi_-^a & = & - \alpha \left(
\overline{\partial} \phi^a \: - \: i g^{a \overline{b}} \partial_{\overline{b}}
\overline{W} \right) \: + \: \mbox{3-fermi terms}
\end{eqnarray*}
(where the $a$ index runs over all holomorphic fields, regardless of twisting)
the topological field theory will localize on solutions of
equation~(\ref{wittenequation}) above.
The same conclusion was also reached in \cite{ito1}, and the same
equation has also previously appeared in \cite{hiv}.

However, there is a simplification.
Physically, there are really two distinct BRST scalars,
which were combined above into the single scalar $\alpha$,
and if we apply the notion of localization to each of them separately,
then we conclude that we must satisfy instead the two partial
differential equations
\begin{eqnarray*}
\overline{\partial} \phi^i & = & 0 \\
d W & = & 0.
\end{eqnarray*}
In other words, the two terms in the Witten equation must vanish
separately.
Furthermore, mathematically it can be shown that all the solutions of the
Witten equation~(\ref{wittenequation}) also necessarily solve the
pair of partial differential equations above -- the two terms
in the Witten equation must vanish separately.
The argument is essentially a repeat of the original derivation from the
kinetic terms above -- if we take the absolute-value-square of the
Witten equation, and integrate over the worldsheet, then we find that
\begin{displaymath}
\int_{\Sigma} g_{i \overline{\jmath}}
\left(
\overline{\partial} \phi^i \: - \: i g^{i \overline{k}} \partial_{\overline{k}}
\overline{W} \right) \left(
\partial \phi^{\overline{\jmath}} \: + \: i g^{\overline{\jmath} m}
\partial_{m} W \right)
\: = \: \int_{\Sigma} \left( 
g_{i \overline{\jmath}} \partial \phi^{\overline{\jmath}}
\overline{\partial} \phi^i \: + \: g^{i \overline{\jmath}}
\partial_i W \partial_{\overline{\jmath}} \overline{W} \right).
\end{displaymath}
Since the right hand side is a sum of absolute squares, the only way that
the Witten equation can be satisfied, the only way that the left hand side
can vanish, is if each of the terms on the right hand side vanishes
separately.  Thus, every solution of the Witten equation must also satisfy
\begin{eqnarray*}
\overline{\partial} \phi^i & = & 0 \\
d W & = & 0,
\end{eqnarray*}
the same equations we found above from the stronger form of topological
field theory localization.

So, in spite of our initial concerns, the nonperturbative sectors will all
arise from holomorphic curves.
In the next few sections, we shall check our methods by computing
nonperturbative corrections to Landau-Ginzburg models in the same
universality classes as NLSMs on nontrivial
Calabi-Yaus, and checking that correlation functions match.

\subsection{Example:  the quintic} 
\label{ex:22quintic}

For our first example, we will study the A twist of the
Landau-Ginzburg model on 
\begin{displaymath}
\mbox{Tot}( {\cal O}(-5) \: \longrightarrow \:
{\bf P}^4 ),
\end{displaymath}
with superpotential $W = p G(\phi)$.
This is the theory that should flow under the renormalization
group to a NLSM on the quintic;
hence, A model correlation functions here should match those
on the quintic, which will provide a strong consistency check
of our methods.  In fact, we will see that the Landau-Ginzburg model
computation gives a physical realization of some tricks for computing
Gromov-Witten invariants due to Kontsevich and others.

The local coordinates on ${\bf P}^4$ are twisted in the usual fashion:
\begin{align*}
\psi_+^i ( \equiv \chi^i ) \: &\in \: \Gamma_{C^{\infty}}\left( \phi^* T^{1,0}{\bf P}^4 \right)                                                                   & \psi_-^i ( \equiv \psi_{\overline{z}}^i ) \: &\in \: \Gamma_{C^{\infty}}\left( \overline{K}_{\Sigma} \otimes (\phi^* T^{0,1}{\bf P}^4)^{\vee} \right) \\
\psi_+^{\overline{\imath}} ( \equiv \psi_z^{\overline{\imath}} ) \: &\in \: \Gamma_{C^{\infty}}\left( K_{\Sigma} \otimes (\phi^* T^{1,0}{\bf P}^4)^{\vee} \right) & \psi_-^{\overline{\imath}} ( \equiv \chi^{\overline{\imath}} ) \: &\in \: \Gamma_{C^{\infty}}\left( \phi^* T^{0,1} {\bf P}^4 \right).
\end{align*}
The fermionic superpartners of the $p$ field, on the other hand,
are twisted differently:
\begin{align*}
\psi_+^p (\equiv \psi_z^p) \:                           & \in \: \Gamma_{C^{\infty}}\left( K_{\Sigma} \otimes \phi^* T^{1,0}_{\pi} \right) & \psi_-^p (\equiv \chi^p) \:                                            & \in \: \Gamma_{C^{\infty}}\left( ( \phi^* T^{0,1}_{\pi} )^{\vee} \right) \\
\psi_+^{\overline{p}} ( \equiv \chi^{\overline{p}} ) \: & \in \: \Gamma_{C^{\infty}}\left( ( \phi^* T^{1,0}_{\pi} )^{\vee} \right)         & \psi_-^{\overline{p}} ( \equiv \psi_{\overline{z}}^{\overline{p}} ) \: & \in \: \Gamma_{C^{\infty}}\left( \overline{K}_{\Sigma}\otimes \phi^* T^{0,1}_{\pi} \right),
\end{align*}
where $T_{\pi}$ denotes the relative tangent bundle of the projection
\begin{displaymath}
\pi: \:
\mbox{Tot}\left( {\cal O}(-5) \: \longrightarrow \: {\bf P}^4 \right)
\: \longrightarrow \: {\bf P}^4.
\end{displaymath}
To be consistent, we also have to twist the $p$ field itself:
\begin{eqnarray*}
p \, (\equiv p_z) 
& \in & \Gamma_{C^{\infty}}\left( K_{\Sigma}
\otimes \phi^* T^{1,0}_{\pi} \right) \\
\overline{p} \, (\equiv \overline{p}_{\overline{z}} ) 
& \in &
\Gamma_{C^{\infty}}\left( \overline{K}_{\Sigma} \otimes
\phi^* T^{0,1}_{\pi} \right),
\end{eqnarray*}
whereas the bosons $\phi^i$ remain untwisted.
(Twisted fermions are slightly unusual to work with;
we shall take their zero mode integration to be an ordinary integral
over the vector space of sections.)

Now, let us check carefully
that the twisting above is consistent with supersymmetry
transformations.
In the A-twisted theory, the BRST transformation parameters will be
$\alpha_-$ and $\tilde{\alpha}_+$, which are Grassmann constants,
whereas $\alpha_+$ is a section of $\overline{K}_{\Sigma}^{-1}$
and $\tilde{\alpha}_-$ is a section of $K_{\Sigma}^{-1}$.

Since the $\phi^i$ multiplets have the standard A twist,
we only need check the $p$ multiplet.
The supersymmetry transformations in the $p$ multiplet are
\begin{eqnarray*}
\delta p & = & i \alpha_- \psi_+^p \: + \: i \alpha_+ \psi_-^p \\
\delta \overline{p} & = & i \tilde{\alpha}_- \psi_+^{
\overline{p}} \: + \: i \tilde{\alpha}_+ \psi_-^{\overline{p}} \\
\delta \psi_+^p & = & - \tilde{\alpha}_- \partial p 
\: - \: i \alpha_+ \psi_-^j \Gamma^p_{j m} \psi_+^m
\: - \: i \alpha_+ \psi_-^p \Gamma^p_{p m} \psi_+^m
\: - \: i \alpha_+ \psi_-^j \Gamma^p_{j p} \psi_+^p
\: - \: i \alpha_+ \psi_-^p \Gamma^p_{p p} \psi_+^p
\\
& &
\: - \:
i \alpha_+ g^{p \overline{\jmath}} \partial_{\overline{\jmath}} \overline{W}
\: - \:
i \alpha_+ g^{p \overline{p}} \partial_{\overline{p}} \overline{W}
\\
\delta \psi_+^{\overline{p}} & = & - \alpha_- \partial 
\overline{p}
\: - \: i \tilde{\alpha}_+ \psi_-^{\overline{\jmath}} \Gamma^{\overline{p}}_{
\overline{\jmath} \overline{m}} \psi_+^{\overline{m}}
\: - \: i \tilde{\alpha}_+ \psi_-^{\overline{p}} \Gamma^{\overline{p}}_{
\overline{p} \overline{m}} \psi_+^{\overline{m}}
\: - \: i \tilde{\alpha}_+ \psi_-^{\overline{\jmath}} \Gamma^{\overline{p}}_{
\overline{\jmath} \overline{p}} \psi_+^{\overline{p}}
\: - \: i \tilde{\alpha}_+ \psi_-^{\overline{\jmath}} \Gamma^{\overline{p}}_{
\overline{p} \overline{p}} \psi_+^{\overline{p}}
\\
& &
\: - \: i \tilde{\alpha}_+ g^{\overline{p} j} \partial_j W 
\: - \: i \tilde{\alpha}_+ g^{\overline{p} p} \partial_p W\\
\delta \psi_-^p & = & - \tilde{\alpha}_+ \overline{\partial} p 
\: - \: i \alpha_- \psi_+^j \Gamma^p_{j m} \psi_-^m
\: - \: i \alpha_- \psi_+^p \Gamma^p_{p m} \psi_-^m
\: - \: i \alpha_- \psi_+^j \Gamma^p_{j p} \psi_-^p
\: - \: i \alpha_- \psi_+^p \Gamma^p_{p p} \psi_-^p
\\ 
& &
\: + \: i \alpha_- g^{p \overline{\jmath}} \partial_{\overline{\jmath}}
\overline{W} 
\: + \: i \alpha_- g^{p \overline{p}} \partial_{\overline{p}}
\overline{W}
\\
\delta \psi_-^{\overline{p}} & = & - \alpha_+ \overline{\partial}
\overline{p} 
\: - \: i \tilde{\alpha}_- \psi_+^{\overline{\jmath}}
\Gamma^{\overline{p}}_{\overline{\jmath} \overline{m}}
\psi_-^{\overline{m}}
\: - \: i \tilde{\alpha}_- \psi_+^{\overline{p}} 
\Gamma^{\overline{p}}_{\overline{p} \overline{m}}
\psi_-^{\overline{m}}
\: - \: i \tilde{\alpha}_- \psi_+^{\overline{\jmath}}
\Gamma^{\overline{p}}_{\overline{\jmath} \overline{p}}
\psi_-^{\overline{p}}
\: - \: i \tilde{\alpha}_- \psi_+^{\overline{p}}
\Gamma^{\overline{p}}_{\overline{p} \overline{p}}
\psi_-^{\overline{p}}
\\
& & 
\: + \: i \tilde{\alpha}_- g^{\overline{p} j} \partial_j W
\: + \: i \tilde{\alpha}_- g^{\overline{p} p} \partial_p W.
\end{eqnarray*}
Now, looking at the transformations above, the reader might be
concerned.  For example, the supersymmetry variation of
$p$, which is a section of $K_{\Sigma} \otimes \phi^* T^{1,0}_{\pi}$,
contains a term proportional to $\alpha_+ \psi_-^p$
 -- but $\alpha_+$ is a section of $\overline{K}_{\Sigma}^{-1}$,
and $\psi_-^p$ is a section of $( \phi^* T^{0,1}_{\pi} )^{\vee}$,
so we appear to have inconsistent bundles.
The fix is that as $C^{\infty}$ bundles, a holomorphic bundle
${\cal E}$ is isomorphic (via a hermitian fiber metric) to the
antiholomorphic bundle $\overline{ {\cal E} }^{\vee}$.
Thus, for example, as $C^{\infty}$ bundles,
\begin{displaymath}
K_{\Sigma} \otimes \phi^* T^{1,0}_{\pi}
\: \cong \:
\overline{K}_{\Sigma}^{-1} \otimes ( \phi^* T^{0,1}_{\pi} )^{\vee}.
\end{displaymath}
Applying this notion to the other supersymmetry transformations,
we find that
the twist of the $p$ fields is consistent with supersymmetry.

The BRST transformations of the fields are given by
\begin{eqnarray*}
\delta \phi^i & = & i \alpha \chi^i \\
\delta \phi^{\overline{\imath}} & = & i \alpha \chi^{\overline{\imath}} \\
\delta \chi^i & = & 0 \\
\delta \chi^{\overline{\imath}} & = & 0 \\
\delta \psi_z^{\overline{\imath}} & = &
- \alpha \partial \phi^{\overline{\imath}} \: - \:
i \alpha \chi^{\overline{\jmath}} \Gamma^{\overline{\imath}}_{\overline{\jmath} 
\overline{m}} \psi_z^{\overline{m}} \: - \:
i \alpha \psi_{\overline{z}}^{\overline{p}} 
\Gamma^{\overline{\imath}}_{\overline{p}
\overline{m}} \psi_z^{\overline{m}} \: - \:
i \alpha \chi^{\overline{\jmath}} \Gamma^{\overline{\imath}}_{\overline{\jmath}
\overline{p}} \chi^{\overline{p}}
\: - \: i \alpha \psi_{\overline{z}}^{\overline{p}} \Gamma^{\overline{\imath}}_{
\overline{p} \overline{p}} \chi^{\overline{p}}
\\
& & 
\: - \:
i \alpha g^{\overline{\imath} j}\partial_j W 
\: - \:
i \alpha g^{\overline{\imath} p} \partial_p W
\\
\delta \psi^i_{\overline{z}} & = &
- \alpha \overline{\partial} \phi^i \: - \:
i \alpha \chi^j \Gamma^i_{j m} \psi_{\overline{z}}^m \: - \:
i \alpha \psi_z^p \Gamma^i_{p m} \psi_{\overline{z}}^m 
\: - \: i \alpha \chi^j \Gamma^i_{j p} \chi^p
\: - \: i \alpha \psi_z^p \Gamma^i_{p p} \chi^p
\\
& &
\: + \:
i \alpha g^{i \overline{\jmath}} \partial_{\overline{\jmath}} 
\overline{W} 
\: + \: i \alpha g^{i \overline{p}} \partial_{\overline{p}}
\overline{W}
\\
\delta p_z & = & i \alpha \psi_z^p \\
\delta \overline{p}_{\overline{z}} & = & 
i \alpha \psi_{\overline{z}}^{\overline{p}} \\
\delta \chi^p & = & - \alpha \overline{\partial} p_z 
\: - \: i \alpha \chi^j \Gamma^p_{j m} \psi_{\overline{z}}^m
\: - \: i \alpha \psi_z^p \Gamma^p_{p m} \psi_{\overline{z}}^m
\: - \: i \alpha \chi^j \Gamma^p_{j p} \chi^p
\: - \: i \alpha \psi_z^p \Gamma^p_{p p} \chi^p
\\
& & \: + \: i \alpha
g^{p \overline{p}} \partial_{\overline{p}} W 
\: + \: i \alpha g^{p \overline{\imath}} \partial_{\overline{\imath}} W
\\
\delta \chi^{\overline{p}} & = &
- \alpha \partial \overline{p}_{\overline{z}} 
\: - \: i \alpha \chi^{\overline{\jmath}} \Gamma^{\overline{p}}_{
\overline{\jmath} \overline{m}} \psi_z^{\overline{m}}
\: - \: i \alpha \psi_{\overline{z}}^{\overline{p}} \Gamma^{\overline{p}}_{
\overline{p} \overline{m}} \psi_z^{\overline{m}}
\: - \: i \alpha \chi^{\overline{\jmath}} \Gamma^{\overline{p}}_{
\overline{\jmath} \overline{p}} \chi^{\overline{p}}
\: - \: i \alpha \psi_{\overline{z}}^{\overline{p}} \Gamma^{\overline{p}}_{
\overline{p} \overline{p}} \chi^{\overline{p}}
\\
& & \: - \:
i \alpha g^{\overline{p} p} \partial_p W 
\: - \: 
i \alpha g^{\overline{p} i} \partial_i W
\\
\delta \psi_z^p & = & 0 \\
\delta \psi_{\overline{z}}^{\overline{p}} & = & 0,
\end{eqnarray*}
where $\alpha = \alpha_- = \tilde{\alpha}_+$.

Let us take a moment to examine the chiral ring in this theory.
The BRST-invariant worldsheet scalars are the $\chi^i$ and
$\chi^{\overline{\imath}}$, and the BRST operator appears to act like
the exterior derivative $d$, so naively one would conclude that
the chiral ring consists of $d$-closed differential forms on 
${\bf P}^4$:
\begin{eqnarray*}
\lefteqn{
b_{i_1 \cdots i_n \overline{\jmath}_1 \cdots
{\jmath}_m}(\phi) \chi^{i_1} \cdots \chi^{i_n}
\chi^{\overline{\jmath}_1} \cdots
\chi^{\overline{\jmath}_m} 
} \\
& \hspace*{1in}  \leftrightarrow &
b_{i_1 \cdots i_n \overline{\jmath}_1 \cdots
{\jmath}_m}(\phi)
dz^{i_1} \wedge \cdots \wedge dz^{i_n} \wedge
d \overline{z}^{\overline{\jmath}_1} \wedge \cdots
\wedge
d \overline{z}^{\overline{\jmath}_m}.
\end{eqnarray*}
However, we need to be slightly careful.  
When constructing such dictionaries in NLSMs,
it is assumed that the bosonic zero modes can wander freely over all
of the space in question, but here, there is a nontrivial superpotential.
Since the superpotential is BRST exact, we can rescale it without changing
the chiral ring, so for simplicity let us rescale $W \mapsto \lambda W$
and take $\lambda \rightarrow \infty$.  In this limit, the bosonic zero modes
are effectively restricted to
live on the quintic.  Therefore, we claim that the correct chiral ring
is not differential forms on ${\bf P}^4$, but rather
the restriction of differential forms on ${\bf P}^4$ to the quintic.
(Indeed, given that this theory RG flows to a NLSM on the
quintic, the chiral ring had better turn out to be the cohomology of the
quintic and not ${\bf P}^4$.)
Further evidence for this interpretation is provided by the discussion
of localization in the topological field theory, in the last section.
We will get further evidence for this in the next section when we study
correlation functions, where we will see that the product structure only
sees the restriction to the quintic.

The full set of interactions derived from the superpotential 
$W = p G(\phi)$
have
the form
\begin{eqnarray*}
\lefteqn{
L_W \: = \: 2 g^{p \overline{p}} |G(\phi)|^2 \: + \:
2 g^{i \overline{\jmath}} p_z \overline{p}_{\overline{z}} D_i G
D_{\overline{\jmath}} \overline{G} } \\
& & \: + \: 2 g^{p \overline{\imath}} G \overline{p}_{\overline{z}}
D_{\overline{\imath}} \overline{G} \: + \:
2 g^{\overline{p} i} \overline{G} p_z D_i G \\
& & \: + \: 
\psi^p_z \psi^i_{\overline{z}} D_i G \: + \:
\chi^i \chi^p D_i G \: + \:
\chi^i \psi^j_{\overline{z}} p_z D_i D_j G \\
& & \: + \:
\chi^{\overline{p}} \chi^{\overline{\jmath}} 
D_{\overline{\jmath}} \overline{G} \: + \:
\psi_z^{\overline{\imath}} \psi^{\overline{p}}_{\overline{z}}
D_{\overline{\imath}} \overline{G} \: + \:
\psi_z^{\overline{\imath}} \chi^{\overline{\jmath}} 
\overline{p}_{\overline{z}}
D_{\overline{\imath}} D_{\overline{\jmath}} \overline{G}.
\end{eqnarray*}
In addition, there are four-fermi interactions that have
the general form
\begin{displaymath}
L_R \: = \: R_{I \overline{J} K \overline{L}} \chi^I \psi_z^{\overline{J}} \psi_{\overline{z}}^K \chi^{\overline{L}},
\end{displaymath}
where the capitalized indices in the expression above run over both base and fiber, not just the base.

We will compute the induced effective interactions in each instanton sector.
In many cases, there will not be enough zero modes to allow 
any contribution from most four-fermi terms, which will greatly
simplify parts of the analysis.

\subsubsection{Classical contribution -- worldsheet genus zero}

Let us now outline how one computes the classical contribution to
correlation functions in this twisted theory.  Since this theory
should be in the same universality class as a NLSM
on the quintic, and topological field theory correlation functions
are independent of universality class representative, we should (and will)
find
the same correlation functions as the classical contribution to
the A-twisted NLSM on the quintic.

For simplicity, we will only work on ${\bf P}^1$.
We can read off from the twist definitions that there are no
$p$, $\overline{p}$, $\psi^i_{\overline{z}}$, $\psi^{\overline{\imath}}_z$,
$\psi^p_z$, or $\psi^{\overline{p}}_{\overline{z}}$ zero modes.
There are as many $\chi^i$ zero modes as the dimension of 
${\bf P}^4$, and as many $\chi^p$ zero modes as the rank of
${\cal O}(-5)$.

Since the Riemann curvature four-fermi terms have factors of
$\psi^i_{\overline{z}}$, $\psi^{\overline{\imath}}_z$,
$\psi_z^p$, and $\psi_{\overline{z}}^{\overline{p}}$,
they do not generate an effective interaction on the classical
component of the bosonic moduli space, with the exception of the
single term that only couples to $\chi$'s:
\begin{displaymath}
R_{i \overline{p} p \overline{k} } \chi^i \chi^{\overline{p}}
\chi^p \chi^{\overline{k}}.
\end{displaymath}

Similarly, most of the terms in $L_W$ do not generate an effective
interaction either, with the exception of the terms
\begin{displaymath}
2 g^{p \overline{p}} |G(\phi)|^2 \: + \:
\chi^i \chi^p D_i G \: + \: 
\chi^{\overline{p}} \chi^{\overline{\jmath}} D_{\overline{\jmath}} 
\overline{G}
\end{displaymath}
(Each of these terms should be evaluated at $p_z = \overline{p}_{\overline{z}}
= 0$, since $p_z$, $\overline{p}_{\overline{z}}$ do not have zero modes.)

There is one other technical remark that should be made, regarding the
normalization of the zero mode part of the path integral measure.
Ordinarily we normalize worldsheet fermion vectors by factors of
$\sqrt{\alpha'}$ so that the path integral measure remains unitless.
Here, we must treat the $\chi^p$ zero modes carefully.
Although they are worldsheet scalars, not vectors, their zero
modes will also require
factors of $\sqrt{\alpha'}$.
We can see this as follows.
To make sense of the BRST
transformations $\delta \chi^p = - \alpha \overline{\partial} p_z + \cdots$
discussed earlier, there is implicitly a factor of the inverse worldsheet
metric $g_{\Sigma}^{z \overline{z}}$ on the right-hand size of the BRST
transformations\footnote{This is also related to the discussion in
the previous section regarding $K_{\Sigma}$ versus
$\overline{K}_{\Sigma}^{-1}$ as smooth bundles.}.  As a result,
the $\chi^p$'s scale nontrivially under worldsheet metric rescalings.
In order to make the path integral measure scale
invariant, we must multiply
the $d \chi^p$ factors in the path integral measure by a factor of
$\sqrt{A}^{-1}$, where $A$ is the worldsheet area.  Once we have done so, we
must then also multiply the $d \chi^p$ factors by $\sqrt{\alpha'}$ to make the
path integral measure unitless.

Correlation functions thus take the form
\begin{eqnarray*}
\lefteqn{
\langle {\cal O}_1 \cdots {\cal O}_n \rangle \: = \:
\int_{ {\bf P}^4 } d^2 \phi^i \int \prod_i d \chi^i d \chi^{
\overline{\imath}} \left(
\sqrt{ \frac{\alpha'}{A \lambda^2} }d \chi^p \right)
\left( \sqrt{ \frac{\alpha'}{A \lambda^2} } d \chi^{\overline{p}} \right)
\,
{\cal O}_1 \cdots {\cal O}_n } \\
& \hspace{1in} & \cdot \exp\left( - 2 \frac{A}{\alpha'} \lambda^2
g^{p \overline{p}} | G(\phi) |^2 \: - \: 
\frac{A}{\alpha'} \lambda^2 \chi^i \chi^p D_i G \: - \:
\frac{A}{\alpha'} \lambda^2 \chi^{\overline{p}} \chi^{\overline{\jmath}}
D_{\jmath} \overline{G} \right. \\
& \hspace{1in} & \hspace*{0.5in} \left.
\: - \: \frac{A}{\alpha'} \lambda^2 
R_{i \overline{p} p \overline{k} } \chi^i \chi^{\overline{p}}
\chi^p \chi^{\overline{k}}
\right),
\end{eqnarray*}
where $A$ is the area of the worldsheet and $\lambda$ is a worldsheet
metric rescaling, just as in our review of the closed string B model.

Let us pause for a moment to examine the results so far.
Omitting factors of $A$, $\lambda$, $\alpha'$, $g^{p \overline{p}}$,
the exponential in the correlation function can be written schematically as
\begin{equation}  \label{mqform}
\exp\left(
- 2 |G|^2 \: - \: \chi^i \chi^p D_i G \: - \: \chi^{\overline{p}}
\chi^{\overline{\jmath}} D_{\overline{\jmath}} \overline{G}
\: - \:
R_{i \overline{p} p \overline{k} } \chi^i \chi^{\overline{p}}
\chi^p \chi^{\overline{k}}
\right),
\end{equation}
which some readers will recognize\footnote{
For the benefit of experts, let us work through the identification in
more detail here.  Since we are examining the classical contribution,
the map $\phi$ is constant, and one can see that the expression in the
path integral determines a representative of the Thom class of the
dual of the vertical subbundle $T_\pi \subset T{\cal O}(-5)$, pulled back to ${\bf P}^4$ by the section
$G(\phi)$.  Here, $\chi^i D_i G$ is the covariant
derivative of the section, while $\chi^p$ plays the role of antighost.  It is
not too hard to see that along the $p=0$ locus (to which we have restricted,
since there are no $p$ zero modes) that $R_{i \overline{p} p \overline{k}} 
\chi^i \chi^{\overline{k}}$ is the curvature two-form associated to
$T_{\pi}$, so that the expression~(\ref{mqform}) is, up to a factor
of $1/\pi$, a normalized Thom form.  Furthermore, by the standard
localization argument ({\it c.f.} section 11.10 of \cite{cmr}), this form
has support along the zero locus of the section ($G(\phi) = 0$).
} as the Mathai-Quillen representative
\cite{mathaiquillen} of the Thom class 
of the vertical subbundle of $T\mathcal O(5)$, pulled back by the
section $G$.
See section 6 of \cite{botttu} for a general discussion of Thom classes,
equation (2.8) of \cite{vw} for an expression nearly identical to
\eqref{mqform}, or alternatively section 11 of \cite{cmr}.  The Thom class
is a cohomology class one can insert in integrals over total spaces of
vector bundles, to give a result equivalent to integrating only over the
base:  if $p: V \rightarrow M$ is a vector bundle over $M$ and $T$ a
differential form representing the Thom class, then for any differential
form $\alpha$ on $M$,
\begin{displaymath}
\int_V p^* \alpha \wedge T \: = \: \int_M \alpha.
\end{displaymath}
Here, the presence of the Thom class means that the correlation function
will be equivalent to one computed by 
integrating only over the locus $\{ G = 0 \}
\subset {\bf P}^4$, exactly as one would expect from the renormalization
group arguments already presented.

Mathai-Quillen representatives of Thom classes are independent of the scale
of the bundle metric, in the sense that the cohomology class does not
change when the metric is scaled \cite{cmr}[section 11.8].  Thus, the
correlators cannot know about the ${\bf P}^4$ metric outside of the
quintic, since they cannot see it in the $\lambda \rightarrow \infty$
limit.  One fully expects this from renormalization group arguments -- the only
metric that correlators should care about is the one on the quintic.

Let us now return to a more concrete analysis of the
correlation function.
Performing the $\chi^p$ and $\chi^{\overline{p}}$ Grassmann integrals
gives
\begin{eqnarray*}
\lefteqn{
\langle {\cal O}_1  \cdots {\cal O}_n \rangle \: = \:
\int_{ {\bf P}^4 } d^2 \phi^i \int \prod_i d \chi^i d \chi^{
\overline{\imath}} {\cal O}_1 \cdots {\cal O}_n
\left(
\frac{A}{\alpha'} \lambda^2 \left| D_i G \chi^i \right|^2 g^{p \overline{p}}
\: + \:
R_{i \overline{p} p \overline{k} } g^{p \overline{p}} \chi^i 
 \chi^{\overline{k}}
\right) 
} \\
& & \hspace*{3.9in} \cdot
\exp\left(  - 2 \frac{A}{\alpha'} \lambda^2
g^{p \overline{p}} | G(\phi) |^2 \right).
\end{eqnarray*}
The $g^{p \overline{p}}$ factors on the $| \chi^i D_i G |^2$ term,
as well as on the $R_{i \overline{p} p \overline{k}}$ term
-- necessary to properly contract the sections of ${\cal O}(5)$ --
arise because of a subtlety in the $\chi^{\overline{p}}$ zero modes.
Strictly speaking, the zero modes are the zero modes of\footnote{
This is discussed in a different context in, for example,
\cite{ks1}.  The point is that in order to make sense of solutions of
$D_{\overline{z}, z} \psi$ as holomorphic, antiholomorphic sections
of suitable (anti)holomorphic vector bundles, if we take
say $\psi^i$ to be a section of $T^{1,0}X$, say, then Riemann-Roch
pairs it with a section of $K \otimes \left( T^{0,1}X \right)^{\vee}$,
and not $T^{0,1}X$ -- the two are related by a choice of metric,
one is $\psi^{\overline{\imath}}$ and the other is $\psi_i$,
but in order to get the zero-mode counting right, we have to be
careful about our conventions.}
$\chi_p = g_{p \overline{p}} \chi^{\overline{p}}$, so 
the Yukawa coupling
\begin{displaymath}
\chi^{\overline{p}} \chi^{\overline{\imath}} D_{\overline{\imath}}
\overline{G}
\end{displaymath}
should be written
\begin{displaymath}
\chi_{p} g^{p \overline{p}}
\chi^{\overline{\imath}} D_{\overline{\imath}}
\overline{G}.
\end{displaymath}
Then, when we integrate out the $\chi^p$, $\chi_p$ zero modes, 
the result is an extra $g^{p \overline{p}}$ factor, exactly as needed for
a consistent result.

The factors of $\chi^i$ in 
$| D_i G  \chi^i |^2$ and $R_{i \overline{p} p \overline{\jmath}}
\chi^i \chi^{\overline{\jmath}}$
are responsible for a selection rule that ensures that the sum of the
$U(1)$ charges of the ${\cal O}_i$, add up to form a top-form on the
quintic $\{ G = 0 \}$, instead of the ambient space ${\bf P}^4$,
exactly as desired.

Since this is a topological field theory, the correlation function
should be independent of $\alpha'$, worldsheet area $A$, and 
worldsheet metric rescaling $\lambda$.
In the two limits $\lambda \rightarrow 0, \infty$ the factors of
$\alpha'$, $A$, and $\lambda$ cancel out:  in the limit $\lambda \rightarrow
0$ there are manifestly no remaining such factors, and in the other
limit the factors multiplying the Gaussian are canceled by a factor
of $( \sqrt{ \alpha' / (A \lambda^2) } )^2$ one gets from performing
the Gaussian.  For intermediate scales, the fact that the correlation
function is independent of those quantities follows mathematically
from the fact \cite{cmr}[section 11.8] that
varying their values just multiplies the Mathai-Quillen form by an
exact form, which then falls out of the integral.  
Physically, the result follows from the fact that the superpotential
is BRST-exact.

Since this Landau-Ginzburg model should be in the same universality
class as a NLSM on the quintic $\{ G = 0 \} \subset
{\bf P}^4$, it should have the same correlation functions as the
A-twisted NLSM on the quintic.
We will check this in the two scaling limits, 
$\lambda \rightarrow \infty$ and $\lambda \rightarrow 0$,
in which $G \mapsto \lambda G$.  Since the superpotential is
BRST-exact, the results should be independent of such rescalings of
$G$. Indeed, we will find the same correlation
function (matching correlation functions of the NLSM),
though realized in two different ways -- the rescaling will interpolate\footnote{
This is a standard characteristic of Mathai-Quillen representatives of
Thom classes, see for example \cite{vw}[section 2.1] or
\cite{cmr}[section 11.10.2], and we will see this again many times in
this paper.
}
between Euler class insertions and computations in the style
of virtual fundamental classes.

First, let us recall the classical contributions to genus-zero
A-twisted NLSM correlation functions on the quintic.
The correlation functions are given by integrals over the quintic,
which can also be written as
integrals over ${\bf P}^4$ with insertions of the Euler class of
${\cal O}(-5)$.

Now, let us compare to the scaling limit $\lambda \rightarrow 0$.
In this limit, the correlation functions can be expressed as
\begin{equation}
\langle {\cal O}_1 \cdots {\cal O}_n \rangle \: = \:
\int_{ {\bf P}^4 } d^2 \phi^i \int \prod d \chi^i d \chi^{\overline{\imath}}
{\cal O}_1 \cdots {\cal O}_n 
\left( R_{i \overline{p} p \overline{\jmath}}
g^{ p \overline{p}} \chi^i \chi^{\overline{\jmath}} \right).
\label{eq:quinticScalingLimit}
\end{equation}

Since $R_{i \overline{p} p \overline{k}} g^{p \overline{p}}$
is the same\footnote{Along the $p=0$ locus, on which we have implicitly
restricted since there are no $p$ zero modes.} 
as $\mbox{tr }F$, where $F$ is the curvature of
${\cal O}(5)$, 
this correlation function is the same as
\begin{displaymath}
\int_{ {\bf P}^4 } \omega_1 \wedge \cdots \wedge \omega_n
\wedge \mbox{tr } F,
\end{displaymath}
where the $\omega_i$ are differential forms on ${\bf P}^4$ corresponding
to the operators ${\cal O}_i$.
In particular, as discussed above, mathematically this is the same
as
\begin{displaymath}
\int_{ Y \equiv \{ G = 0 \} } \omega_1 |_Y \wedge \cdots
\wedge \omega_n |_Y.
\end{displaymath}
Thus,
in this limit, the correlation function is the same as
the NLSM
correlation function -- we integrate over ${\bf P}^4$ and
wedge with the Euler class, which is equivalent to integrating just
over the quintic $Y = \{ G = 0 \}$.

Next let us consider the scaling limit
$\lambda \rightarrow
\infty$, in which the four-fermi term drops out.
For the same reasons as discussed for the closed string B model,
the method of steepest descent will give an exact answer for
this integral, so for example instead of integrating over
${\bf P}^4$ we could just integrate over the total space of the
normal bundle ${\cal N}$ to $Y = \{ G = 0 \}$ in ${\bf P}^4$.
Suppressing the factors of $\alpha'$, $A$, $\lambda$,
and $g^{p \overline{p}}$,
the correlation function can be written as
\begin{equation}
\langle {\cal O}_1 \cdots {\cal O}_n \rangle \: = \:
\int_{ {\cal N} } d^2 \phi \int \prod d \chi^i
d \chi^{\overline{\imath}} {\cal O}_1 \cdots {\cal O}_n
\left| D_i G  \chi^i \right|^2
\exp\left( - | G |^2 \right)
\end{equation}
(omitting an irrelevant factor of $2$).

As a consistency check, note that this result is independent of
the detailed form of $G(\phi)$ --  we effectively integrate 
along the zero section of ${\cal N}$
by virtue of the $\exp( - |G|^2 )$ factor,
and the factors\footnote{
To get factors of $D_i G$ rather than $\partial_i G$ from expanding
the exponential, we must remember to not only expand 
$G$ about the zero locus, but also the implicit metric factor
$g^{p \overline{p}}$.
} of $D_i G$ that one picks up from integrating over
transverse directions are canceled out by the
$| \chi^i D_i G |^2$ factor.  One may also argue this point more
abstractly: since the superpotential terms are BRST exact,
the theory should be independent of the detailed form of $G$.

As another consistency check, note that these correlation functions
are defining a product structure
on the correlators that only sees the restriction to the quintic.
Aside from the selection rule discussed earlier, the
Gaussian exponential is implicitly restricting the integration to the
quintic.  Furthermore, the $\chi^i D_i G$ insertions have the effect
of killing off any part of the ${\cal O}_i$'s that is normal to the quintic,
in addition to giving the correct selection rules.
After all, the tangent bundle to the hypersurface
$\{ G = 0 \}$ is defined by vectors $\chi^i$ such that
$\chi^i D_i G = 0$ \cite{gh}[section I.3],
so there is a natural mesh between the fact that any correlator containing a $\chi^i D_i G$ factor
would be annihilated by the factors brought down when we integrated out
$\chi^p$ and $\chi^{\overline{p}}$ and the statement
that the chiral ring is defined by cohomology of the quintic,
and not of ${\bf P}^4$.

To summarize, in the $\lambda \rightarrow \infty$ scaling limit,
correlation functions involve integrating over the total space of the
normal bundle to the quintic and intersecting with the zero section
of the bundle -- the effect of the $\exp( - |G|^2)$ Gaussian.
Instead of inserting factors of the Euler class, we are
intersecting with the zero section of the bundle, which is well-known
to be equivalent \cite{botttu}[chapter 11].

This method of computation is a very simple example of the form of a virtual
fundamental class computation; in effect, replacing the A-twisted
NLSM with the A-twisted Landau-Ginzburg model in the same universality
class gives us an alternative computational of the correlation functions,
an alternative computation that matches virtual fundamental class computations.
See \cite{coxkatz}[example 7.1.6.1, p. 184] for more details on this 
computation (and its degree $> 0$ counterparts) as virtual fundamental
class computations.  We emphasize this because this is, to our knowledge,
the first physical realization of virtual fundamental class computations of
A model correlation functions.
We will see in further examples that this is a common property of 
these Landau-Ginzburg model computations, that they physically realize
simple examples of virtual fundamental class computations,
as one limit of a scaling that interpolates between
virtual-fundamental-class-style computations and insertions of
Euler classes.

\subsubsection{Maps of degree greater than zero}

Next, let us consider correlation functions in a sector of maps of degree
$d > 0$, on a genus zero worldsheet.  Here, the $\phi^i$ zero modes map out
a moduli space ${\cal M}_d$ of maps into ${\bf P}^4$ of degree $d$.
Such moduli spaces are not compact, reflecting an IR divergence physically;
the path integral is regularized by compactifying the moduli spaces.
We shall use the GLSM compactification described in \cite{daveronen}, which
in the present case will be ${\cal M}_d = {\bf P}^{5d+4}$.  Similarly,
the $\chi^i$ zero modes are now holomorphic sections of $\phi^*(T{\bf
P}^4)$, of which there will be $5d+4$.  There will be no
$\psi_z^{\overline{\imath}}$ or $\psi_{\overline{z}}^i$ zero modes.  There
will be no $p_z$ zero modes\footnote{ If there were $p_z$ zero modes, if
they were sections of $K_{\Sigma} \otimes \phi^* {\cal O}(5) = {\cal
O}(5d-2)$, for example, then the complete moduli space would be the total
space of a bundle over the ${\cal M}_d$ above.   The bundle could be
computed by either computing the GLSM moduli space for the total space of
${\cal O}(5)$ using the methods of \cite{daveronen}, or alternatively
applying methods of \cite{ks1} to find the induced bundle, taking into
account the minor difference that here there is a $K$ twisting.  }, as they
are sections of $K_{\Sigma} \otimes {\cal O}(-5d)$, but there will be zero
modes of $\chi^p$, as they are holomorphic sections of $\phi^* {\cal O}(5)
= {\cal O}(5d)$, of which there will be $5d+1$ sections.  For the same
reason as the $p_z$'s, there will be no $\psi_z^p$ or
$\psi_{\overline{z}}^{\overline{p}}$ zero modes.

In addition, the original section $G$ will induce a section 
of the bundle $R^0 \pi_* \alpha^* {\cal O}(5)$ over
the moduli space.  As written,
that induced bundle is only defined over the part of the moduli space
${\cal M}_d$
described by honest maps, and must be extended over the compactification.
Methods to perform such an extension over GLSM moduli spaces
were described in \cite{ks1} (see \cite{bchir,ksa,ksb} for more information);
applying those methods here, one finds that the
extension (denoted with a tilde) is given by
\begin{equation}
\widetilde{ R^0 \pi_* \alpha^* {\cal O}(5) } \: 
\cong \:
{\cal O}(5)^{5d+1},
\label{eq:moduliSpaceInducedBundle}
\end{equation}
over the GLSM moduli space, whose components $\tilde{G}_a$ naturally pair
with the $\chi^p$.  
Here, in the language of \cite{ks1}, we have simply
identified ${\cal O}(5)$ with a single field and taken advantage of the
fact that ${\bf P}^4$ is a toric variety, so we expand the free field in
its zero modes and think of each element of a basis as a generator of a
copy of a ${\cal O}(5)$ over the GLSM moduli space.  We shall denote
the induced bundle above $\widetilde {{\cal O}(5)}$. 

In general, the bundles one obtains by such methods are only uniquely
defined over that part of the moduli space describing honest maps; there
are several inequivalent extensions over the compactification locus.  In
\cite{ks1}, where this was applied to understand (0,2) quantum cohomology,
the choice of extension was a function of which (0,2) GLSM one was working
with.

Putting this together, we find that correlation functions should be
given by
\begin{eqnarray*}
\lefteqn{
\langle  {\cal O}_1 \cdots {\cal O}_n \rangle \: = \:
\int_{ {\cal M}_d } d \phi^i
\int d \chi^i d \chi^{\overline{\imath}} d \chi^p_a d \chi^{\overline{p}}_b 
\, {\cal O}_1 \cdots {\cal O}_n 
} \\
& & \cdot \exp\left(
- \: 2 \sum_a | \tilde{G}_a(\phi) |^2 
\: - \:
\chi^i \chi^p_a D_i \tilde{G}_a \: - \:
\chi^{\overline{p}}_a \chi^{\overline{\imath}} D_{ \overline{\imath}}
\overline{\tilde{ G} }_a
\: - \:
\tilde{R}_{i \overline{a} b \overline{k} }
\chi^i \chi^{\overline{p}}_a \chi^p_b \chi^{\overline{k}}
\right).
\end{eqnarray*}
Here, factors such as $g^{p \overline{p}}$ have been
suppressed, and the $\tilde{R}$ indicates the 
curvature of the induced metric\footnote{
Let $\tilde{g}_{\mu \nu}$ denote the moduli space metric.
Formally, this is induced as follows.  Let $( \delta \phi )^i_{\mu}$
denote the $\mu$th component, on the moduli space, of the infinitesimal
deformation of $\phi^i$, then
\begin{displaymath}
\tilde{g}_{\mu \nu} \: = \: \int_{\Sigma} g_{i \overline{\jmath}}
( \delta \phi )^i_{\mu} ( \delta \phi )^{\overline{\jmath}}_{\nu}
\end{displaymath}
In any event, since we are working in the A model, the precise metric
is irrelevant, so long as it is reasonably generic.
} on
${\cal M}_d$.
Since there are no $\psi_z^{\overline{\imath}}$,
$\psi_{\overline{z}}^i$, $\psi_z^p$, or $\psi_{\overline{z}}^{\overline{p}}$
zero modes, there will be no contribution to this correlation function
from any other (Riemann curvature) four-fermi terms beyond the one
shown.  Readers familiar with Mathai-Quillen forms should note that,
as expected, we again have a Mathai-Quillen form.

Next, we integrate out the $\chi^p_a$'s, to get
\begin{eqnarray*}
\lefteqn{
\langle  {\cal O}_1 \cdots {\cal O}_n  \rangle \: = \:
\int_{ {\cal M}_d } d \phi^i  
\int d \chi^i d \chi^{\overline{\imath}}  
\, {\cal O}_1 \cdots {\cal O}_n \,
\prod_a \left( \left| \chi^i D_i \tilde{G}_a \right|^2 
\: + \: \tilde{R}_{i \overline{a} b \overline{k}} \chi^i
\chi^{\overline{k}}
\right)
} \\
& & \hspace*{3in} \cdot \exp\left(
- \: 2 \sum_a | \tilde{G}_a(\phi) |^2 
\right).
\end{eqnarray*}

Since this Landau-Ginzburg model should lie in the same universality
class as a NLSM on the quintic, let us check that the
correlation functions match.

First, note that the expression above implicitly encodes the
correct selection rule.  The sum of the degrees of the correlators
${\cal O}_i$ should be the same as the dimension of the moduli space
${\cal M}_d$
minus the number of $\tilde{G}_a$'s, as the moduli space in the NLSM is more nearly the complete intersection $\{ \tilde{G}_a = 0 \}
\subset {\cal M}_d$.  

Next, let us examine the two scaling limits $\lambda \rightarrow 0$,
$\lambda \rightarrow \infty$ of the superpotential
$G \mapsto \lambda G$.  As before, we will get the same result from
both limits -- necessarily so, since the superpotential is BRST exact.

In the $\lambda \rightarrow 0$ limit, the correlation function above
becomes
\begin{displaymath}
\langle  {\cal O}_1 \cdots {\cal O}_n \rangle \: = \:
\int_{ {\cal M}_d } d^2 \phi^i  
\int d \chi^i d \chi^{\overline{\imath}}  
\, {\cal O}_1 \cdots {\cal O}_n \,
\prod_a \left(  
\tilde{R}_{i \overline{a} a \overline{k}} \chi^i
\chi^{\overline{k}}
\right).
\end{displaymath}
After integrating out the $\chi^i$, this becomes
\begin{displaymath}
\int_{ {\cal M}_d } \omega_1 \wedge \cdots \wedge \omega_n \wedge
\mbox{Eul}(\mbox{$\widetilde {{\cal O}(5)}$}),
\end{displaymath}
where the $\omega_i$ are differential forms corresponding to the
operators ${\cal O}_i$.  As we have seen elsewhere, this is computing
an integral over the moduli space of maps into the quintic
by computing an integral over the moduli space ${\cal M}_d$ of maps
into ${\bf P}^4$ and inserting the Euler class of the induced 
bundle to localize, which is well-known (see for example
\cite{coxkatz}[example 7.1.6.1] and references therein) to be equivalent
to the NLSM computation.

Next, let us consider the opposite scaling limit $\lambda \rightarrow
\infty$.  Here, the correlation function becomes
\begin{displaymath}
\langle  {\cal O}_1 \cdots {\cal O}_n \rangle \: = \:
\int_{ {\cal M}_d } d^2 \phi^i  
\int d \chi^i d \chi^{\overline{\imath}}  
\, {\cal O}_1 \cdots {\cal O}_n \,
\prod_a \left( \left| \chi^i D_i \tilde{G}_a \right|^2 
\right)
 \exp\left(
- \: 2 \sum_a | \tilde{G}_a(\phi) |^2 
\right).
\end{displaymath}
Since the Gaussian localizes on $\{ \tilde{G}_a = 0 \}$
(in the sense that the method of steepest descent now gives exact results),
we could replace ${\cal M}_d$ by the total space ${\cal N}$ of the
normal bundle to the $\{ \tilde{G}_a = 0 \}$ locus, to get
\begin{displaymath}
\langle  {\cal O}_1 \cdots {\cal O}_n \rangle  \: = \:
\int_{ {\cal N} } d^2 \phi^i  
\int d \chi^i d \chi^{\overline{\imath}}  
\, {\cal O}_1 \cdots {\cal O}_n \,
\prod_a \left( \left| \chi^i D_i \tilde{G}_a \right|^2 
\right)
 \exp\left(
- \: 2 \sum_a | \tilde{G}_a(\phi) |^2 
\right).
\end{displaymath}

The result is independent of the detailed form of
$\tilde{G}$, for the same reasons as discussed in previous
examples.
Furthermore, the presence of
$\chi^i D_i \tilde{G}_a$ factors ensures that correlators that
are `transverse' to the moduli space are annihilated, as one would
expect from previous examples.

We see that, as before, this furnishes a very simple example of a virtual
fundamental class computation -- instead of wedging with the Euler
class of the induced bundle, we intersect with the vanishing locus
of a section of an induced bundle.

Thus, as before, we see that by varying the scale of the superpotential
(which leaves the correlation functions unchanged as the
superpotential is BRST exact), we interpolate between
Euler class insertions and virtual-fundamental-class-style computations
of the correlation functions in the quintic.

\subsubsection{Classical contribution -- genus greater than zero}

In this section let us briefly outline the analysis of Landau-Ginzburg
models corresponding to A-twisted NLSMs on worldsheets
of genus $g > 0$.  We shall work at fixed complex structure on the worldsheet
 -- we shall not couple to topological gravity, for simplicity.
This will give us a simple sanity check of the manner in which
we deal with zero modes of twisted bosons.

To be specific, let us consider an A-twisted NLSM on
a quintic in ${\bf P}^4$, again.
For maps of degree zero, there are as many $\chi^i$ as the dimension
of the target space -- $n$, say -- and also $ng$ many 
$\psi_z^{\overline{\imath}}$.
The correlators only contain $\chi$'s, so the
$\psi_z^{\overline{\imath}}$ zero modes must be absorbed via
bringing down copies of the four-fermi term.
However, if $g>1$, then bringing down four-fermi terms will always
generate more $\chi$'s than available $\chi$ zero modes,
regardless of the correlators, and so all correlation functions
vanish for $g>1$.
At genus one, there is precisely one nonvanishing correlator, given by
\begin{displaymath}
\langle 1\rangle \: = \: \int_X \mbox{Eul} \: = \: \mbox{Euler characteristic
of the quintic}.
\end{displaymath}

Next, let us perform the corresponding 
Landau-Ginzburg computation, to check that
we get comparable results.
The $\phi^i$ are now maps into ${\bf P}^4$, and the $p_z$ map into the
fibers of ${\cal O}(-5)$ -- 
but the $p_z$ are also twisted.  On a genus $g$ worldsheet,
the vector space of $p_z$ zero modes is $g$-dimensional.
There are also $4g$ $\psi_z^{\overline{\imath}}$ and
$\psi_{\overline{z}}^i$ zero modes, $g$ $\psi_z^p$ and
$\psi_{\overline{z}}^{\overline{p}}$ zero modes, as well as
$\chi^p$ and $\chi^{\overline{p}}$ zero modes.

Putting this together, a correlation function will have the form
\begin{eqnarray*}
\lefteqn{
\langle {\cal O}_1 \cdots {\cal O}_n \rangle \: = \:
\int_{ {\bf P}^4 } d \phi^i \int_{ {\bf C}^g } d^2 p_z 
\int d \chi^i d \chi^{\overline{\imath}}
d \psi_z^{\overline{\imath}} d \psi_{\overline{z}}^i
\int d\chi^p d \chi^{\overline{p}}
d \psi_z^p d \psi_{\overline{z}}^{\overline{p}} 
\: {\cal O}_1 \cdots {\cal O}_n 
} \\
& & \hspace*{0.5in} \cdot \exp\left(
- \: 2 |G|^2 \: - \: 2 g^{i \overline{\jmath}} |p_z|^2 D_i G
D_{\overline{\imath}} \overline{G} 
\: - \: 2 g^{p \overline{\imath}} G \overline{p}_{\overline{z}} D_{
\overline{\imath}} \overline{G} 
\: - \: 2 g^{\overline{p} i} \overline{G} p_z D_i G
\right. \\
& & \hspace*{1.0in} \left.
\: - \:
\psi_z^p \psi_{\overline{z}}^i D_i G \: - \:
\chi^i \chi^p D_i G 
\: - \: \chi^i \psi_{\overline{z}}^j p_z D_i D_j G
\right. \\
& & \hspace*{1.0in}
\left.
\: - \: \chi^{\overline{p}} \chi^{\overline{\jmath}} 
D_{ \overline{\jmath}} \overline{G} \: - \:
\psi_z^{\overline{\imath}} \psi_{\overline{z}}^{\overline{p}}
D_{\overline{\imath}} \overline{G} 
\: - \: \psi_z^{\overline{\imath}} \chi^{\overline{\jmath}}
\overline{p}_{\overline{z}} D_{\overline{\imath}} D_{\overline{\jmath}}
\overline{G} \right. \\
& & \hspace*{1.0in} \left.
\: - \:
\mbox{four-fermi terms}
\right).
\end{eqnarray*}

To do this computation in complete generality -- working with all four-fermi
terms at the same time as all superpotential terms -- could become extremely
tedious.  For example, there are six four-fermi terms which are nonvanishing
when $p=0$, namely
\begin{displaymath}
\begin{array}{c}
R_{i \overline{\imath} j \overline{\jmath}} \chi^i
\psi_z^{\overline{\imath}} \psi_{\overline{z}}^j \chi^{\overline{\jmath}}
\: + \:
R_{p \overline{p} j \overline{\jmath}} \psi_z^p \chi^{\overline{p}}
\psi_{\overline{z}}^j \chi^{\overline{\jmath}}
\: + \:
R_{i \overline{\imath} p \overline{p}}
\chi^i \psi_z^{\overline{\imath}} \chi^p \psi_{\overline{z}}^{
\overline{p}} \\
\: + \: R_{i \overline{p} p \overline{k}} \chi^i \chi^{\overline{p}}
\chi^p \chi^{\overline{k}} \: + \:
R_{p \overline{\jmath} k \overline{p}}
\psi_z^p \psi_z^{\overline{\jmath}} \psi_{\overline{z}}^{k}
\psi_{\overline{z}}^{\overline{p}}
\: + \: R_{p \overline{p} p \overline{p}}
\psi_z^p \chi^{\overline{p}} \chi^p \psi_{\overline{z}}^{\overline{p}},
\end{array}
\end{displaymath}
plus numerous others which are nonvanishing when $p \neq 0$.
Rather than try to keep track of all of them, we shall instead use the
following simplification:  since the superpotential is BRST-exact,
we shall work in a large scaling limit of $G$, in which the four-fermi
terms are insignificant and can be omitted.

Now let us turn to a subtlety in the bosonic integration.
Strictly speaking, we should combine the $\phi^i$ and $p_z$ zero-mode
integrals into a single integral over the total space of
a rank $g$ vector bundle over ${\bf P}^4$, with fibers the
vector space $p_z$ zero modes ${\bf C}^g$.
One way to understand this bundle is via the construction of GLSM
moduli spaces in \cite{daveronen}, that proceeds by describing the
total space of ${\cal O}(-5)$ as a toric variety.
The only modification is that when expanding physical fields of the
GLSM in a basis of zero modes, one should take the $p_z$'s to be
sections of $K_{\Sigma}\otimes \phi^* {\cal O}(-5)$ rather than
merely $\phi^* {\cal O}(-5)$.  For constant maps into the target,
that means there are as many $p_z$ zero modes as sections of
$K_{\Sigma}$.  Each such zero mode is then interpreted as a 
homogeneous coordinate on the moduli space of the same
${\bf C}^{\times}$ weights as the original $p_z$.
In this fashion, applying the moduli space construction of \cite{daveronen}
then gives us the total space of the bundle
\begin{displaymath}
{\cal O}(-5)^{\oplus g} \: \longrightarrow \: {\bf P}^4.
\end{displaymath}

There is also another way to understand that bundle.
Let ${\cal M}_d$ denote a moduli space of worldsheet instantons in
${\bf P}^4$ of
degree $d$, and let $\alpha: \Sigma \times {\cal M}_d \rightarrow {\bf P}^4$
denote the universal instanton (for the moment, for the purposes
of our formal illustration, we will work with ${\cal M}_d$ such that
$\alpha$ exists).  Let $\pi_{1,2}$ denote the projections from
$\Sigma \times {\cal M}_d$ to $\Sigma$, ${\cal M}_d$,
respectively.
Then the $p$ zero modes live in the sheaf
\begin{displaymath}
\pi_{2 *} \left( \left( \pi_1^* K_{\Sigma} \right) \otimes
\alpha^* {\cal O}(-5) \right)
\end{displaymath}
In terms of the present case of degree zero maps,
${\cal M}_0 = X$ so $\alpha$ is the same as the projection map
$\pi_2$.  By the projection formula
$\pi_{2 *} \pi_2^* {\cal L} = {\cal L} \otimes \pi_{2 *} {\cal O}_{
\Sigma \times {\cal M}_0}$ for any line bundle ${\cal L}$, 
but because $\pi_2$ is a trivial projection,
$\pi_{2 *} {\cal O}_{ \Sigma \times {\cal M}_0 } = {\cal O}_{ {\cal M}_0 }$.
Thus, we will have that 
\begin{displaymath}
\pi_{2 *} \left( \left( \pi_1^* K_{\Sigma} \right) \otimes
\alpha^* {\cal O}(-5) \right)
\: \cong \:
{\cal O}(-5) \otimes_{ {\bf C} } H^0(\Sigma, K_{\Sigma} ) 
\: \cong \:
{\cal O}(-5)^{\oplus g}.
\end{displaymath}
Yet another way to get this result is to apply the methods of
\cite{ks1} on induced bundles over toric moduli spaces.
Using these techniques, one expands the $p_z$ field in its zero modes (taking in
to
account the $K_{\Sigma}$ twisting), and identifies elements of a basis
of the space of zero modes with generators of copies of ${\cal O}(-5)$,
much as in the GLSM moduli space construction of \cite{daveronen} described
above.  The result is the same.

In any event, now that we understand the global structure of the
space of bosonic zero modes, let us turn to computing correlation
functions.
First, we integrate out the $\psi_z^p$'s:
\begin{eqnarray*}
\lefteqn{
\langle {\cal O}_1 \cdots {\cal O}_n \rangle \: = \:
\int_{ {\bf P}^4 } d \phi^i \int_{ {\bf C}^g } d^2 p_z 
\int d \chi^i d \chi^{\overline{\imath}}
d \psi_z^{\overline{\imath}} d \psi_{\overline{z}}^i
\int d\chi^p d \chi^{\overline{p}}
\: {\cal O}_1 \cdots {\cal O}_n 
} \\
& & \cdot
\left( \psi_{\overline{z}}^i D_i G 
\right)^g
\left( \psi_z^{\overline{\imath}} D_{\overline{\imath}} 
\overline{G} 
\right)^g
\\
& & \cdot \exp\left(
- \: 2 |G|^2 \: - \: 2 g^{i \overline{\jmath}} |p_z|^2 D_i G
D_{\overline{\imath}} \overline{G} 
\: - \: 2 g^{p \overline{\imath}} G \overline{p}_{\overline{z}} D_{
\overline{\imath}} \overline{G}
\: - \: 2 g^{\overline{p} i} \overline{G} p_z D_i G
\: - \:
\chi^i \chi^p D_i G \right. \\
& & \hspace*{0.5in} \left.
\: - \: \chi^i \psi_{\overline{z}}^j p_z D_i D_j G
\: - \: \chi^{\overline{p}} \chi^{\overline{\jmath}} 
D_{ \overline{\jmath}} \overline{G} 
\: - \: \psi_z^{\overline{\imath}} \chi^{\overline{\jmath}}
\overline{p}_{\overline{z}} D_{\overline{\imath}} D_{\overline{\jmath}}
\overline{G} \right).
\end{eqnarray*}

Next, we integrate out the $\chi^p$'s:
\begin{eqnarray*}
\lefteqn{
\langle {\cal O}_1 \cdots {\cal O}_n \rangle \: = \:
\int_{ {\bf P}^4 } d \phi^i \int_{ {\bf C}^g } d^2 p_z 
\int d \chi^i d \chi^{\overline{\imath}}
d \psi_z^{\overline{\imath}} d \psi_{\overline{z}}^i
\: {\cal O}_1 \cdots {\cal O}_n 
} \\
& & \cdot 
\left( \psi_{\overline{z}}^i D_i G \right)^g
\left( \chi^{\overline{\imath}} D_{\overline{\imath}}
\overline{G} \right)
\left( \psi_z^{\overline{\imath}} D_{\overline{\imath}} 
\overline{G} \right)^g
\left( \chi^i D_i G \right)
\\
& & \cdot \exp\left(
- \: 2 |G|^2 \: - \: 2 g^{i \overline{\jmath}} |p_z|^2 D_i G
D_{\overline{\imath}} \overline{G} 
\: - \: 2 g^{p \overline{\imath}} G \overline{p}_{\overline{z}} 
D_{\overline{\imath}} \overline{G}
\: - \: 2 g^{\overline{p} i} \overline{G} p_z D_i G 
\right. \\
& & \hspace*{0.5in} \left.
\: - \: \chi^i \psi_{\overline{z}}^j p_z D_i D_j G
\: - \: \psi_z^{\overline{\imath}} \chi^{\overline{\jmath}}
\overline{p}_{\overline{z}} D_{\overline{\imath}} D_{\overline{\jmath}}
\overline{G} \right).
\end{eqnarray*}

Finally, we integrate out the $p_z$'s
(more properly, perform a pushforward along the fibers of the
vector bundle ${\cal O}(-5)^{\oplus g}$).  
Expanding out the exponential in factors of $p_z$
will yield several possible terms, but only one will have enough
$\psi_z^{\overline{\imath}}$ and $\psi_{\overline{z}}^i$ factors to
match the corresponding fermionic zero mode integrals:
\begin{eqnarray*}
\lefteqn{
\langle {\cal O}_1 \cdots {\cal O}_n \rangle \: = \:
\int_{ {\bf P}^4 } d \phi^i  
\int d \chi^i d \chi^{\overline{\imath}}
d \psi_z^{\overline{\imath}} d \psi_{\overline{z}}^i 
\: {\cal O}_1 \cdots {\cal O}_n
} \\
& \hspace*{1.0in} & \cdot 
\left( \psi_{\overline{z}}^i D_i G \right)^g
\left( \chi^{\overline{\imath}} D_{\overline{\imath}}
\overline{G} \right)
\left( \psi_z^{\overline{\imath}} D_{\overline{\imath}} 
\overline{G} \right)^g
\left( \chi^i D_i G \right)
\\
& \hspace*{1.0in} & \cdot
\left( \chi^i \psi_{\overline{z}}^j D_i D_j G \right)^{(n-1)g}
\left( \chi^{\overline{\imath}} \psi_z^{\overline{\jmath}}
D_{\overline{\imath}} D_{\overline{\jmath}} \overline{G}
\right)^{(n-1)g} 
 \\
& \hspace*{1.0in} & \cdot 
\left( f(G, \overline{G}, D_i G, D_{\overline{\imath}} \overline{G})
\right)^{-ng}
 \\
& \hspace*{1.0in} & \cdot \exp\left(
- \: 2 |G|^2 
\right),
\end{eqnarray*}
where $n = \mbox{dim }{\bf P}^4 = 4$, and $f$ is a function whose
details will not be relevant for our analysis.

From this expression, we can read off the following facts immediately.
If $g>1$, then there are more $\chi$ insertions than
$\chi$ zero modes, and the correlation function must necessarily vanish,
matching the NLSM result.
In the special case that $g=1$, there are exactly as many 
$\chi$ insertions (outside of the correlators ${\cal O}_i$ themselves)
as there are $\chi$ zero modes, so the only possible nonvanishing
correlation function is $\langle 1 \rangle $, also matching the NLSM
result.

\subsubsection{The quintic GLSM -- classical contribution}

In this section we shall compute A-twisted correlation functions from
first principles in GLSMs\footnote{
We have not seen this particular computation anywhere else in the
literature, though certainly related matters have been considered
by others.  See \cite{daveronen} for an excellent discussion of
many issues in A-twisted GLSMs.
Direct GLSM computations of the sort considered here have also been
considered by A.~Adams and J.~McGreevy (Aspen, 2004), as well as
by M.~Plesser and I.~Melnikov, though to our
knowledge none of these computations were ever written up.
}.
In principle, such GLSM computations should be more difficult than the
A-twisted Landau-Ginzburg computations that we have described so far,
and so our A-twisted Landau-Ginzburg computations ought to
be a good stepping-stone to understanding analogous GLSM computations.
Therefore, by checking that we can set up basic GLSM computations,
we get a good robustness check of our methods. 

In this section, we shall only consider degree zero maps.

To describe a hypersurface of degree $d$ in ${\bf P}^n$,
the GLSM has $n+1$ chiral superfields with
components $(\phi^i, \psi_{\pm}^i)$, which are A-twisted in the same way
as the corresponding coordinates discussed above for the Landau-Ginzburg
model on the quintic.  The GLSM also has a $(p, \psi_{\pm}^p)$ chiral 
superfield, which is also A-twisted in the same way as the corresponding
field for the Landau-Ginzburg model on the quintic.
In addition, the GLSM adds a bosonic scalar field $\sigma$ 
(invariant under A-twisting), and gauginos $\lambda_{\pm}$, which the
A-twisting treats as follows:
\begin{alignat*}{6}
   & \lambda_+ (\equiv \lambda_z)                     &  & \in \Gamma_{C^{\infty}}\left( K_{\Sigma} \right) & \hspace*{1cm} & \lambda_- (\equiv \lambda)                                      & &\in \Gamma_{C^{\infty}}\left( {\cal O} \right) \\
   & \overline{\lambda}_+ (\equiv \overline{\lambda}) &  & \in \Gamma_{C^{\infty}}\left( {\cal O} \right)   &               & \overline{\lambda}_- (\equiv \overline{\lambda}_{\overline{z}}) & &\in \Gamma_{C^{\infty}}\left( \overline{K}_{\Sigma} \right).
\end{alignat*}
In these conventions, the supersymmetry transformations are
of the form
\begin{alignat*}{4}
   & \delta \sigma               &  & ={} & & i \tilde{\alpha}_+ \lambda_- \: + \: i \alpha_- \overline{\lambda}_+ \\
   & \delta \overline{\sigma}    &  & =   & & i \alpha_+ \overline{\lambda}_- \: + \: i \tilde{\alpha}_- \lambda_+ \\
   & \delta \lambda_+            &  & =   & & i \alpha_+ D \: + \: \partial \overline{\sigma} \alpha_- \: - \: F_{z \overline{z}} \alpha_+ \\
   & \delta \lambda_-            &  & =   & & i \alpha_- D \: + \: \overline{\partial} \sigma \alpha_+ \: + \: F_{z \overline{z}} \alpha_- \\
   & \delta \overline{\lambda}_+ &  & =   & & -i \tilde{\alpha}_+ D \: + \: \partial \sigma \tilde{\alpha}_- \: - \: F_{z \overline{z}} \tilde{\alpha}_+ \\
   & \delta \overline{\lambda}_- &  & =   & & -i \tilde{\alpha}_- D \: + \: \overline{\partial} \overline{\sigma} \tilde{\alpha}_+ \: + \: F_{z \overline{z}} \tilde{\alpha}_-.
\end{alignat*}
As a result, the BRST transformations are given by
\begin{alignat*}{4}
   & \delta \sigma                            &  & = {} & & i \tilde{\alpha}_+ \lambda \: + \: i \alpha_- \overline{\lambda} \\
   & \delta \overline{\sigma}                 &  & =    & & 0 \\
   & \delta \lambda_z                         &  & =    & & \partial \overline{\sigma} \alpha_-  \\
   & \delta \lambda                           &  & =    & & i \alpha_- D \: + \: F_{z \overline{z}} \alpha_- \\
   & \delta \overline{\lambda}                &  & =    & & -i \tilde{\alpha}_+ D \: - \: F_{z \overline{z}} \tilde{\alpha}_+ \\
   & \delta \overline{\lambda}_{\overline{z}} &  & =    & & \overline{\partial} \overline{\sigma} \tilde{\alpha}_+,
\end{alignat*}
since the BRST transformation parameters are given by
$\tilde{\alpha}_+$, $\alpha_-$.

Thus, as in \cite{daveronen} (modulo a convention change), the only
gauge-invariant and BRST-invariant operator is $\overline{\sigma}$,
so we will be interested in correlation functions of the form
$\langle  \overline{\sigma}^k \rangle $.

Following \cite{witphases}, the interaction terms from the
superpotential $W = p G(\phi)$ are of the form
\begin{eqnarray*}
\lefteqn{
L_W \: = \:
 2 | G(\phi) |^2 \: + \: 2 p_z \overline{p}_{\overline{z}} \sum_i \left|
\partial_i G \right|^2 } \\
& & \: + \: \psi_z^p \psi_{\overline{z}}^i \partial_i G
 \: + \:
\chi^i \chi^p \partial_i G \: + \: 
\chi^i \psi_{\overline{z}}^j p_z \partial_i \partial_j G \\
& & \: + \: i \chi^{\overline{p}} \chi^{\overline{\jmath}} \partial_{
\overline{\jmath}} \overline{G} \: + \:
\psi_z^{\overline{\imath}} \psi_{\overline{z}}^{\overline{p}} 
\partial_{\overline{\imath}} \overline{G} \: + \:
\psi_z^{\overline{\imath}} \chi^{\overline{\jmath}} 
\overline{p}_{\overline{z}} \partial_{\overline{\imath}} \partial_{
\overline{\jmath}} \overline{G},
\end{eqnarray*}
just as in the Landau-Ginzburg models discussed previously.
There is no four-Fermi curvature coupling, unlike the Landau-Ginzburg models
discussed previously, but instead there are Yukawa couplings of the form
\begin{eqnarray*}
\lefteqn{
L_{Yuk} \: = \:
\overline{\sigma} \psi_z^{\overline{\imath}} \psi_{\overline{z}}^i \: + \:
\sigma \chi^{\overline{\imath}} \chi^i } \\
& & \: + \:
\phi^{\overline{\imath}} \psi_{\overline{z}}^i \lambda_z \: + \:
\phi^{\overline{\imath}} \chi^i \lambda
\: + \:
\phi^i \psi_z^{\overline{\imath}} \overline{\lambda}_{\overline{z}}
\: + \: \phi^i \chi^{\overline{\imath}} \overline{\lambda},
\end{eqnarray*}
as well as additional bosonic couplings of the form
\begin{displaymath}
- | \sigma |^2 \sum_i | \phi^i |^2 \: - \: \frac{e_0^2}{2}D^2,
\end{displaymath}
where
\begin{displaymath}
D \: = \: \sum_i | \phi^i|^2 \: - \: r.
\end{displaymath}

As discussed above,
in the A-twisted theory
one computes correlation functions of the form $\langle  \overline{\sigma}^k\rangle $ 
for some 
positive integer $k$.
Using the data above, the contribution to such a correlation function
from degree zero maps on a genus zero worldsheet $\Sigma$ is
\begin{eqnarray*}
\lefteqn{
\langle  \overline{\sigma}^k \rangle  \: = \: \int \prod_i d^2 \phi^i d^2 \sigma
\int \left( \frac{1}{\sqrt{A}}d \chi^p \right)
\left( \frac{1}{\sqrt{A}} d \chi^{\overline{p}}\right)
 \prod_i d \chi^i d \chi^{\overline{\imath}}
\left( \frac{1}{\sqrt{A}} d \lambda\right)
\left( \frac{1}{\sqrt{A}} d \overline{\lambda}\right) 
\: \overline{\sigma}^k } \\
& & \cdot
\exp\left( \, - \: A |G|^2 \: - \: A |\sigma|^2 \sum_i | \phi^i |^2 
\: - \: \frac{e_0^2}{2} A D^2 
\right. \\
& & \hspace*{0.5in} \left. \: - \: A \chi^i \chi^p \partial_i G
\: - \: A \chi^{\overline{p}} \chi^{\overline{\jmath}} 
\partial_{\overline{\jmath}} \overline{G} \: - \: 
A \sigma \chi^{\overline{\imath}} \chi^i 
\: - \: A \phi^{\overline{\imath}} \chi^i \lambda
\: - \: A \phi^i \chi^{\overline{\imath}} \overline{\lambda}
\, \right)
\end{eqnarray*}
after reducing to zero modes, where $A$ is the worldsheet area
and factors of $2$ have generally been eliminated.
We are using the normalization of the $\chi^p$ zero modes discussed
previously (with $\alpha'$ factors suppressed -- we are only retaining
worldsheet area factors here).  Also, the $\lambda$ zero modes have the
same normalization as the $\chi^p$ zero modes, and for the same reason:
BRST transformations such as $\delta \lambda = F_{z \overline{z}} \alpha_-
 +  \cdots$
require a factor of the inverse worldsheet metric $g_{\Sigma}^{z \overline{z}}$
implicitly on the right-hand side, and so the $\lambda$'s rescale under
worldsheet metric rescalings.  By adding the $1/\sqrt{A}$ factor to the
$\lambda$ zero mode measure, we insure the path integral measure is invariant
under worldsheet metric rescalings.
The integral over the $\phi^i$'s should only integrate over a gauge
slice; we have omitted that detail to leave the expression above
in a gauge-invariant form.

Integrating out the $\lambda$, $\overline{\lambda}$ zero modes yields
\begin{eqnarray*}
\lefteqn{
\langle  \overline{\sigma}^k \rangle \: = \: \int \prod_i d^2 \phi^i d^2 \sigma
\int \left( \frac{1}{\sqrt{A}} d \chi^p \right)
\left( \frac{1}{\sqrt{A}} d \chi^{\overline{p}} \right)
\prod_i d \chi^i d \chi^{\overline{\imath}}
\, \overline{\sigma}^k 
\frac{ A^2}{A} \left( \phi^{\overline{\imath}} \chi^i \right)
\left( \phi^j \chi^{\overline{\jmath}} \right)
} \\
& & \cdot
\exp\left( \, - \: A |G|^2 \: - \: A |\sigma|^2 \sum_i | \phi^i |^2 
\: - \: \frac{e_0^2}{2} A D^2 
\: - \: A \chi^i \chi^p \partial_i G
\: - \: A \chi^{\overline{p}} \chi^{\overline{\jmath}} 
\partial_{\overline{\jmath}} \overline{G} \: - \: 
A \sigma \chi^{\overline{\imath}} \chi^i 
\, \right).
\end{eqnarray*}

Next, let us integrate out the $\sigma$ zero modes.
The $d^2 \sigma$ integral of terms with different numbers of $\sigma$
and $\overline{\sigma}$ factors necessarily vanishes.
The only nonzero contributions to that integral must have
equal factors of $\sigma$ and $\overline{\sigma}$ in the integrand.
Thus, we must expand out $\exp( A \sigma \chi^{\overline{\imath}}
\chi^i )$ to $k$th order in the integrand.
When we do so, we will be left with an integral of the form
\begin{displaymath}
\int d^2 \sigma | \sigma |^{2k} \exp\left( - \alpha | \sigma |^2 \right)
\: \propto \:
\alpha^{-k-1},
\end{displaymath}
where in this case $\alpha = A \sum_i | \phi^i |^2$.

Therefore, we are now left with
\begin{eqnarray*}
\lefteqn{
\langle  \overline{\sigma}^k \rangle  \: = \: \int \prod_i d^2 \phi^i 
\int \left( \frac{1}{\sqrt{A}} d \chi^p \right)
\left( \frac{1}{\sqrt{A}} d \chi^{\overline{p}} \right)
\prod_i d \chi^i d \chi^{\overline{\imath}}
\, 
\frac{ A^2 }{A} \left( \phi^{\overline{\imath}} \chi^i \right)
\left( \phi^j \chi^{\overline{\jmath}} \right)
\frac{ \left( A \chi^{\overline{n}} \chi^n \right)^k }{
\left( A \sum_m | \phi^m |^2 \right)^{k+1} }
} \\
&& \hspace*{2.0in} \cdot 
\exp\left( \, - \: A |G|^2 
\: - \: \frac{e_0^2}{2} A D^2
\: - \: A \chi^i \chi^p \partial_i G
\: - \: A \chi^{\overline{p}} \chi^{\overline{\jmath}} 
\partial_{\overline{\jmath}} \overline{G} 
\, \right),
\end{eqnarray*}
where we have omitted numerical factors, as elsewhere.

This expression now begins to resemble the expression we found
when working in the Landau-Ginzburg model on the total space of the
bundle ${\cal O}(-5) \rightarrow {\bf P}^4$.
Integrating out the $\chi^p$, $\chi^{\overline{p}}$ zero modes yields
\begin{eqnarray*}
\lefteqn{
\langle  \overline{\sigma}^k \rangle  \: = \: \int \prod_i d^2 \phi^i 
\int \prod_i d \chi^i d \chi^{\overline{\imath}}
\, 
\frac{ A^2 }{A^2}\left( \phi^{\overline{\imath}} \chi^i \right)
\left( \phi^j \chi^{\overline{\jmath}} \right)
\frac{ \left( A \chi^{\overline{n}} \chi^n \right)^k }{
\left( A \sum_m | \phi^m |^2 \right)^{k+1} }
\left| A \chi^i \partial_i G \right|^2
} \\
& & \hspace*{2.5in} \cdot
\exp\left( \, - \: A |G|^2 
\: - \: \frac{e_0^2}{2} A D^2
\, \right).
\end{eqnarray*}

At this point we can now read off a selection rule.
If the GLSM were describing a degree $d$ hypersurface in 
${\bf P}^n$, then there would be $n+1$ $\phi^i$s and $\psi_{\pm}^i$'s,
and in principle the correlation function should only be nonzero when
$k = n+1-2 = n-1$ -- since the hypersurface has dimension $n-1$.
We can see that same selection rule in the structure of the
result above.  When we perform the $\chi^i$,
$\chi^{\overline{\imath}}$ integrals, there must be a matching number
of $\chi^i$'s, $\chi^{\overline{\imath}}$'s in order to get a nonvanishing
result.  Indeed, there are $k+2$ factors of each $\chi^i$ and
$\chi^{\overline{\imath}}$ above, so the correlation function above
will be nonvanishing only when
$k+2 = n+1$ or $k=n-1$, exactly as needed.

Similarly, the factors of $A$ also cancel out:
outside of the exponential, there is an overall factor of
$A$, for any $k$.  When we perform the bosonic Gaussian integral,
integrating over normal directions, we will get a factor of
$( 1/\sqrt{A} )^2 = A^{-1}$, exactly right to cancel out the factor of
$A$ in the integrand.  Thus, the expression above is independent of the
area of the worldsheet, as must be true in a topological field theory.

Eliminating the factors of $A$ we can write the expression more cleanly as
\begin{eqnarray*}
\lefteqn{
\langle \overline{\sigma}^k \rangle \: = \: \int \prod_i d^2 \phi^i 
\int \prod_i d \chi^i d \chi^{\overline{\imath}}
\, 
\left( \phi^{\overline{\imath}} \chi^i \right)
\left( \phi^j \chi^{\overline{\jmath}} \right)
\left(  \chi^{\overline{n}} \chi^n \right)^k
\left|  \chi^i \partial_i G \right|^2
} \\
& & \hspace*{2.5in} \cdot
\exp\left( \, - \:  |G|^2 
\: - \: \frac{e_0^2}{2}  D^2
\, \right),
\end{eqnarray*}
where we have also used the $D=0$ constraint implicit in the Gaussian
to turn the denominator into a constant factor, which has been absorbed.
We have already discussed how the selection rule on $k$ arises,
let us now discuss the factors above in more detail.
Just as in the Landau-Ginzburg models discussed previously,
the factors of $\chi^i \partial_i G$ help to constrain correlators to the
hypersurface $\{ G = 0 \}$, by killing off any normal vectors.
The $\phi^{\overline{\imath}} \chi^i$ factors also annihilate components
parallel to the ${\bf C}^{\times}$ quotient, as needed.
Finally, for each $\overline{\sigma}$ there is a factor of
$\chi^i \chi^{\overline{\imath}}$, exactly as originally argued in
\cite{edver}.

The rest of the expression should look closely comparable to our results
for the Landau-Ginzburg model on the total space of the line
bundle ${\cal O}(-5) \rightarrow {\bf P}^1$.
We will not push this computation any further, but, we thought it important
to observe the analogue for GLSMs of the Landau-Ginzburg computations
that we are describing in this paper.

Elsewhere in this paper, we have seen Mathai-Quillen forms arise.
Presumably in this GLSM, there is an equivariant Mathai-Quillen form
present.  We have not worked it out explicitly, though it would be interesting
to do so.

\subsection{Example:  small resolution of the conifold}
\label{ssec:smallResolution}

Let us next examine a Landau-Ginzburg model in the same
universality class as a small
resolution of the conifold, 
{\it i.e.} the total space of the vector bundle
${\cal O}(-1) \oplus {\cal O}(-1) \rightarrow {\bf P}^1$.
We will see that the multicover computation of \cite{davepaul} is replaced
in the Landau-Ginzburg model by, for example, a physical realization of
a very simple version of a virtual fundamental class computation.

To build a Landau-Ginzburg model in the same universality class,
we need to describe the small resolution as a complete intersection.
If the original conifold is defined by the equation
\begin{displaymath}
xy \: - \: zt \: = \: 0
\end{displaymath}
in ${\bf C}^4 = \mbox{Spec }{\bf C}[x,y,z,t]$,
then a small resolution can be described as the complete intersection
\begin{eqnarray*}
G_1 & \equiv & xu \: - \: vz \: = \: 0 \\
G_2 & \equiv & tu \: - \: vy \: = \: 0
\end{eqnarray*}
in ${\bf C}^4 \times {\bf P}^1 
= \mbox{Spec }{\bf C}[x,y,z,t] \times \mbox{Proj }{\bf C}[u,v]$.

Our Landau-Ginzburg model is then defined over
\begin{displaymath}
X \: = \: \mbox{Tot} \left(
{\cal O}(-1) \oplus {\cal O}(-1) \: \longrightarrow \: {\bf P}^1 
\times {\bf C}^4
\right)
\end{displaymath}
with superpotential
\begin{displaymath}
W \: = \: p_1 G_1 \: + \: p_2 G_2,
\end{displaymath}
where the $p_i$ are local coordinates on the fibers of each ${\cal O}(-1)$
and the $G_i$ are the two sections of ${\cal O}(1) \rightarrow
{\bf P}^1 \times {\bf C}^4$ defined above.

Following the same notation as before, let $i$'s index affine coordinates
on ${\bf P}^1 \times {\bf C}^4$, so that the fermions tangent to the
base are twisted as
\begin{align*}
\psi_+^i ( \equiv \chi^i ) \: &\in \: \Gamma_{C^{\infty}}\left( \phi^* T^{1,0}{\bf P}^4 \right)                                                                   & \psi_-^i ( \equiv \psi_{\overline{z}}^i ) \: &\in \: \Gamma_{C^{\infty}}\left( \overline{K}_{\Sigma} \otimes (\phi^* T^{0,1}{\bf P}^4)^{\vee} \right) \\
\psi_+^{\overline{\imath}} ( \equiv \psi_z^{\overline{\imath}} ) \: &\in \: \Gamma_{C^{\infty}}\left( K_{\Sigma} \otimes (\phi^* T^{1,0}{\bf P}^4)^{\vee} \right) & \psi_-^{\overline{\imath}} ( \equiv \chi^{\overline{\imath}} ) \: &\in \: \Gamma_{C^{\infty}}\left( \phi^* T^{0,1} {\bf P}^4 \right),
\end{align*}
just as for the last example.
Similarly, the $p_i$ and their superpartners receive an asymmetric
twist, for the same reasons as before:
\begin{eqnarray*}
p \, (\equiv p_z) 
& \in & \Gamma_{C^{\infty}}\left( K_{\Sigma}
\otimes \phi^* T^{1,0}_{\pi} \right) \\
\overline{p} \, (\equiv \overline{p}_{\overline{z}} ) 
& \in &
\Gamma_{C^{\infty}}\left( \overline{K}_{\Sigma} \otimes
\phi^* T^{0,1}_{\pi} \right),
\end{eqnarray*}
with fermions
\begin{align*}
\psi_+^p (\equiv \psi_z^p) \: &\in \: \Gamma_{C^{\infty}}\left( K_{\Sigma} \otimes \phi^* T^{1,0}_{\pi} \right)                   & \psi_-^p (\equiv \chi^p) \: &\in \: \Gamma_{C^{\infty}}\left( ( \phi^* T^{0,1}_{\pi} )^{\vee} \right) \\
\psi_+^{\overline{p}} ( \equiv \chi^{\overline{p}} ) \: &\in \: \Gamma_{C^{\infty}}\left( ( \phi^* T^{1,0}_{\pi} )^{\vee} \right) & \psi_-^{\overline{p}} ( \equiv \psi_{\overline{z}}^{\overline{p}} ) \: &\in \: \Gamma_{C^{\infty}}\left( \overline{K}_{\Sigma}\otimes \phi^* T^{0,1}_{\pi} \right).
\end{align*}
Here, $T_{\pi}$ denotes the relative tangent bundle of the projection
\begin{displaymath}
\pi: \:
\mbox{Tot}\left( {\cal O}(-1)^{\oplus 2} \: \longrightarrow \: {\bf P}^1 
\times {\bf C}^4 \right)
\: \longrightarrow \: {\bf P}^1 \times {\bf C}^4.
\end{displaymath}
The BRST transformations of the fields are just as in previous examples, so we
omit them for brevity.  Also as before, the chiral ring may naively appear to
consist of differential forms on ${\bf P}^1 \times {\bf C}^4$, but since there
is a potential the bosonic zero modes can really only run freely over the
vanishing locus $\{ G_1 = G_2 = 0 \}$. The chiral ring is therefore the
restriction of differential forms on the ambient to the vanishing locus, just
as before.

\subsubsection{Classical contribution -- worldsheet genus zero}

Let us now outline how one computes the classical contributions to
correlation functions in this twisted theory.  This will be very
closely related to the analogous computation for the quintic.

For simplicity, we will only work on ${\bf P}^1$.
For the same reasons as in the quintic example, there are no
$p_a$, $\overline{p}_a$, $\psi_{\overline{z}}^i$,
$\psi_z^{\overline{\imath}}$, $\psi_z^p$, or 
$\psi_{\overline{z}}^{\overline{p}}$
zero modes.  There are as many $\chi^i$ zero modes as the dimension of
${\bf P}^1 \times {\bf C}^4$, and as many $\chi^p$ zero modes as the
rank of ${\cal O}(-1)^{\oplus 2}$.
The Riemann curvature four-fermi terms with factors of 
$\psi_{\overline{z}}^i$, $\psi_z^{\overline{\imath}}$, $\psi_z^p$,
and $\psi_{\overline{z}}^{\overline{p}}$, namely, all but one,
do not generate
any effective interactions on the classical component of the bosonic
moduli space.
Similarly, the only interactions generated by the superpotential are those
involving only $\chi$'s.
Hence, correlation functions have the form
\begin{eqnarray*}
\lefteqn{
\langle  {\cal O}_1 \cdots {\cal O}_n \rangle 
\: = \:
\int_{ {\bf P}^1 \times {\bf C}^4 } d^2 \phi^i
\int \prod_i d \chi^i d \chi^{\overline{\imath}} d \chi^{p_a} 
d \chi^{\overline{p}_a} \, {\cal O}_1 \cdots {\cal O}_n
} \\
& & \cdot \exp\left(
- \: 2 | G_1 |^2 \: - \: 2 | G_2 |^2 \: - \:
\chi^i \chi^{p_a} D_i G_a \: - \: \chi^{\overline{p}_a} 
\chi^{\overline{\jmath}} D_{\overline{\jmath}} \overline{G}_a 
\: - \: R_{i \overline{p}_a p_b \overline{k}}
\chi^i \chi^{\overline{p_a}} \chi^{p_b} \chi^{\overline{k}}
\right).
\end{eqnarray*}
Readers well acquainted with Mathai-Quillen forms will note that,
again, we have a Mathai-Quillen form, and can anticipate the result.

Performing the $\chi^{p_a}$ and $\chi^{\overline{p}_a}$ Grassmann integrals
gives
\begin{eqnarray*}
\lefteqn{
\langle {\cal O}_1 \cdots {\cal O}_n \rangle \: = \:
\int_{ {\bf P}^1 \times {\bf C}^4 } d^2 \phi^i
\int \prod_i d \chi^i d \chi^{\overline{\imath}}
{\cal O}_1 \cdots {\cal O}_n \prod_a \left(
\left| D_i G_a \chi^i \right|^2 
\: + \: R_{i \overline{p}_a p_a \overline{k} } \chi^i \chi^{\overline{k}}
\right)
} \\
& & \hspace*{3.5in} \cdot
\exp\left( - \: |G_1|^2 \: - \: |G_2|^2 \right),
\end{eqnarray*}
where irrelevant factors of $2$ have been omitted.

As before, let us examine this in the two scaling limits
$\lambda \rightarrow 0$ and $\lambda \rightarrow \infty$ where
$G_i \mapsto \lambda G_i$.  Since the superpotential is BRST-exact,
we should get the same result for all $\lambda$, and since this
theory should be in the same universality class as a NLSM on the small resolution of the conifold, we should also get
(for all $\lambda$) the same correlation functions as in the NLSM.  Mathematically, this is guaranteed by the fact that the
general expression, valid for all scales, is the Mathai-Quillen form.

In the $\lambda \rightarrow 0$ limit, the expression above becomes
\begin{displaymath}
\langle {\cal O}_1 \cdots {\cal O}_n  \rangle \: = \:
\int_{ {\bf P}^1 \times {\bf C}^4 } d^2 \phi^i
\int \prod_i d \chi^i d \chi^{\overline{\imath}}
{\cal O}_1 \cdots {\cal O}_n \prod_a \left(
R_{i \overline{p}_a p_a \overline{k} } \chi^i \chi^{\overline{k}}
\right).
\end{displaymath}
After integrating out the $\chi^i$, this can be expressed as
\begin{displaymath}
\int_{ {\bf P}^1 \times {\bf C}^4 } 
\omega_1 \wedge \cdots \wedge \omega_n
\wedge \mbox{Eul}.
\end{displaymath}
As discussed previously, this is equivalent to
\begin{displaymath}
\int_Y \omega_1 |_Y \wedge \cdots \wedge
\omega_n |_Y,
\end{displaymath}
where $Y$ is the complete intersection $\{ G_1 = G_2 = 0 \}$
defining the small resolution.
This is obviously the classical contribution to the corresponding
correlation function in the A-twisted NLSM on the
small resolution, exactly as expected.

Next, let us consider the $\lambda \rightarrow \infty$ scaling limit.
Here, we could just as well integrate only
over the total space of the normal bundle ${\cal N}$ to
the vanishing locus $\{ G_1 = G_2 = 0 \}$, so this becomes
\begin{displaymath}
\langle  {\cal O}_1 \cdots {\cal O}_n \rangle  \: = \:
\int_{ {\cal N} } d^2 \phi^i
\int \prod_i d \chi^i d \chi^{\overline{\imath}}
{\cal O}_1 \cdots {\cal O}_n \prod_a \left(
\left| D_i G_a \chi^i \right|^2 \right)
\exp\left( - \: |G_1|^2 \: - \: |G_2|^2 \right).
\end{displaymath}

First, note that as before, the correlation function will be independent
of the detailed form of the $G_a$:  integrating over the bosonic
zero modes normal to the vanishing locus $\{ G_1 = G_2 = 0 \}$
will generate factors of $D_i G$ that exactly cancel those appearing
in the integrand above.

The factors of $\chi^i$ in the integrand are responsible for a selection
rule that insures that the sum of the $U(1)_R$ charges of the ${\cal O}_i$
adds up to a top-form on $\{ G_1 = G_2 = 0 \}$, instead of the ambient
space ${\bf P}^1 \times {\bf C}^4$, just as in previous examples.
As before, these correlation functions define a product structure on the
correlators that only sees the restriction to the vanishing locus,
in agreement with our general remarks on the chiral ring.
Finally, the $\chi^i D_i G_a$ insertions have the effect
of killing off any part of the ${\cal O}_i$'s that is normal
to the quintic, just as before.

As expected, these computations match the classical contributions
to correlation functions of the A-twisted
NLSM on the small resolution.  In effect,
we are working on the normal bundle of the embedding and restricting to the
vanishing locus, another computation very much in the spirit of
virtual fundamental classes.

In particular, we see that rescaling
the superpotential interpolates between Euler class insertions
and virtual-fundamental-class-type computations of the correlation functions.

\subsubsection{Maps of degree greater than zero}

Next, let us consider correlation functions in a sector of maps of degree
$d > 0$, on a genus zero worldsheet.
Here, the $\phi^i$ zero modes map out a moduli space
${\cal M}_d$ of maps into ${\bf P}^1 \times {\bf C}^4$ of degree $d$,
with ${\cal M}_d = {\bf P}^{2d+1} \times {\bf C}^4$.
The $\chi^i$ zero modes are now holomorphic sections of
$\phi^* T( {\bf P}^1 \times {\bf C}^4)$, of which there are
$4 + 2d+1 = 2d+5$.  There are no 
$\psi_z^{\overline{\imath}}$, $\psi_{\overline{z}}^i$,
$\psi_z^p$ or $\psi_{\overline{z}}^{\overline{p}}$ zero modes,
nor\footnote{
If there were $p_z$ zero modes, then we would treat them by doing an
ordinary integral over the vector space of $p_z$ zero modes.
In fact, those zero modes would themselves comprise fibers of a vector
bundle over ${\cal M}_d$.  Suppose, for the sake of argument, that
the zero modes of $p_z^a$
were sections of $K_{\Sigma} \otimes \phi^* {\cal O}(1)^2$,
of which there are $2(d-1)$.  Strictly speaking, the zero modes of the
$p_z^a$ fiber over the bosonic moduli space ${\cal M}_d$,
to form a rank-$2(d-1)$ vector bundle.
We can derive the exact form of that vector bundle in several
ways, as discussed previously.  For example, since the total space of
${\cal O}(1)^2 \rightarrow {\bf P}^1 \times {\bf C}^4$
is a toric variety, we can modify the methods of
\cite{daveronen} (to take into account the twisting of the $p$ field)
to construct the GLSM moduli space, which will itself have the form
of a bundle over ${\cal M}_d$ above.  Alternately, we can apply
a slight modification of the results of \cite{ks1} on induced
bundles over toric variety moduli spaces, as discussed previously.  
In any event, the result
of either computation is that the vector bundle is
\begin{displaymath}
{\cal O}(1)^{2(d-1)} \: \longrightarrow \: {\cal M}_d
\end{displaymath}.
} $p_z$ zero modes, though there are $\chi^p$ zero modes.

Putting this together, we find that correlation functions are
given by
\begin{eqnarray*}
\lefteqn{
 \langle {\cal O}_1 \cdots {\cal O}_n \rangle  \: = \:
\int_{ {\cal M}_d } d^2 \phi^i 
\int d \chi^i d \chi^{\overline{\imath}} d \chi^p_a d \chi^{\overline{p}}_b
\, {\cal O}_1 \cdots {\cal O}_n
} \\
& & \hspace*{0.5in} \cdot \exp\left(
- 2 \sum_a \left| \tilde{G}_a \right|^2 
\: - \:
\chi^i \chi^p_a D_i \tilde{G}_a \: - \:
\chi^{\overline{p}}_a \chi^{\overline{\jmath}} D_{\overline{\jmath}}
\overline{\tilde{G}}_a
\: - \:
\tilde{R}_{i \overline{p}_a p_b \overline{k}} \chi^i
\chi^{\overline{p}}_a \chi^p_b \chi^{\overline{k}}
\right).
\end{eqnarray*}
The $\tilde{G}_a$ are the sections induced by the $G_a$;
there are as many $\tilde{G}_a$ induced sections as $\chi^p_a$ 
zero modes.
(Mathematically-inclined readers will note that, again, we have a
Mathai-Quillen form.)

Integrating out the $\chi^p_a$'s yields
\begin{eqnarray*}
\lefteqn{
\langle  {\cal O}_1 \cdots {\cal O}_n \rangle  \: = \:
\int_{ {\cal M}_d } d^2 \phi^i 
\int d \chi^i d \chi^{\overline{\imath}} 
\, {\cal O}_1 \cdots {\cal O}_n 
\, \prod_a \left( \left| \chi^i D_i \tilde{G}_a \right|^2 
\: + \: \tilde{R}_{i \overline{p}_a p_a \overline{k}}
\chi^i \chi^{\overline{k}}
\right)
} \\
& & \hspace*{3.5in} \cdot \exp\left(
-  \sum_a \left| \tilde{G}_a \right|^2 
\right)
\end{eqnarray*}
(omitting irrelevant factors of $2$).

Since this Landau-Ginzburg theory should be in the same universality
class as a NLSM on the small resolution, we should
get matching correlation functions.
To that end, note that
the $\chi^i$ factors in the integrand enforce the correct
selection rule:
the correlation function will only be nonvanishing when
the sum of the degrees of the correlators ${\cal O}_i$
equals $(2d+5)-(2d+2) = 3$, which is precisely the usual selection
rule for Calabi-Yau threefold targets in NLSMs.

Next, let us perform a more detailed analysis for the two scaling
limits $\lambda \rightarrow 0, \infty$ of the superpotential:
$G_i \mapsto \lambda G_i$, $\tilde{G}_a \mapsto
\lambda \tilde{G}_a$.

First, consider the $\lambda \rightarrow 0$ scaling limit.
Here, the correlation function reduces to
\begin{displaymath}
 \langle {\cal O}_1 \cdots {\cal O}_n  \rangle \: = \:
\int_{ {\cal M}_d } d^2 \phi^i 
\int d \chi^i d \chi^{\overline{\imath}} 
\, {\cal O}_1 \cdots {\cal O}_n 
\, \prod_a \left( 
\tilde{R}_{i \overline{p}_a p_a \overline{k}}
\chi^i \chi^{\overline{k}}
\right).
\end{displaymath}
Just as in all the previous examples we have seen,
this is
\begin{displaymath}
\int_{ {\cal M}_d } \omega_1 \wedge \cdots \wedge \omega_n \wedge
\mbox{Eul}(\mbox{$\widetilde {{\cal O}(5)}$}),
\end{displaymath}
which is realizing localization via Euler classes,
and in particular matches the result for the contributions for degree
$d$ curves to correlation functions in the A-twisted NLSM.

Next, let us consider the $\lambda \rightarrow \infty$ scaling limit.
Here, the correlation function reduces to
\begin{displaymath}
\langle  {\cal O}_1 \cdots {\cal O}_n  \rangle \: = \:
\int_{ {\cal M}_d } d^2 \phi^i 
\int d \chi^i d \chi^{\overline{\imath}} 
\, {\cal O}_1 \cdots {\cal O}_n 
\, \prod_a \left( \left| \chi^i D_i \tilde{G}_a \right|^2 
\right)
\exp\left(
- \sum_a \left| \tilde{G}_a \right|^2 
\right).
\end{displaymath}

Just as in previous expressions of this form, the resulting
correlation function does not depend upon the explicit details of
$\tilde{G}_a$, as the factor one gets from integrating out the
normal directions in the Gaussian cancels the $D \tilde{G}$ factors.
The factors of $\chi^i D_i \tilde{G}_a$ will have the effect
of killing off any correlators in which ${\cal O}_i$'s have factors
normal to the complete intersection, exactly as expected.

Again, we have found a physical realization of a very simple example of
a virtual fundamental class computation, matching the result for
the correlation function in the NLSM.
In this case, the virtual fundamental class does something intriguing:
in the ordinary NLSM on the small resolution,
there are multicovers, which means that to evaluate correlators in the
A model one must pull down factors of four-fermi terms, generating
the Euler class of the obstruction bundle.
Here, by contrast, there are no four-fermi terms in the computation: 
we have replaced the multicovers with the vanishing locus of
a section of some bundle.
The results are the same, but they are realized in different ways.

Also as before, the scaling $\lambda$ interpolates between
Euler class insertions and computations in the style
of virtual fundamental classes,
as is typical of Mathai-Quillen forms.

\subsubsection{The corresponding GLSM}

Let us now discuss a corresponding GLSM.
For simplicity, as this section is merely illustrative and not essential
for the rest of the paper, we will only consider the special case of
degree zero maps.

We will describe small resolutions of the conifold by a  
GLSM with fields with $U(1)$ charges as listed below:
\begin{center}
\begin{tabular}{cccccccc}
$x$ & $y$ & $z$ & $t$ & $u$ & $v$ & $p_1$ & $p_2$ \\ \hline
$0$ & $0$ & $0$ & $0$ & $1$ & $1$ & $-1$ & $-1$
\end{tabular}
\end{center}
and superpotential $W = p_1 G_1 + p_2 G_2$.
The fields above are twisted in the same fashion as the affine coordinates
in the Landau-Ginzburg model above -- most one way, the $p$'s differently.
The GLSM adds a bosonic scalar field $\sigma$, which is invariant under
the twisting, plus gauginos $\lambda_{\pm}$, which under the A-twisting
behave as
\begin{alignat*}{4}
   & \lambda_+ (\equiv \lambda_z)                     &  & \in \Gamma_{C^{\infty}}\left( K_{\Sigma} \right) & \hspace*{1cm} & \lambda_- (\equiv \lambda)                                      &  & \in \Gamma_{C^{\infty}}\left( {\cal O} \right) \\
   & \overline{\lambda}_+ (\equiv \overline{\lambda}) &  & \in \Gamma_{C^{\infty}}\left( {\cal O} \right)   &               & \overline{\lambda}_- (\equiv \overline{\lambda}_{\overline{z}}) &  & \in \Gamma_{C^{\infty}}\left( \overline{K}_{\Sigma} \right).
\end{alignat*}
As for the quintic, the only gauge-invariant BRST-invariant operator
is $\overline{\sigma}$, 
so we will be interested in correlation functions of the
form $ \langle \overline{\sigma}^k \rangle $.

The interaction terms from the superpotential are of the form
\begin{eqnarray*}
\lefteqn{
L_W \: = \:
- 2 | G_1 |^2  \: - \: 2 | G_2 |^2 \: - 
\:  2 \sum_i \left|
p_{a z} \partial_i G_a \right|^2 } \\
& & \: - \: \psi_z^{p a} \psi_{\overline{z}}^i \partial_i G_a
 \: - \:
\chi^i \chi^{p a} \partial_i G_a \: - \: 
\chi^i \psi_{\overline{z}}^j p_{a z} \partial_i \partial_j G_a \\
& & \: - \: \chi^{\overline{p} a} \chi^{\overline{\jmath}} \partial_{
\overline{\jmath}} \overline{G}_a \: - \: 
\psi_z^{\overline{\imath}} \psi_{\overline{z}}^{\overline{p} a} 
\partial_{\overline{\imath}} \overline{G}_a \: - \:
\psi_z^{\overline{\imath}} \chi^{\overline{\jmath}} 
\overline{p}_{a \overline{z}} \partial_{\overline{\imath}} \partial_{
\overline{\jmath}} \overline{G}_a,
\end{eqnarray*}
much as in the Landau-Ginzburg model discussed previously.
In addition, just as in the GLSM for the quintic, there are Yukawa couplings
of the form
\begin{eqnarray*}
\lefteqn{
L_{Yuk} \: = \:
- \overline{\sigma} \psi_z^{\overline{\imath}} \psi_{\overline{z}}^i \: - \:
\sigma \chi^{\overline{\imath}} \chi^i } \\
& & \: - \:
\phi^{\overline{\imath}} \psi_{\overline{z}}^i \lambda_z \: - \:
\phi^{\overline{\imath}} \chi^i \lambda
\: - \:
\phi^i \psi_z^{\overline{\imath}} \overline{\lambda}_{\overline{z}}
\: - \: \phi^i \chi^{\overline{\imath}} \overline{\lambda},
\end{eqnarray*}
where here the $\phi^i$ only range over charged base fields, 
{\it i.e.} $u$, $v$, but not $x$, $y$, $z$, $t$.
In addition, there are
bosonic couplings of the form
\begin{displaymath}
- | \sigma |^2 \sum_i | \phi^i |^2 \: - \: \frac{e_0^2}{2}D^2,
\end{displaymath}
where the $\phi^i$ range only over $u$, $v$, and
\begin{displaymath}
D \: = \: \sum_i | \phi^i|^2 \: - \: r,
\end{displaymath}
where again the $\phi^i$ only range over $u$, $v$.

Putting this together, we see that after reducing to zero modes, the contribution to the
correlation function $ \langle \overline{\sigma}^k \rangle$ from degree zero maps on
a genus zero worldsheet $\Sigma$ is
\begin{eqnarray*}
\lefteqn{
\langle  \overline{\sigma}^k \rangle \: = \: \int \prod_i d^2 \phi^i d^2 \sigma
\int \left( \frac{1}{\sqrt{A}} d \chi^p \right)
\left( \frac{1}{\sqrt{A}} d \chi^{\overline{p}} \right)
\prod_i d \chi^i d \chi^{\overline{\imath}}
\left( \frac{1}{\sqrt{A}} d \lambda \right)
\left( \frac{1}{\sqrt{A}} d \overline{\lambda} \right)
\:  \overline{\sigma}^k } \\
& & \cdot
\exp\left( \, - \: A |G_1|^2 \: - \: A |G_2|^2 
\: - \: A |\sigma|^2 \sum_i | \phi^i |^2 
\: - \: \frac{e_0^2}{2} A D^2 
\right. \\
& & \hspace*{0.5in} \left. \: - \: A \chi^i \chi^p_a \partial_i G_a
\: - \: A \chi^{\overline{p}}_a \chi^{\overline{\jmath}} 
\partial_{\overline{\jmath}} \overline{G}_a \: - \: 
A \sigma \chi^{\overline{\imath}} \chi^i 
\: - \: A \phi^{\overline{\imath}} \chi^i \lambda
\: - \: A \phi^i \chi^{\overline{\imath}} \overline{\lambda}
\, \right).
\end{eqnarray*}
Here $A$ is the worldsheet area
and factors of $2$ have generally been eliminated.
The sums in the terms involving $\sigma$ and $\lambda$
should only sum over the $u$, $v$ fields and their superpartners.
The integral over the $\phi^i$'s should only integrate over a gauge
slice; we have omitted that detail to leave the expression above
in a gauge-invariant form.

The analysis now proceeds in the same fashion as before.  
Integrating out the $\lambda$, $\overline{\lambda}$ zero modes yields
\begin{eqnarray*}
\lefteqn{
\langle  \overline{\sigma}^k \rangle \: = \: \int \prod_i d^2 \phi^i d^2 \sigma
\int \left( \frac{1}{\sqrt{A}} d \chi^p \right)
\left( \frac{1}{\sqrt{A}} d \chi^{\overline{p}} \right)
\prod_i d \chi^i d \chi^{\overline{\imath}}
\: \overline{\sigma}^k 
\frac{ A^2}{A} \left( \phi^{\overline{\imath}} \chi^i \right)
\left( \phi^j \chi^{\overline{\jmath}} \right)
} \\
& & \hspace*{0.5in} \cdot
\exp\left( \, - \: A |G_1|^2 \: - \: A |G_2|^2
\: - \: A |\sigma|^2 \sum_i | \phi^i |^2 
\: - \: \frac{e_0^2}{2} A D^2 
\right. \\
& & \hspace*{1.25in} \left.
\: - \: A \chi^i \chi^p_a \partial_i G_a
\: - \: A \chi^{\overline{p}}_a \chi^{\overline{\jmath}} 
\partial_{\overline{\jmath}} \overline{G}_a \: - \: 
A \sigma \chi^{\overline{\imath}} \chi^i 
\, \right),
\end{eqnarray*}
where in the $\phi^{\overline{\imath}} \chi^i$ factors in the integrand,
one should only sum over the $u$, $v$ chiral superfields and their
superpartners.

Next, let us integrate out the $\sigma$ zero modes.
The only nonzero contributions to that integral must have
equal factors of $\sigma$ and $\overline{\sigma}$ in the integrand
-- hence, we must expand out $\exp( A \sigma \chi^{\overline{\imath}}
\chi^i )$ to $k$th order in the integrand.
When we do so, an integral of the form
\begin{displaymath}
\int d^2 \sigma | \sigma |^{2k} \exp\left( - \alpha | \sigma |^2 \right)
\: \propto \:
\alpha^{-k-1},
\end{displaymath}
remains, where in this case $\alpha = A \sum_i | \phi^i |^2$.

Therefore, we are now left with
\begin{eqnarray*}
\lefteqn{
\langle  \overline{\sigma}^k \rangle  \: = \: \int \prod_i d^2 \phi^i 
\int \left( \frac{1}{\sqrt{A}} d \chi^p \right)
\left( \frac{1}{\sqrt{A}} d \chi^{\overline{p}} \right)
\prod_i d \chi^i d \chi^{\overline{\imath}}
\, 
\frac{ A^2 }{A} \left( \phi^{\overline{\imath}} \chi^i \right)
\left( \phi^j \chi^{\overline{\jmath}} \right)
\frac{ \left( A \chi^{\overline{n}} \chi^n \right)^k }{
\left( A \sum_m | \phi^m |^2 \right)^{k+1} }
} \\
&& \hspace*{1.0in} \cdot 
\exp\left( \, - \: A |G_1|^2 \: - \: |G_2|^2
\: - \: \frac{e_0^2}{2} A D^2
\: - \: A \chi^i \chi^p_a \partial_i G_a
\: - \: A \chi^{\overline{p}}_a \chi^{\overline{\jmath}} 
\partial_{\overline{\jmath}} \overline{G}_a 
\, \right).
\end{eqnarray*}
Here, we have omitted numerical factors, as elsewhere,
and in the new $\chi^{\overline{n}} \chi^n$ and
$\sum_i |\phi^i|^2$ factors in the integrand, the sums are only over
the $u$, $v$ chiral superfields and their superpartners.

Finally, integrating out the $\chi^p$, $\chi^{\overline{p}}$ zero modes yields
\begin{eqnarray*}
\lefteqn{
\langle \overline{\sigma}^k \rangle  \: = \: \int \prod_i d^2 \phi^i 
\int \prod_i d \chi^i d \chi^{\overline{\imath}}
\, 
\frac{A^2}{A^2} \left( \phi^{\overline{\imath}} \chi^i \right)
\left( \phi^j \chi^{\overline{\jmath}} \right)
\frac{ \left( A \chi^{\overline{n}} \chi^n \right)^k }{
\left( A \sum_m | \phi^m |^2 \right)^{k+1} }
\left( \prod_a \left| A \chi^i \partial_i G_a \right|^2 \right)
} \\
& & \hspace*{2.5in} \cdot
\exp\left( \, - \: A |G_1|^2 \: - \: A |G_2|^2
\: - \: \frac{e_0^2}{2} A D^2
\, \right).
\end{eqnarray*}

As in our previous GLSM computation, and also our LG computations,
the overall factor of $A$ in the integrand is canceled out
by the $(1/\sqrt{A})^2$ factor that arises after performing the bosonic 
Gaussian integral over normal directions.  In this fashion we see the
correlation function above is independent of the area $A$ of the worldsheet,
as must be true in a topological field theory. 
Eliminating the factors of $A$, we can rewrite the expression above in the
simpler form
\begin{eqnarray*}
\lefteqn{
\langle  \overline{\sigma}^k \rangle \: = \: \int \prod_i d^2 \phi^i 
\int \prod_i d \chi^i d \chi^{\overline{\imath}}
\, 
\left( \phi^{\overline{\imath}} \chi^i \right)
\left( \phi^j \chi^{\overline{\jmath}} \right)
\frac{ \left(  \chi^{\overline{n}} \chi^n \right)^k }{
\left(  \sum_m | \phi^m |^2 \right)^{k+1} }
\left( \prod_a \left|  \chi^i \partial_i G_a \right|^2 \right)
} \\
& & \hspace*{2.5in} \cdot
\exp\left( \, - \:  |G_1|^2 \: - \:  |G_2|^2
\: - \: \frac{e_0^2}{2}  D^2
\, \right).
\end{eqnarray*}

As before, the $\chi^i$ factors in the numerator yield the desired
selection rule on correlators.
Similarly, the detailed $G_a$ dependence is also canceler out when one
performs the bosonic Gaussian, so that the final correlation function
is independent of the detailed form of $G_a$ -- as expected, since this is
an A model correlation function.

\subsection{Remarks on virtual fundamental classes}
\label{ssec:VFCRemarks}

\subsubsection{General observations}

In several examples so far, we have seen how Landau-Ginzburg model
computations in a scaling limit realize some extremely simple virtual
fundamental class computations.  In simpler language, computations, which
in a NLSM would involve inserting copies of the Euler class of a bundle
into correlators, are replaced in Landau-Ginzburg models by computations in
which one restricts to the zero locus of a section of a bundle.  Replacing
the NLSM with another QFT in the same universality class has changed the
details of the computation of A-model correlation functions, and created a
physical realization for some alternative computations.

More generally \cite{katzpriv}, the computation of the virtual fundamental
classes uses the tangent and obstruction bundles\footnote{
In general, the obstruction `bundle' is actually a sheaf, and so the details
are more technical than we shall describe here.  In the examples we
shall work with, involving toric varieties, the obstruction sheaf 
will always be a bundle.
} to construct a cone over
the moduli space (a space with linear fibers, not necessarily a bundle),
which embeds into the obstruction bundle.  One then intersects the cone
with the zero section of the bundle (and counts with multiplicity).  The
computations we have seen so far correspond to the prototypical examples of
this construction, in which the moduli space in question is defined by the
zero locus of a section $s$ of a vector bundle
\cite{coxkatz}[example 7.1.4.1 p. 177].

Let us outline briefly how this arises. 
Let $s$ be a generic section of a vector bundle $\mathcal E \rightarrow X$,  
with its zero locus defined as
\begin{equation*}
	Z := \{ p \in X \, \vert \, s(p) = 0 \}.
	\label{eq:vfc:introduction:zeroLocus}
\end{equation*}
Since the section is generic, we will have that the $\dim_{\bf C} Z = n - r$,
where $n$ is the dimension of $X$ and $r$ is the rank of the bundle $\mathcal
E$.  Such a generic section induces a short exact sequence of bundles on
$Z$:
\begin{equation*}
  0 \: \longrightarrow \: T_Z \: \longrightarrow \: T_X\vert_Z \:
\stackrel{ds}{\longrightarrow} \: {\cal E} \vert_Z \: \longrightarrow \: 0.
  \label{eq:vfc:introduction:genericSectionSES}
\end{equation*}
Here, $ds$ is the differential of the section.  One can roughly say that
since $s$ is generic, it ``fills out'' ${\cal E}$ and the map $ds$ is
surjective.

The next simplest example of a virtual fundamental class concerns a smooth
section of ${\cal E} \rightarrow X$ that only fills out a subbundle
${\cal E}^\prime \subset {\cal E}$.  In this case, when ${\cal
E}^\prime$ has rank $r^\prime < r$, the zero locus is of dimension $n -
r^\prime > n - r$: it is said to have excess dimension. This is because if
$s$ as a section of ${\cal E}$ were generic, $Z$ would have dimension
$n-r$ as before.  Furthermore, the morphism induced by the differential of
the section  is no longer surjective.  We can construct an exact sequence
of bundles by taking the cokernel of ${\cal E}^\prime$ in ${\cal E}$
and restricting it to $Z$\footnote{
An exact sequence of bundles when the cokernel is locally-free, that is.
}.  This is known as the obstruction bundle, and
fits in the exact sequence
\begin{equation}
  0 \: \longrightarrow \: T_Z \: \longrightarrow \: T_X\vert_Z \:
\stackrel{ds}{\longrightarrow} \: {\cal E} \vert_Z \:
\longrightarrow \: \text{Ob} \: \longrightarrow \: 0,
  \label{eq:vfc:introduction:genericSubSectionSES}
\end{equation}
with $\text{Ob} = ({\cal E} \slash {\cal E}^\prime) \vert_Z$.  Note that
the Euler class of the obstruction bundle is of the `expected' or 
`virtual' dimension:
\begin{equation*}
  \begin{split}
	 \dim_{\bf C} (\text{Eul(Ob)} \cap [Z]) &\: = \: 
(n - r^\prime) - (r - r^\prime) \\
	 &\: = \:n -r.
	 \label{eq:vfc:introduction:obstructedCase}
  \end{split}
\end{equation*}

More generally, the virtual fundamental class may be realized using
so-called ``tangent-obstruction functors,'' which we can take to be sheaves
${\cal T}_1$ and ${\cal T}_2$ on some space $Z$.  Then, we construct a 
two-term locally-free resolution ${\cal E}_1 \rightarrow {\cal E}_2$ of
these sheaves:
\begin{equation*}
  0 \: \longrightarrow \: {\cal T}_1 \: \longrightarrow \: {\cal E}_1 
\: \longrightarrow \: {\cal
  E}_2 \: \longrightarrow \: {\cal T}_2 \: \longrightarrow \: 0.
  \label{eq:vfc:introduction:generalResolution}
\end{equation*}
For an extremely readable introduction to virtual moduli cycles
couched in this type of language, see \S 3 of \cite{Thomas:1998uj}.

In our obstructed example, $T_X\vert_Z$ and $E\vert_Z$ play the role of the
two term resolution, with the morphism induced by Yukawa couplings such as
\begin{displaymath}
\psi_+^i \psi_-^j D_i \partial_j W.
\end{displaymath}
We think of $dW$ as a section of $T^*X$, so that one obtains a section of
$TX$ by application of the target-space metric.  Then, the differential of
this section -- in our language, $D_i \partial_j W$ -- realizes the
morphism.  Thus, the superpotential gives mass to the fermions and defines
the tangent and obstruction bundles respectively as the kernel and cokernel
of this morphism.  We will give an explicit example of our formalism that 
realizes an
obstructed version of these constructions in the next section.

Finally, because the method of steepest descent gives exact answers in
these topological field theories, bosonic moduli space integrals can
equivalently integrate only over the total space of normal bundles
to the loci $\{ dW = 0 \}$, and exponential weightings of the
form $\exp(-|dW|^2)$ realize an intersection with the zero locus of
a section of that bundle.  At least at this general level,
A-twisted Landau-Ginzburg computations should therefore always 
have a presentation that looks like some form of a simple
virtual fundamental class computation.

Granted, the examples of this phenomenon studied in this paper
are all simple special cases of a much more general computational tool.
However, we emphasize them nonetheless because to our knowledge,
physical realizations of virtual fundamental class computations have
not appeared previously in the physics literature.
After the early 1990s, when basics were first developed, mathematicians
developed many techniques to study Gromov-Witten theory independently of
input from physicists, and as a result, now have many very powerful tools,
whose physical realization is largely unknown.
Virtual fundamental classes are one example of such a mathematical tool,
and as we have seen a physical realization of such computations,
we feel compelled to point it out explicitly, even though the examples
considered are very simple ones.

It would be very interesting to understand the physical realization of
more general virtual fundamental class computations, but it seems that
will require working in A-twisted Landau-Ginzburg theories
coupled to topological gravity, which we do not consider here.
A simple example of a virtual fundamental class computation
that is not equivalent to inserting factors of Euler classes,
involves \cite{katzpriv}
genus one Gromov-Witten computations on the quintic.  
There, the obstruction ``bundle'' is actually a sheaf, not a bundle.

We shall briefly return to the issue of virtual fundamental
class realizations in \cite{jstoappear} when studying heterotic versions of
Landau-Ginzburg models.  There, when studying simple examples, we shall
again find that Landau-Ginzburg model computations realize correlation functions
in a virtual-fundamental-class-type fashion.  
The correlation functions therein are computing the (0,2) version of
quantum cohomology rings \cite{ks1}, and so presumably physics is
telling us about simple versions of the heterotic analogue of
virtual fundamental class computations.

\subsubsection{An obstructed example}
\label{app:obstructed}

So far, we have seen that A-twisted Landau-Ginzburg models construct the
simplest version of a virtual fundamental class.  Let us now see how to
construct an example of the obstructed form -- \`a la equation
  \eqref{eq:vfc:introduction:genericSubSectionSES} in section
\ref{ssec:VFCRemarks} -- an example of a virtual fundamental
class construction with one more degree of complexity than those seen so far.

Here, we examine a model in the same universality class as the A model on
${\bf P}^2$.  Consider the A twist of the Landau-Ginzburg model on the total
space of ${\cal E} = {\cal O}(-1) \oplus {\cal O}(-1) \longrightarrow 
{\bf P}^3$, whose
superpotential is defined with the aid of the section 
\begin{equation*}
G =
        \left ( \begin{matrix} G_1 \\ G_2 \end{matrix} \right ) =
        \left ( \begin{matrix} \phi_1 \\ 0 \end{matrix} \right ) 
\end{equation*}
of the dual bundle ${\cal O}(1) \oplus {\cal O}(1) \longrightarrow 
{\bf P}^3$.
Clearly,
$G$ is a section of the subbundle ${\cal O}(1)_1$, where the subscript
denotes the first factor in ${\cal E}$.  In terms of the fiber coordinates
$\{p_1, p_2\}$ and the components $\{G_1,G_2\}$ of the section, the
superpotential is 
\[ 
W = p^\alpha G_\alpha = p_1 G_1.
\]
We utilize the notation in section \ref{ssec:smallResolution}, where fermions 
tangent to the
base are twisted as 
\begin{align*} 
  \psi_+^i (\equiv \chi^i)                         & \in \Gamma_{C^\infty} 
\left(
  \phi^* T^{1,0} {\bf P}^3\right)
  &
  \psi_-^i (\equiv \psi^i_{\bar z})                  & \in \Gamma_{C^\infty} 
\left(
 \overline K_\Sigma \otimes \left(\phi^* T^{0,1} {\bf P}^3\right)^\vee
\right)
  \\
  % line 2
  \psi_+^{\bar \imath} (\equiv \psi^{\bar \imath}_z) & \in \Gamma_{C^\infty} 
\left(
 K_\Sigma \otimes \left(\phi^* T^{1,0} {\bf P}^3\right)^\vee \right)
  &
  \psi_-^{\bar \imath} (\equiv \chi^{\bar \imath}) & \in \Gamma_{C^\infty} 
\left(
  \phi^* T^{0,1} {\bf P}^3\right).
\end{align*} 

Similarly, the fiber multiplet is twisted as 
\begin{align*}
  p^a (\equiv p^a_z)          \:\:                & \in \Gamma_{C^\infty} 
\left(
 K_\Sigma \otimes \phi^* T_\pi^{1,0}\right)                        & p^{\bar b}
 (\equiv p^{\bar b}_{\bar z}) \:\: & \in \Gamma_{C^\infty} \left ( \overline 
K_{\Sigma}
 \otimes \phi^* T_\pi^{0,1}\right)\\
  \psi_+^a (\equiv \psi^a_z)                      & \in \Gamma_{C^\infty} 
\left(
 K_{\Sigma} \otimes \phi^* T_\pi^{1,0} {\bf P}^3\right)            & \psi_-^a 
(
\equiv \chi^a)                     & \in \Gamma_{C^\infty} \left ( \left(\phi^* 
T_\pi^{0,1} {\bf P}^3\right)^\vee\right)\\
  % line 2
  \psi_-^{\bar a} (\equiv \psi^{\bar a}_{\bar z}) & \in \Gamma_{C^\infty} 
\left(
 \overline K_\Sigma \otimes \phi^* T_\pi^{0,1} {\bf P}^3 \right) & \psi_+^{\bar
 a} (\equiv \chi^{\bar a})       & \in \Gamma_{C^\infty} \left ( \left(\phi^* 
T^{1,0} {\bf P}^3\right)^\vee \right).
\end{align*} 

Here, as before, $T_\pi$ denotes the vertical subbundle of $T{\cal E}$.
Again, there no surprises in the BRST transformations of the fields, so we
omit them for brevity.

\textbf{Classical contribution in genus zero.}

Let us now outline how one computes the classical contributions to
correlation functions in this twisted theory.  Again, these computations
will be very closely analogous to the ones for the quintic.

As before, since we are working on ${\bf P}^1$, there are no
$p_z^a$, $p_{\bar z}^{\bar a}$, $\psi^a_z$, $\psi^{\bar a}_{\bar z}$, 
$\psi^i_{\bar z}$,
or $\psi^{\bar \jmath}_z$ zero modes. There are three $\chi^i$ zero modes,
and two $\chi^a$ zero modes. There is exactly one non-vanishing Riemann
curvature term, so that altogether, an arbitrary correlation function takes
the form 
\begin{multline*}
\langle {\cal O}_1 \cdots {\cal O}_n \rangle = \int_{{\bf P}^3} d^2 \phi^i
 \int
\prod_i d\chi^i d\chi^{\bar \imath} d\chi^a d\chi^{\bar a} {\cal O}_1
\cdots {\cal O}_n \\
\cdot \exp \left ( - 2 \vert G_1 \vert^2 - \chi^i \chi^{p_1} D_i G_1 -
\chi^{\bar p_1} \chi^{\bar \jmath}  D_{\bar \jmath} \overline G_1  - R_{i
\bar \jmath a \bar b} \chi^i \chi^{\bar \jmath} \chi^a \chi^{\bar b} \right).
\end{multline*}

Note that there are no $\chi^{p_2}$ factors in the Yukawa terms, so that in
order to saturate the integrals, we must bring down a curvature factor:
\begin{multline*}
\langle {\cal O}_1 \cdots {\cal O}_n \rangle = \int_{{\bf P}^3} d^2 \phi^i
 \int
\prod_i d\chi^i d\chi^{\bar \imath} d\chi^1 d\chi^{\bar p_1}
(R_{i\bar \jmath p_2 \bar p_{2}} \chi^i \chi^{\bar \jmath} g^{p_2 p_{\bar
2}})
{\cal O}_1 \cdots {\cal O}_n\\ 
\cdot \exp \left ( - 2 \vert G_1 \vert^2 - \chi^i \chi^{p_1} D_i G_1 -
\chi^{\bar p_{1}} \chi^{\bar \jmath}  D_{\bar \jmath} \overline G_1  - R_{i
\bar \jmath p_1 \bar p_{1}} \chi^i \chi^{\bar \jmath} \chi^{p_1} \chi^{\bar p_{1
}} \right).
\end{multline*}

Note that we cannot bring down factors of $R_{i\bar \jmath p_1 p_{\bar
2}}R_{i\bar \jmath p_2 \bar p_{1}}$, since these terms are respectively
proportional to $\bar p_{1} p_2$ and $p_1 \bar p_{2}$, and thus vanish for
classical contributions at genus zero.

We now consider the infinite scaling limit; we scale the section $G$ by a
factor $\lambda$, and take $\lambda \mapsto \infty$.  As $\lambda$ becomes
very large, the remaining four-fermi term in the exponential is suppressed,
so that the remaining $\chi^p$ integrals must be saturated by the Yukawa
interactions:
\begin{equation*}
\langle {\cal O}_1 \cdots {\cal O}_n \rangle = \int_{{\bf P}^3} d^2 \phi^i
d^2\chi^i 
(R_{i\bar \jmath p_2 \bar p_{2}} \chi^i \chi^{\bar \jmath} g^{p_2 p_{\bar
2}})( \vert \lambda \chi^i D_i G \vert^2)
{\cal O}_1 \cdots {\cal O}_n e^{- 2 \vert \lambda G_1 \vert^2}.
\end{equation*}

The Gaussian causes the integral to be supported on an infinitesimal
neighborhood of the zero locus of $G$, ${\bf P}^2 \subset {\bf P}^3$,
so we can take the integral to be over the normal bundle $N_{{\bf P}^2
\slash {\bf P}^3}$.  With this identification, and the interpretation of
$\vert \chi^i D_i G \vert^2$ 
as eating up the $d\chi^a$ integration along the fiber
directions, we see that the correlation function becomes 
\begin{equation*}
  \langle {\cal O}_1 \cdots {\cal O}_n \rangle \sim \int_{{\bf P}^2}
  d^2\phi^i d^2 \chi^i \; \text{Eul}(\text{Ob}) \; {\cal O}_1 \cdots 
{\cal O}_n.
\end{equation*}
Thus, we can explicitly see the emergence of $\text{tr }F$, 
for $F$ the curvature of the
subbundle ${\cal O}(1)_2 \subset {\cal E}$,  as the first Chern class of the
rank one obstruction bundle (the Euler class).

Similarly, we can take the $\lambda \mapsto 0$ limit, wherein the
section-dependent terms vanish, and we are left with only four-fermi
curvature terms to soak up the fermionic integrals.  Correlation functions
then become
\begin{equation*}
  \langle {\cal O}_1 \cdots {\cal O}_n \rangle  = \int_{{\bf P}^3}
  d^2\phi^i d^2 \chi^i {\cal O}_1 \cdots {\cal O}_n R_{i\bar \jmath p_1
  \bar p_{1}} R_{k \bar \ell p_2 \bar p_2}  g^{p_1 \bar p_1} g^{p_2 \bar
  p_2} \chi^i \chi^{\bar \jmath} \chi^k \chi^{\bar \ell}.
\end{equation*}

We see that the path integral has inserted a factor of 
\[ c_1 ({\cal O}(1)_1) \cup c_1 ({\cal O}(1)_2),\]
which by the short exact sequence 
\[ 0 \rightarrow {\cal O}(1)_1 \rightarrow {\cal O}(1)_1 \oplus {\cal O}(1
)_2 \rightarrow 
{\cal O}(1)_2 \rightarrow 0\]
of bundles on ${\bf P}^3$ is equal to $c_2({\cal E}) = \text{Eul}({\cal
E})$.

\section{Landau-Ginzburg models on stacks and hybrid GLSM phases}
\label{lgstx}

In examples studied earlier in this paper, we showed how, for example, a
Landau-Ginzburg model on the total space of the line bundle ${\cal O}(-5)
\rightarrow {\bf P}^4$ with suitable superpotential is in the same
universality class (and hence has the same A model correlation functions)
as a NLSM on the quintic.  More generally, it should be clear that all
large-radius phases of GLSMs have a representative in the same universality
class given by a Landau-Ginzburg model on the total space of some
noncompact toric variety.

The non-geometric phases at other ends of GLSM K\"ahler moduli spaces also
have a very simple description as Landau-Ginzburg models, but now typically
Landau-Ginzburg models on stacks.  This has already been discussed in a few
examples in \cite{cdhps}, and is worth repeating here.

The simplest example is the Landau-Ginzburg point of the GLSM for the
quintic in ${\bf P}^4$:  there, the Landau-Ginzburg theory is defined by a
quintic superpotential over the orbifold $[ {\bf C}^5/{\bf Z}_5]$.  A
little more generally, if one has a complete intersection of hypersurfaces
of degree $d_1, \cdots, d_r$, then the Landau-Ginzburg point will be
defined by a superpotential on the total space of a vector bundle over the
weighted projective stack $WP_{[d_1, \cdots, d_r]}$.  If the $d_i$ have a
greatest common divisor greater than one, then this weighted projective
stack will be a gerbe, and the GLSM physics will (nonperturbatively) see
the difference between the gerbe and the space defined by dividing out the
gcd \cite{ps3}.

Another example is furnished by the GLSM for the complete intersection
${\bf P}^5[3,3]$.  Here, the D-terms are of the form
\begin{displaymath}
\sum_i | \phi_i |^2  \: - \: 3 |p_1|^2 \: - \: 3 | p_2|^2 \: = \: r.
\end{displaymath}
For $r>0$, the $\phi_i$ cannot all be zero, whereas for
$r<0$, the $p_a$ cannot both be zero.

Thus, the theories for $r>0$ look like a family of Landau-Ginzburg models
over 
\begin{displaymath}
\mbox{Tot}\left( {\cal O}(3) \oplus {\cal O}(3) \: \longrightarrow \:
{\bf P}^5\right).
\end{displaymath}

For $r\ll 0$, the theory looks naively like some sort of ${\bf Z}_3$
orbifold of ${\bf C}^6$, fibered over ${\bf P}^1$.
We can make this much more precise as follows.
First, just as an abelian gauge theory with fields of charge,
say, $-k$, $+1$, $+1$ describes the total space of the line
bundle ${\cal O}(-k)$ over ${\bf P}^1$, 
an abelian gauge theory with charges $-1$, $+3$, $+3$
describes the total space of a line bundle over a ${\bf Z}_3$ gerbe
on ${\bf P}^1$, using the fact that one of the ${\bf Z}_3$ gerbes on
${\bf P}^1$ can be described by an abelian gauge theory with two fields
of (nonminimal) charge $+3$.
This particular line bundle is commonly denoted ``${\cal O}(-1/3)$,''
and its total `space' (more accurately, total stack) is 
a $[{\bf C}/{\bf Z}_3]$ bundle over ${\bf P}^1$.
Similarly, the orbifold $[{\bf C}/{\bf Z}_k]$ is the same thing
as the total space of the line bundle ${\cal O}(-1/k) \rightarrow
B{\bf Z}_k$.  Such bundles on gerbes will be discussed in more
detail in \cite{metonytoappear}.  

In particular, the $r \ll 0$ limit of the GLSM for ${\bf P}^5[3,3]$
can now be trivially seen to be a Landau-Ginzburg model over
\begin{displaymath}
\mbox{Tot}\left( {\cal O}(1/3)^{\oplus 6} 
\: \longrightarrow \: G {\bf P}^1 \right),
\end{displaymath} 
where $G {\bf P}^1$ denotes a ${\bf Z}_3$ gerbe
on ${\bf P}^1$.

Examples of this form were also recently discussed in
\cite{cdhps}, such as the GLSM for the complete intersection
${\bf P}^7[2,2,2,2]$.  The Landau-Ginzburg point of this GLSM
defined by a superpotential on the total space of
the bundle ${\cal O}(-1/2)^8 \rightarrow G{\bf P}^3$, where
$G{\bf P}^3$ denotes a ${\bf Z}_2$ gerbe on ${\bf P}^3$.
The analysis here is closely related to the example above.
Just as the total space of the line bundle ${\cal O}(-n) \rightarrow
{\bf P}^N$ can be described by a GLSM with $N+1$ superfields of
charge 1 (for the base ${\bf P}^N$) and one superfield of
charge $-n$, the total space of the line bundle
${\cal O}(-1/2) \rightarrow G{\bf P}^N$ can be described by a GLSM
with $N+1$ superfields of charge 2 (for the base $G {\bf P}^N$) and
one superfield of charge $-1$.
In particular, the total space of
${\cal O}(-1/2)^8 \rightarrow G{\bf P}^3$
describes a $[{\bf C}^8/{\bf Z}_2]$ orbifold fibered nontrivially
over ${\bf P}^3$.

Much of \cite{cdhps} was devoted to understanding this Landau-Ginzburg model:
since the superpotential describes degree {\it two} hypersurfaces,
at first glance it might appear that all of the fields along fiber directions
are massive, leaving one with merely a NLSM on 
${\bf P}^3$.  Since the large-radius limit is a Calabi-Yau, the
$r \ll 0$ phase also ought to describe a Calabi-Yau, and further
analysis reveals that this is exactly what happens.
First, the fact that $p$ fields have nonminimal charges means they 
describe a gerbe, not a space, and physics sees gerbes as multiple covers
\cite{ps4}.  So, at minimum, we have a double cover of ${\bf P}^3$,
not ${\bf P}^3$ itself.  
Furthermore, the mass terms degenerate along a degree 8 
divisor, which defines a branch locus for the double cover.  
The conclusion is that this Landau-Ginzburg model flows to\footnote{
Well, nearly to the branched double cover.  In this particular example,
the branched double cover has singularities, mathematically, but the
GLSM does not have any noncompact branches.  Physics actually sees
a ``noncommutative'' resolution of the branched double cover,
as described in \cite{cdhps}.
}
a NLSM on a branched double cover of ${\bf P}^3$,
branched over a degree 8 locus, which is a Calabi-Yau,
known as Clemens' octic double solid.

In any event, we see here that the ``hybrid Landau-Ginzburg'' phases
appearing in GLSM K\"ahler moduli spaces are precisely 
Landau-Ginzburg models over stacks which are total spaces of
vector bundles on gerbes.
These bundles on gerbes will be described in greater detail in
\cite{metonytoappear}; however, for the moment this should help the
reader have a more precise understanding of the nature of the theories
appearing in various limits of GLSMs.

We shall return to the analysis of Landau-Ginzburg models on stacks
in future work.

\section{Conclusions}

After a review of B-twisted Landau-Ginzburg models on general spaces,
we described A-twisted Landau-Ginzburg models.  There are at least two
different notions of A twist; we focus on one in particular.  We check our
methods by using pairs of Landau-Ginzburg models and NLSMs that are in the
same universality class, and so should have the same topological subsector.
We find that the Landau-Ginzburg computations give matching results, though
the details of the computations differ.  Our methods give (the first)
physical realizations of some old mathematical
tricks from the Gromov-Witten literature.
Finally, we briefly
outline how the hybrid Landau-Ginzburg models appearing at limits of GLSMs
are Landau-Ginzburg models on stacks, though the bulk of the analysis is
left for future work.

One direction that would be interesting to explore would be matrix
factorizations in general B-twisted Landau-Ginzburg models.  For example,
in a Landau-Ginzburg model over the total space of a vector bundle ${\cal
E} \rightarrow X$, with suitable superpotential, the matrix factorizations
should be equivalent to sheaf theory on the zero locus of the section
defining the superpotential, as discussed earlier.  Furthermore,
Born-Oppenheimer analyses of matrix factorizations should imply that matrix
factorizations behave well in families.  To the best of our knowledge,
these physically-motivated statements about matrix factorizations have not
yet been proven mathematically.

In this spirit, one also wonders if Landau-Ginzburg descriptions of
universality classes containing NLSMs could shed any light on Kuznetsov's
homological projective duality \cite{kuz1,kuz2,kuz3}.  This duality is
conjectured \cite{cdhps,ps6} to describe the relationship between different
K\"ahler ``phases'' of GLSMs.  Unfortunately, at present Kuznetsov's
duality is described in a fashion that is difficult to work with.  The idea
here is that by replacing NLSMs with Landau-Ginzburg models, and sheaf
theory with matrix factorizations, one might be able to get a more nearly
symmetric description of the K\"ahler phases of GLSMs which might lend
itself to a simplified description of Kuznetsov's duality.  Furthermore,
given a full understanding of Landau-Ginzburg models on stacks, it should
be possible to compute Gromov-Witten invariants of the noncommutative
spaces that sometimes arise as homological projective duals by computing
A-twisted Landau-Ginzburg model correlation functions.

One direction that would be very interesting to explore would be whether
these ideas could assist in physical constructions of mirror-symmetric
theories.  The methods described in the paper \cite{horivafa}, for example,
produce physical theories in which the A twist of one is equivalent to the
B twist of the other, but, which are usually {\it not} the same CFT, and
hence are not mirrors in the full physics sense.  One might speculate that
by replacing NLSMs with Landau-Ginzburg models in the same universality
class, one might be able to recast mirror symmetry as a relationship
between different Landau-Ginzburg models in the same universality class.
(This was the spirit of the work \cite{daveronen2}, though with GLSMs
instead of Landau-Ginzburg models.) One might further speculate that a
modified version of \cite{horivafa} might produce physical theories which
do share the same CFT.

\section{Acknowledgments}

We would like to thank P.~Clarke, R.~Donagi, T.~Jarvis,
S.~Katz, E.~Witten, 
and especially I.~Melnikov and T.~Pantev for useful conversations.
In particular, we would like to thank I.~Melnikov for initial collaboration
and many useful conversations, and T.~Pantev for providing the
hypercohomology computations used in this paper.
J.G. was partially supported by NSF grant DMS-02-44412.
E.S. was partially supported by NSF grant DMS-0705381.

\appendix

\section{Alternate A twist}   \label{alta}

In section~\ref{atwiststart} we mentioned that, because the Yukawa
terms break worldsheet Lorentz invariance of the naive A twist,
there are multiple notions of A twist one could apply to Landau-Ginzburg
models.

In this appendix, we are going to outline a different notion of A twist
than what we have used in the vast majority of this paper.
The alternative twist we shall discuss in this section
was first used in another
context by Witten in \cite{ed4dsym}.
There, he was considering four-dimensional super-Yang-Mills theories
on four-manifolds.  Starting with an ${\cal N}=2$ theory, he would break
the ${\cal N}=2$ to ${\cal N}=1$ by adding a mass term for the adjoint-valued
scalar, and then topologically twist the ${\cal N}=1$ theory.
Unfortunately, the topological twist of the ${\cal N}=1$ theory was
incompatible with the mass term, in exactly the same sense as we
see here in our attempt to A twist Landau-Ginzburg models.

Witten's solution to this problem was to promote the superspace integral
to a section of the canonical bundle.  The original mass terms appeared
in the action multiplied by a section of the canonical bundle, which restored
Lorentz invariance.  Away from the zeroes of that section, one had
an ${\cal N}=1$ theory, which developed a gap in the IR, but along the
zero locus of that distinguished section, the ${\cal N}=1$ was
effectively restored to ${\cal N}=2$, giving rise to a `cosmic string'
in the four-dimensional gauge theory. 

The two-dimensional analogue of Witten's solution is to
multiply the superpotential terms by a section $\omega$ of the
worldsheet canonical bundle, effectively leading to 
worldsheet-position-dependent mass terms.
In this section we will describe the resulting action and some of its
features in greater detail.

First, let us enumerate below which bundles the fermions couple to,
and their new names, after the A twist in the NLSM:
\begin{align*}
\psi_+^i ( \equiv \chi^i ) \: &\in \: \Gamma_{C^{\infty}}\left( \phi^* T^{1,0}X \right)                                                                   & \psi_-^i ( \equiv \psi_{\overline{z}}^i ) \: &\in \: \Gamma_{C^{\infty}}\left( \overline{K}_{\Sigma} \otimes (\phi^* T^{0,1}X)^{\vee} \right) \\
\psi_+^{\overline{\imath}} ( \equiv \psi_z^{\overline{\imath}} ) \: &\in \: \Gamma_{C^{\infty}}\left( K_{\Sigma} \otimes (\phi^* T^{1,0}X)^{\vee} \right) & \psi_-^{\overline{\imath}} ( \equiv \chi^{\overline{\imath}} ) \: &\in \: \Gamma_{C^{\infty}}\left( \phi^* T^{0,1} X \right).
\end{align*}
Thus, in order to make the superpotential Yukawa terms Lorentz-invariant
after the topological twist, we must multiply the $\psi_+^i \psi_-^i$
term by a holomorphic section of $K_{\Sigma}$,
and the $\psi_+^{\overline{\imath}} \psi_-^{\overline{\jmath}}$ term by
a holomorphic section of $\overline{K}_{\Sigma}$.
(We are required to use a holomorphic section, not merely a meromorphic
section, in order for the supersymmetry transformations to close.
As a result, this twist can not be done on ${\bf P}^1$, but rather
only on Riemann surfaces of genus $g \geq 1$.)
Let $\omega$ denote a holomorphic section of $K_{\Sigma}$,
then following Witten's analysis in \cite{ed4dsym}, we replace
\begin{displaymath}
\int d^2 \theta W(\Phi) \: \mapsto \:
\omega \wedge \int d^2 \theta W(\phi).
\end{displaymath}

After making this change, the action has the form
\begin{eqnarray*}
\lefteqn{ 
\frac{1}{\alpha'} \int_{\Sigma} d^2z \left(
 g_{\mu \nu} \partial \phi^{\mu} \overline{\partial} \phi^{\nu}
\: + \: i B_{\mu \nu} \partial \phi^{\mu} \overline{\partial}
\phi^{\nu} \: + \:
\frac{i}{2} g_{\mu \nu} \psi_-^{\mu} D_z \psi_-^{\nu} \: + \:
\frac{i}{2} g_{\mu \nu} \psi_+^{\mu} D_{\overline{z}} \psi_+^{\nu}
\: + \:
R_{i \overline{\jmath} k \overline{l}} \psi_+^i \psi_+^{\overline{\jmath}} 
\psi_-^k
\psi_-^{\overline{l}}
\right. } \\
& & \hspace*{2.0in}  \left. 
2 \left( \omega \wedge \overline{\omega} \right)
g^{i \overline{\jmath}} \partial_i W \partial_{\overline{\jmath}}
\overline{W} \: + \: \omega \chi^i \psi_{\overline{z}}^j 
D_i \partial_j W \: + \:
\overline{\omega} 
\psi_z^{\overline{\imath}} \chi^{\overline{\jmath}} D_{\overline{\imath}}
\partial_{\overline{\jmath}} \overline{W} \right),
\end{eqnarray*}
with supersymmetry transformations
\begin{eqnarray*}
\delta \phi^i & = & i \alpha_- \psi_+^i \: + \: i \alpha_+ \psi_-^i \\
\delta \phi^{\overline{\imath}} & = & i \tilde{\alpha}_- \psi_+^{
\overline{\imath}} \: + \: i \tilde{\alpha}_+ \psi_-^{\overline{\imath}} \\
\delta \psi_+^i \: = \: \delta \chi^i & = & - \tilde{\alpha}_- \partial \phi^i 
\: - \:
i \alpha_+ \psi_-^j \Gamma^i_{j m} \psi_+^m \: - \:
i \alpha_+ g^{i \overline{\jmath}} \partial_{\overline{\jmath}} \overline{W}
\overline{\omega}\\
\delta \psi_+^{\overline{\imath}} \: = \: \delta \psi_z^{\overline{\imath}}
& = & - \alpha_- \partial 
\phi^{\overline{\imath}}
\: - \: i \tilde{\alpha}_+ \psi_-^{\overline{\jmath}} 
\Gamma^{\overline{\imath}}_{\overline{\jmath} \overline{m} }
\psi_+^{\overline{m}} 
\: - \: i \tilde{\alpha}_+ g^{\overline{\imath} j} \partial_j W \omega \\
\delta \psi_-^i \: = \: \delta \psi_{\overline{z}}^i 
& = & - \tilde{\alpha}_+ \overline{\partial} \phi^i \: - \:
i \alpha_- \psi_+^j \Gamma^i_{j m} \psi_-^m 
\: + \: i \alpha_- g^{i \overline{\jmath}} \partial_{\overline{\jmath}}
\overline{W} \overline{\omega}\\
\delta \psi_-^{\overline{\imath}} \: = \: \delta \chi^{\overline{\imath}}
& = & - \alpha_+ \overline{\partial}
\phi^{\overline{\imath}} \: - \: 
i \tilde{\alpha}_- \psi_+^{\overline{\jmath}} 
\Gamma^{\overline{\imath}}_{\overline{\jmath} \overline{m}}
\psi_-^{\overline{m}} 
\: + \: i \tilde{\alpha}_- g^{\overline{\imath} j} \partial_j W \omega.
\end{eqnarray*}
A careful reader will note that in order for the $\tilde{\alpha}_+$
and $\alpha_-$ transformations to close, we must require
$\overline{\partial} \omega = 0$ -- so $\omega$ must be a holomorphic
section of the canonical bundle, not just a meromorphic section.
In addition, in the $\alpha_+$, $\tilde{\alpha}_-$
transformations, there is a potential difficulty, in that for the 
transformations to close, one must commute $\partial$ past $\omega$
and $\tilde{\alpha}_-$, which are neither constants nor antiholomorphic,
and $\overline{\partial}$ past $\overline{\omega}$ and $\alpha_+$,
which are neither constant nor holomorphic.
The fix is that 
$\alpha_+ \in \overline{\Gamma}\left( \overline{K}_{\Sigma}^{-1}\right)$,
$\tilde{\alpha}_- \in \Gamma\left( K_{\Sigma}^{-1} \right)$,
and in the relevant terms, one has the products $\omega \tilde{\alpha}_-$,
$\overline{\omega} \alpha_+$, so that for suitable choices,
$\omega \tilde{\alpha}_-$ and $\overline{\omega} \alpha_+$ are constant.

Defining $\alpha = \alpha_-$,
$\tilde{\alpha} = \tilde{\alpha}_+$, we find that the BRST transformations
of the fields are given by
\begin{eqnarray*}
\delta \phi^i & = & i \alpha \chi^i \\
\delta \phi^{\overline{\imath}} & = & i \tilde{\alpha} \chi^{
\overline{\imath}} \\
\delta \chi^i & = & 0 \\
\delta \chi^{\overline{\imath}} & = & 0 \\
\delta \psi_{\overline{z}}^i & = & 
- \tilde{\alpha} \overline{\partial} \phi^i \: - \: i \alpha
\chi^j \Gamma^i_{j m} \psi_{\overline{z}}^m \: + \: i \alpha
g^{i \overline{\jmath}} \partial_{\overline{\jmath}}\overline{W} 
\overline{\omega}\\
\delta \psi_z^{\overline{\imath}} & = & 
- \alpha \partial \phi^{\overline{\imath}} \: - \:
i \tilde{\alpha} \chi^{\overline{\jmath}} \Gamma^{\overline{\imath}}_{
\overline{\jmath} \overline{m}} \psi_z^{\overline{m}} \: - \:
i \tilde{\alpha} g^{\overline{\imath} j} \partial_j W \omega.
\end{eqnarray*}

\section{A hypercohomology computation}   \label{hypbmodel}

In this appendix we will argue a result on hypercohomology that was
used in section~\ref{genlbtwiststudy}.  We would like to thank R.~Donagi
and T.~Pantev for providing the argument which we
repeat here.

Let $X$ be a variety and $E$ an algebraic vector bundle over $X$ of rank $n$.
Let $\alpha \in H^{0}(X,E)$ be a section whose          
scheme-theoretic zero locus $Y$ is smooth.
We want to describe
the                                                                             
hypercohomology of the complex $(\wedge^{\bullet}E^{\vee},i_{\alpha})$ on       
$X$ in                                                                          
terms of data on $Y$.
The answer is as follows:

\begin{quotation}
Let $k = \dim Y$, $N \to Y$ be the normal bundle of $Y$ in    
$X$, and let $i_{Y} :Y \hookrightarrow X$ be the embedding map. Then there      
is                                                                              
a natural injective map                                                         
\[                                                                              
i : N \: \longrightarrow \: E_{|Y}                                                                
\]                                                                              
and if we write $F \equiv E_{|Y}/N$ for the quotient bundle, then the 
restriction to                                                                  
$Y$ followed by projection to $F$ gives a quasi-isomorphism                     
\[                                                                              
(\wedge^{\bullet}E^{\vee},i_{\alpha}) \: \cong \: 
i_{Y*}(\wedge^{\bullet}F^{\vee},0)\otimes                                       
\wedge^{n-k}N^{\vee}.                                                           
\]                    
\end{quotation}

In particular if we apply this claim to the situation $E = \Omega^{1}_{X}$      
and                                                                             
$\alpha = dW$ we get an isomorphism                                             
\[                                                                              
{\bf H}^{i}(X,(\wedge^{\bullet}T_{X},i_{dW})) \: \cong \: \oplus_{a+b = i}         
H^{a}(Y,\wedge^{b}T\otimes \wedge^{n-k}N^{\vee}),
\]                                                                              
which is the result cited in section~\ref{genlbtwiststudy}.                                                        
                                                                                
Next we shall prove the claim.  First let us describe the 
map     
$i$.                                                                            
It is made out of derivatives of the section $\alpha$ and it can be described  
in                                                                           
local coordinates. The slick way of doing this is to use the jet bundle
$J^{1}(E)$                                                                      
of $E$.                                                                         
The section $\alpha$ of $E$ has a first jet $j^{1}(\alpha)$ which is a          
section                                                                         
in $J^{1}(E)$. The first jet bundle sits in a short exact sequence              
\[                                                                              
0 \: \longrightarrow \: E\otimes \Omega^{1}_{X} \:
\longrightarrow \:  J^{1}(E) \: \longrightarrow \:
 E \: \longrightarrow \:  0                          
\]                                                                              
and $j^{1}(\alpha)$ maps to $\alpha$. 
So when we restrict to $Y$ we get that    
$j^{1}(\alpha)_{|Y}$ maps to zero in $E_{|Y}$ and so is a section in            
$E_{|Y}\otimes \Omega^{1}_{X|Y}$, in fact a section in the subbundle         
$E_{|Y}\otimes N^{\vee}$. 
This section gives the desired map $i : N \to         
E_{|Y}$                                                                         
and the assumption that $Y$ is smooth guarantees that $i$ is injective.

As an aside, note that in the case of
main interest $E = \Omega^{1}_{X}$,     
$\alpha = dW$, the jet $j^{1}(\alpha)$ is just the Hessian of $W$. In fact      
in                                                                              
this case the subvariety $Y$ has virtual dimension zero, and we have a          
symmetric perfect obstruction theory for it given by $T_{X} \to                 
\Omega^{1}_{X}$                                                                 
where the map is the contraction with the Hessian of $\alpha$. When          
restricted 
to $Y$, the image of this map is precisely the normal bundle to $Y$, sitting    
inside $\Omega^{1}_{X|Y}$. 
This also gives an isomorphism of       
the                                                                             
normal and the conormal bundle of $Y$.                    

Returning to the proof, we have a natural isomorphism of
complexes                                                                       
\[                                                                              
(\wedge^{\bullet}E^{\vee},i_{\alpha})\otimes \wedge^{n}E \: \cong \:
(\wedge^{\bullet}E,\alpha\wedge).                                               
\]                                                                              
The restriction to $Y$ map composed with projection to $F$ gives a map of 
complexes                                                                       
\[                                                                              
(\wedge^{\bullet}E,\alpha\wedge) \: \longrightarrow \:
 i_{Y*}(\wedge^{\bullet}F,0).               
\]                                                                              
Now tensor this map with $\wedge^{n}E^{\vee}$ and note that                     
$\wedge^{n}E^{\vee}_{|Y} = \wedge^{n-k}N^{\vee} \otimes \wedge^{k}F^{\vee}$.    
Thus we get a map                                                               
\[                                                                              
r : (\wedge^{\bullet}E^{\vee},i_{\alpha}) \: \longrightarrow \:
i_{Y*}(\wedge^{\bullet}F^{\vee},0)\otimes \wedge^{n-k}N^{\vee}                  
\]                                                                              
The claim is that $r$ is quasi-isomorphism. 
This is easy to check since it      
is a                                                                            
local question. 
Locally on $X$ we can split $E$ as a direct sum of vector       
bundles $E \cong V\oplus Q$, where $V$ has rank $k$, $Q$ has rank $n-k$, the    
section $\alpha$ of $E$ is actually a section of $V$ and as a section of $V$    
is                                                                              
regular. 
Then $F = Q_{|Y}$ and the Koszul complex                               
$(\wedge^{\bullet}E^{\vee},i_{\alpha})$ is isomorphic to a tensor product of    
the                                                                             
Koszul complexes $(\wedge^{\bullet}V^{\vee},i_{\alpha})$ and                    
$(\wedge^{\bullet}Q^{\vee},0)$.                                                 
The map $r$ is just coming from the restriction map                             
\[                                                                              
(\wedge^{\bullet}V^{\vee},i_{\alpha}) \: \longrightarrow \:
 i_{Y*}\mathcal{O}_{Y}                 
\]                                                                              
which by regularity of $\alpha$ as a section of $V$ is a quasi-isomorphism.

\end{document}